\documentclass[12pt,preprint]{aastex}
\shorttitle{Jet Break Candidates in X-ray and Optical Afterglow Lightcurves
}\shortauthors{Liang et al.} \slugcomment{Accepted for publication in ApJ}
\begin{document}

\title{A Comprehensive Analysis of {\em Swift}/XRT Data:
\\ III. Jet Break Candidates in X-ray and Optical Afterglow Lightcurves}
\author{En-Wei Liang\altaffilmark{1,2}, Judith L. Racusin \altaffilmark{3}, Bing Zhang\altaffilmark{1}, Bin-Bin
  Zhang\altaffilmark{1,4}, David N. Burrows\altaffilmark{3}
} \altaffiltext{1}{Department of Physics and Astronomy, University of Nevada, Las
Vegas, NV 89154; lew@physics.unlv.edu} \altaffiltext{2}{Department of Physics,
Guangxi University, Nanning 530004, China} \altaffiltext{3}{Department of
Astronomy and Astrophysics, Pennsylvania State University, University Park, PA
16802.}\altaffiltext{4}{National Astronomical Observatories/Yunnan Observatory,
CAS, Kunming 650011, China}

\begin{abstract}
The {\em Swift}/XRT data of 179 GRBs (from 050124 to 070129) and the optical
afterglow data of 57 pre- and post-{\em Swift} GRBs are analyzed, in order to
investigate jet-like breaks in the afterglow lightcurves. Using progressively
rigorous definitions of jet breaks, we explore whether the observed breaks in the
X-ray and optical lightcurves can be interpreted as a jet break, and their
implications in understanding GRB energetics if these breaks are jet breaks. We
find that not a single burst can be included in the ``Platinum'' sample, in which
the data satisfy all the criteria needed to define a jet break, i.e. a clear
achromatic break observed in both the X-ray and optical bands, and that the pre-
and post-break decay segments satisfy the closure relations in the same, simplest
jet model. However, by releasing one or more requirements to define a jet break,
some jet-break candidates of various degrees could be identified. In the X-ray
band, 42 out of the 103 well-sampled X-ray lightcurves have a decay slope of the
post-break segment $\gtrsim 1.5$ (``Bronze'' sample), and 27 of them also satisfy
the closure relations of the forward shock models for both the pre- and post-
break segments (``Silver'' sample). The numbers of the ``Bronze'' and ``Silver''
candidates in the optical lightcurves are 27 and 23, respectively. Thirteen
bursts have well-sampled optical and X-ray lightcurves, but only seven cases are
consistent with an achromatic break, but even in these cases only one band
satisfies the closure relations (``Gold'' sample). The breaks in other GRBs are
all chromatic. The observed break time in the XRT lightcurves is statistically
earlier than that in the optical bands. All these raise great concerns in
interpreting jet-like breaks as jet breaks and further inferring GRB energetics
from these breaks. On the other hand, if one {\em assumes} that these breaks are
jet breaks, one can proceed to perform a similar analysis as previous work to
study GRB collimation and energetics. We have performed such an analysis with the
``Silver'' and ``Gold'' jet break candidates. We calculate the jet opening angle
($\theta_j$) and kinetic energy ($E_K$) or their lower limits with the ISM
forward shock models using the X-ray afterglow data. The derived $E_K$
distribution reveals a much larger scatter than the pre-{\em Swift} sample. A
tentative anti-correlation between $\theta_j$ and $E_{\rm K,iso}$ is found for
both the pre-{\em Swift} and {\em Swift} GRBs, indicating that the $E_K$ could
still be quasi-universal, if the breaks in discussion are indeed jet breaks.
\end{abstract}


\section{Introduction\label{sec:intro}}
{\em Swift}, a multi-wavelength gamma-ray burst (GRB) mission (Gehrels et al.
2004), has led to great progress in understanding the nature of the GRB
phenomenon (see recent reviews by M\'{e}sz\'{a}ros 2006; Zhang 2007). One
remarkable advance from {\em Swift} is that the on-board X-ray telescope (XRT;
Burrows et al. 2005a) has established a large sample of X-ray lightcurves from
tens of seconds to days, sometimes even months (e.g. GRB 060729, Grupe et al.
2006) after the GRB triggers, and revealed a canonical X-ray lightcurve that is
composed of four successive power-law decaying segments (Zhang et al. 2006;
Nousek et al. 2006; O'Brien et al. 2006a) with superimposing erratic flares
(Burrows et al. 2005b). These segments include a GRB tail segment (Segment 1,
with a decay slope\footnote{Throughout, we use the convention that the X-ray flux
evolves as $f\propto t^{-\alpha}\nu^{-\beta}$, where $\alpha$ is the decay slope,
$\beta$ is the spectral index, and the subscript of $\alpha$ and $\beta$ marks
the segment of the lightcurve.} $\alpha_1>2$), a shallow decay segment (Segment
2, $\alpha_2<0.75$), a normal decay segment (Segment 3, $\alpha_3\sim 1$), and a
jet-like decay segment (Segment 4, $\alpha_4>1.5$). The GRB tail and the shallow
decay segment are usually seen in the XRT lightcurves (O'Brien et al. 2006a;
Liang et al. 2006; Willingale et al. 2007; Zhang et al. 2007c, hereafter Paper I;
Liang et al. 2007, hereafter Paper II). The jet-like decay segment, however, has
occasionally been observed, but only for a small fraction of bursts (Burrows \&
Racusin 2007; Covino et al. 2006). Some bursts were observed with {\em Swift}/XRT
and/or {\em Chandra} for weeks and even months after the GRB triggers, with no
evidence of detecting a jet break in their X-ray lightcurves (Grupe et al. 2006;
Sato 2007).

The jet models had been extensively studied in the pre-{\em Swift} era (e.g.,
Rhoads 1999, Sari et al. 1999; Huang et al. 2000; see reviews by M\'{e}sz\'{a}ros
2002; Zhang \& M\'{e}sz\'{a}ros 2004; Piran 2005). An achromatic break is
expected to be observed in multi-wavelength afterglow lightcurves at a time when
the ejecta are decelerated by the ambient medium down to a bulk Lorentz factor
$\sim 1/\theta_j$, where $\theta_j$ is the jet opening angle (Rhoads 1999; Sari
et al. 1999). Most GRBs localized in the pre-{\em Swift} era with deep and long
optical monitoring have a jet-like break in their optical afterglow lightcurves
(see Frail et al. 2001; Bloom, Frail, \& Kulkarni 2003; Liang \& Zhang 2005 and
the references therein), but the achromaticity of these breaks was not confirmed
outside of the optical band. Panaitescu (2007) and Kocevski \& Butler (2007)
studied the jet breaks and the jet energy with the XRT data. However, the lack of
detection of a jet-like break in most XRT lightcurves challenges the jet models,
if both the optical and X-ray afterglows are radiated by the forward shocks.
Multiwavelength observational campaigns raise the concerns that some jet-break
candidates may not be achromatic (Burrows \& Racusin 2007; Covino et al. 2006,
cf. Dai et al. 2007; Curran et al. 2007). Issues regarding the nature of previous
``jet breaks'' have been raised (e.g. Zhang 2007).

The observational puzzles require a systematical analysis on both the X-ray and
the optical data. This is the primary goal of this paper. We analyze the {\em
Swift}/XRT data of 179 GRBs (from 050124 to 070129) and the optical afterglow
data of 57 pre-{\em Swift} and {\em Swift} GRBs, in order to systematically
investigate the jet-like breaks in the X-ray and optical afterglow lightcurves
(\S 2). We measure a jet break candidate from the data with a uniform method and
grade the consistency of these breaks with the forward shock models (\S 3), then
compare these breaks observed in the X-ray and optical lightcurves (\S 4).
Assuming that these breaks are real jet breaks, we revisit the GRB jet energy
budget (Frail et al. 2001; Bloom et al. 2003; Berger et al. 2003) with the
conventional jet models (\S 5). Conclusions and discussion are presented in \S 6.
Throughout this paper the cosmological parameters $H_0 = 71$ km s$^{-1}$
Mpc$^{-1}$, $\Omega_M=0.3$, and $\Omega_{\Lambda}=0.7$ are adopted.

\section{Data\label{sec:data}}
The XRT data are taken from the {\em Swift} data archive. We have developed a
script to automatically download and maintain all the XRT data. The {\em HEAsoft}
packages, including {\em XSPEC}, {\em XSELECT}, {\em XIMAGE}, and {\em Swift}
data analysis tools, are used for the data reduction. We have developed an IDL
code to automatically process the XRT data for a given burst in any
user-specified time interval. For details of our code please see Papers I and II.

We process all the XRT data (179 bursts) observed between 2005 January and 2007
January with our tools. We are only concerned with the power-law afterglow
segments 2, 3, \& 4 without considering the steep decay segment (1) and the
flares in the lightcurves. Since the flares are generally superimposed upon the
underlying afterglows (Chincarini et al. 2007) and their spectral properties are
different from those of the power-law decaying afterglows (Falcone et al. 2007),
we do not consider the afterglow phases with significant flares. First, we
inspect the XRT lightcurve of each burst and specify the time interval(s) that we
use to derive the spectral and temporal properties. Then, we fit the lightcurve
in this time interval with a power-law-like model as presented below. We regard
that a lightcurve in the specified time interval does not have significant
flares, if the reduced $\chi^2$ of the power law fits is less than 2. We obtain a
sample of 103 XRT lightcurves that have a good temporal coverage without
significant flares.

We fit the lightcurve in the specified time interval to derive the decay slopes
of the three segments and the two breaks, $t_{b,1}$ (the shallow to normal
transition break, possibly due to cessation of energy injection in the forward
shock; Paper II) and $t_{b,2}$ (the normal to steep transition break, possibly a
jet break). Physically, these breaks should be smooth (e.g., Panaitescu \&
M\'{e}sz\'{a}ros 1999; Moderski et al. 2000; Kumar \& Panaitescu 2000; Wei \& Lu
2000). As shown in Paper II, the energy injection break is usually seen in the
XRT lightcurves, and a smoothly broken power law (SBPL) model fits most XRT
lightcurves well, which is defined as
\begin{equation}\label{SBPL}
f=f_0\left[\left(\frac{t}{t_{\rm
b,1}}\right)^{\omega_1\alpha_2}+\left(\frac{t}{t_{\rm
b,1}}\right)^{\omega_1\alpha_3}\right]^{-1/\omega_1},
\end{equation}
where $\omega_1$ describes the sharpness of the break at $t_{\rm b,1}$, with a
larger value corresponding to a sharper break. If the jet-like decay segment is also
observed, the lightcurve break near $t_{b,2}$ evolves as
\begin{equation}\label{PL}
f_j=f_0\left[\left(\frac{t_j}{t_{\rm
b,1}}\right)^{\omega_1\alpha_2}+\left(\frac{t_j}{t_{\rm
b,1}}\right)^{\omega_1\alpha_3}\right]^{-1/\omega_1}
\left(\frac{t}{t_{b,2}}\right)^{-\alpha_4}.
\end{equation}
Therefore, a three-segment XRT afterglow lightcurve should be fitted with a
smoothed triple power law (STPL) model,
\begin{equation}\label{STPL}
F=(f^{-\omega_2}+f_j^{-\omega_2})^{-1/\omega_2}
\end{equation}
where $\omega_2$ is the sharpness factor of the jet break at $t_{b,2}$. At $t\ll
t_{\rm b, 1}$, the lightcurve is dominated by the shallow decay phase, $F\sim
(t/t_{\rm b,1})^{-\alpha_2}$, and at $t\gg t_{\rm b, 2}$, the lightcurve decays
as $F\sim (t/t_{\rm b,2})^{-\alpha_4}$. As shown in Paper II, $t_{b,1}$ and
$t_{b,2}$ are not significantly affected by $\omega_1$ and $\omega_2$, but
$\alpha_3$ is. The normal decay segment can be smeared by both the pre- and post-
segments if $\omega_1$ and $\omega_2$ are small ($< 1$). We find that
$\omega_1=\omega_2=3$ can well identify the breaks in the lightcurves. Taking the
XRT lightcurve for GRB 060814 as an example, in Fig.1 we compare the fit curve of
the STPL model with a simpler fit by a joint triple power law (JTPL) model, which
is defined as
\begin{equation}
\label{JPL} F=f_0\cases{ t^{-\alpha_2},
            & $t<t_{b,1}$, \cr
 t_{b,1}^{\alpha_3-\alpha_2} t^{-\alpha_3},
            & $t_{b,1}\leqslant t\leqslant t_{b,2}$, \cr
 t_{b,1}^{\alpha_3-\alpha_2}t_{b,2}^{\alpha_4-\alpha_3}t^{-\alpha_4},
            & $t>t_{b,2}$. \cr
            }
\end{equation}
We find that the breaks at $\sim3.4$ ks and $\sim40$ ks are well identified in
both models, and the results are consistent with each other.
On the other hand, the STPL model is smooth
without sharp breaks (Fig.1), coinciding more within the physical context of
these breaks. The fitting result of the JTPL model strongly depends on the
initial values of the two breaks. The results may be misleading, especially when
the normal decay phase lasts only a very short time (in log-scale). Therefore, we
use the STPL model and fix $\omega_1=\omega_2=3$ throughout this analysis.

The jet break signature may not be obvious, therefore we use the following
strategy to select the best model among the STPL, SBPL, and single power law
(SPL) models to fit the XRT lightcurves. In the sense of Occam's Razor, the
simplest model should be adopted. On ther other hand,
in order to avoid missing a jet break in the
lightcurves, we accept a fit model as the best one when the derived breaks are
sufficiently constrained by the data (i.e. $\delta t_b<t_b$, where $\delta t_b$
is the fitting error of $t_b$, even if the $\chi^2$ is not significantly improved
when compared to a simpler model). We thus first fit the lightcurves with the
STPL model (Eq. [\ref{STPL}]). This model is a reasonable fit to all of the
lightcurves. In case of $\delta t_b<t_b$, we suggest that such a lightcurve has
three segments and we adopt the STPL model fit. We find that only 6 lightcurves
satisfy this criterion (see Table 1). We fit the remaining lightcurves with the
SBPL model (Eq. [\ref{SBPL}]), and similarly we examine whether or not $t_{b,1}$
is sufficiently constrained. The SBPL fits are adopted for 78
lightcurves. We fit the remaining lightcurves (26 bursts) with the SPL model.
Please note that, as shown in Paper II, the sharp breaks in GRBs 060413, 060522,
060607A, and 070110 are possibly not of external origin (see also Troja et al.
2007). We do not include these sharp breaks in this analysis. GRB 060522 and
070110 have a normal decay segment after an abnormally sharp lightcurve break,
therefore we fit this post-break region to a simple power law. GRB 061202 shows
significant spectral evolution throughout its lightcurve, so we do not consider
this burst either. Our full resulting fits are summarized in Table 1. Using the
time intervals defined by the fitting results, we extract the spectrum of each
segment, and fit it with a simple power law model with absorption by both our
Galaxy and the host galaxy. The spectral fitting results are also reported in
Table 1.

In order to compare the X-ray break candidates with the optical lightcurves, we
also perform an extensive analysis of the optical lightcurves for both pre-{\em Swift}
and {\em Swift} bursts. We search for the optical afterglow data in the literature
and compile a sample of 57 optical lightcurves that have a good temporal coverage.
These lightcurves are fit with the same strategy as that for the XRT lightcurves.
The fitting results are reported in Table 2.

\section{Jet Break Candidates in the X-Ray and Optical Lightcurves\label{sec:normal}}
A break with $\Delta \alpha \sim 1$ is predicted by the forward shock jet models.
Since it is purely due to dynamic effects, it should be achromatic with no
spectral evolution across the break, and both the pre- and post-break segments
should also be consistent with the forward shock models. As shown in Table 2, no
significant spectral evolution in the segments 3 and 4 is found for most bursts,
and the X-ray spectral index is $\sim 1$ (see also O'Brien et al. 2006b).
Assuming that both the optical and the X-ray afterglows are produced by the
forward shocks, we select jet break candidates from the results shown in Tables 1
and 2, and grade these candidates as ``Bronze'', ``Silver'', ``Gold'', and
``Platinum'' based on the consistency of data with the models. The definitions of
these grades are summarized in Table 3. A break with a post-break segment being
steeper than 1.5 is selected as ``Bronze''. It is promoted to ``Silver'', if both
pre- and post-break segments are consistent with the closure relations of the
models\footnote{Notice that the ``Bronze' and ``Silver'' samples also include
bursts that are detected in both X-ray and optical bands. We include them as
long as one band satisfies the listed criteria, even if the breaks are
chromatic.}. If multiwavelength data are consistent with an achromatic break with only
one band satisfies the jet models, a ``Silver'' Candidate is elevated to ``Gold''
candidate. If an achromatic break can be established independently at least in
two bands with both bands satisfying the jet models, this break is termed as a
``Platinum'' jet break candidate.

\subsection{``Bronze'' Jet Break Candidates}
We first select the ``Bronze'' jet break candidates from both the X-ray and
optical data shown in Tables 1 and 2. Without multiple wavelength modelling, the
closure relations between the spectral index ($\beta=\Gamma-1$) and temporal
decay slope of the GRB afterglows present an approach to verify whether or not
the data satisfy the models (see Table 1 of Zhang \& M\'{e}sz\'{a}ros 2004 and
references therein, in particular Sari et al. 1998; Chevalier \& Li 2000; Dai \&
Cheng 2001). As shown in our Tables 1 and 2, the observed X-ray and optical
spectral indices are larger than 0.5 (except for the optical data of GRB 021004),
indicating that the observed X-ray and optical afterglows are usually in the
spectral regime of $\nu_X>\max(\nu_m,\nu_c)$ (Regime I) or $\nu_m<\nu_X<\nu_c$
(Regime II), where $\nu_c$ and $\nu_m$ are the cooling frequency and the typical
frequency of the synchrotron radiation, respectively. In the standard forward
shock models, the decay slope of the pre-break segment is
$\alpha_3=(3\beta_3-1)/2$ for emission in the spectral regime I (both ISM and
wind) and $\alpha_3=3\beta_3/2$ (ISM) or $\alpha_3=(3\beta_3+1)/2$ (wind) for
emission in Regime II. After the jet break and assuming maximized sideways
expansion of jets, the lightcurve evolves as $\alpha_4=2\beta$ (spectral regime
I) or $\alpha_4=2\beta+1$ (spectral regime II). If the jet sideways expansion
effect can be negligible, the post-break decay index $\alpha_4$ is shallower,
i.e. $\alpha_4=\alpha_3+0.75$ (ISM) and $\alpha_4=\alpha_3+0.5$ (wind)
(Panaitescu 2005). The observed X-ray spectral indices are greater than 0.5.
Therefore, within the ISM forward shock jet model the decay slopes of the pre-
and post-break segments of the X-rays in the spectral regime II should be greater
than 0.75 and 1.5, respectively (even without significant sideways expansion).
The wind model (regime II) and the jet model with maximum sideways expansion
would make the slopes even steeper. We therefore pick 1.5 as the critical slope
to define the ``Bronze'' jet break sample. As shown in Tables 1 and 2, 42 breaks
of the XRT lightcurves and 27 of the optical lightcurves satisfy the ``Bronze''
jet break candidate criterion. These lightcurves are shown in Fig. 2. We
summarize the data of these breaks in Table 4. Our ``Bronze'' jet break candidate
sample is roughly consistent with that reported by Panaitescu (2007). The jet
breaks in the radio afterglow lightcurve of GRBs 970508 (Frail et al.2000) and
000418 (Berger et al. 2001) are also included in our ``Bronze'' sample.

The criterion of the ``Bronze'' jet break candidate concerns only the decay slope
of the post-break segment. We notice that the lightcurve of the normal-decaying
phase declines as $\alpha_3=(3\beta+1)/2$ in wind medium, i.e., $\alpha_3\sim 2$
for $\beta\sim 1$. As shown in Paper II, some jet-like breaks have a pre-break
segment much shallower than that expected from the jet models. A reasonable
possibility would be that they are due to the energy injection effect in the
forward shock model in the wind medium. Therefore, some ``Bronze'' jet break
candidates may be fake energy injection breaks instead.

\subsection{``Silver'' Jet Break Candidates}
We promote a ``Bronze'' jet break candidate to the ``Silver'' sample if both the
pre- and post-break segments are consistent with the models in at least one band.
The decay slope of the pre-break segment of a jet break for the bursts in our
sample should be steeper than 0.75. Fifty-two out of the 71 ``Bronze'' jet break
candidates in Table 4 agree with the ``Silver'' candidate criterion (29 in the
X-ray lightcurves and 23 in the optical light curves). The X-ray light curve of
GRB 970828 has a break at $\sim 2.2$ days, with $\alpha_{3}=1.44$,
$\alpha_{4}=2.6$, and $\beta_{X}\sim 1$ (Djorgovski et al. 2001). The X-ray
lightcurve of GRB 030329 has a break at $0.52\pm 0.05$ days, with
$\alpha_{3}=0.87\pm 0.05$, $\alpha_{4}=1.84\pm 0.07$, $\beta_{3} =1.17$,  and
$\beta_{4}=0.8\pm 0.3$ (Willingale et al. 2004). We include these two pre-{\em
Swift} GRBs in the X-ray jet break candidate ``Silver'' sample.

Figure 3 shows the distribution of these bursts in the ($\alpha, \beta$)-plane
combined with the closure relations for the models (ISM and wind medium). The
X-ray data of the ``Silver'' jet break candidates are shown in Figs. 3(a) and
3(b). It is found that the X-rays are consistent with the models in the spectral
regime I, although the decay slopes of both the pre- and post-break segments are
slightly shallower than the model predictions according to the observed spectral
indices (see also Willingale et al. 2007; Paper II). As argued in Paper II, this
may be due to the simplification of the models. Simulations considering more
realistic physical effects, such as energy transitions between different epoches
(Kobayashi \& Zhang 2007), evolution of microphysics parameters (Panaitescu et
al. 2006; Ioka et al. 2005), and jet profiles (Zhang et al. 2004; Yamazaki et al.
2006), could expand the model lines into broad bands, which could
accommodate the observational data better.

The data for the ``Silver'' jet break candidates in the optical band (15 pre-{\em
Swift} GRBs and 8 {\em Swift} GRBs) are shown in Figs. 3(c) and 3(d). Since no
time-resolved spectral analysis for the optical data is available, we take the
same spectral index for both the pre- and post-break segments. Differing from the
X-rays, the optical emission of the post-break segment is consistent with the jet
model in the spectral regime II for most bursts. However, the pre-break segment
is also shallower than that predicted by the models in this spectral regime.

\subsection{``Gold'' Jet Break Candidates}
A ``Gold'' jet break candidate requires that the break is achromatic at least in
two bands, and that the break should satisfy the criteria of a ``Silver''
candidate at least in one band. Inspecting the data in Table 4 and the
lightcurves in Fig. 2, one approximately achromatic break is observed in both
X-ray and the optical lightcurves of GRBs 030329, 050730, 050820A, 051109A, and
060605. The optical afterglows of GRBs 050525A, 060206, 060526, and 060614 are
bright, and a jet-like break is clearly observed in their optical lightcurves.
Guided by the optical breaks, some authors argued for achromatic breaks in the
XRT lightcurves of these GRBs. Without the guidance of the optical lightcurves,
one cannot convincingly argue a break in the XRT lightcurves of these GRBs, but
the data may be still consistent with the existence of an achromatic break. Both
the optical and radio data of GRB 990510 are consistent with the jet models. We
inspect the data of these bursts case by case, and finally identify 7 ``Gold''
candidates, as discussed below.
\begin{itemize}

{\item GRB 990510}: The pre-Swift jet break candidate of GRB 990510 is an
exemplar of a jet break (Harrison et al. 1999). The break is achromatic in
different colors in the optical band. The radio data post the break is also
consistent with the jet models. However, with the radio data alone, one cannot
independently claim this break (D. Frail, 2006, personal communication).
Therefore, we include this jet candidate in the ``Gold'' but not ``Platinum''
category.

{\item GRB 030329}: Its X-ray lightcurve has only five data points. Fitting with
the SBPL model shows that $\alpha_1=0.96\pm 0.56$, $\alpha_2=1.81\pm 0.05$, and
$t_{b,X}=30.6\pm 19.3$ ks with $\chi^2/{\rm dof}=4.2/1$. The break is consistent
with the closure relations, and the $t_{b,X}$ agrees with the $t_{b,O}$ within
error scope (see also Willingale et al. 2004). However, the achromaticity of this
break is somewhat questionable. The break in the X-ray lightcurve has a great
uncertainty since the fit has only one degree of freedom. On the other hand, the
fit to the SPL model yields $\alpha=1.72\pm 0.01$ and $\chi^2=9.0/3$, indicating
that the SBPL fit is required by the data. We cautiously grade this burst to the
``Gold'' sample with the caveat of sparse X-ray data in mind.

{\item GRB 050525A}: Blustin et al. (2006) fitted the XRT data of GRB 050525A and
derived a break at $13.726^{+7.469}_{-5.123}$ ks, with $\alpha_3=1.20\pm 0.03$
and $\alpha_4=1.62^{+0.11}_{-0.16}$. The break is consistent with being achromatic
with the break identified in the optical band. The issue with their fitting is that
the $\chi^2$ is too small (reduced $\chi^2_r$ = 0.50 (25 dof)). We increase the
signal-to-noise ratio of the data by rebinning the lightcurve, and fit it from
5.94ks to  157.85 ks. We find that a simple power law is the best fit to the
data, with a decay slope $1.40\pm 0.05$ ($\chi^2_r\sim 1$, 11 dof). The decay
slope is much larger than that of the normal decay segment($\sim 1$). A jet-like
break is likely embedded in the data. We thus adopt the fitting by Blustin et al.
(2006) and cautiously include this burst in the ``Gold'' sample.

{\item GRB050820A}:  Its optical lightcurve traces the XRT lightcurves after
$10^4$ seconds post-burst. An achromatic break at $\sim 4$ days post-burst is
observed. Its pre-break segments are well consistent with the models, but the
post-break segments are slightly shallower than the prediction of the jet models.
We cautiously promote this break to the ``Gold'' sample.

{\item GRB 051109A}: The break time in both the X-ray and optical lightcurves is
$\sim 25$ ks. Although the decay slopes are $\alpha_{X, 4}=1.53\pm 0.08$ and
$\alpha_{O,4}=1.42\pm 0.12$, slightly shallower than those predicted by the
no-spreading jet models according to the observed spectral index, this burst is
included in the ``Gold'' candidate sample, similar to GRB 050820A.

{\item GRB 060526}: The optical lightcurve of GRB 060526 has a significant break
at $\sim 1$ day post-burst (Dai et al. 2007). Its X-ray flux after $10^2$ ks is
very low (with a significance level of detection being lower than 3 $\sigma$).
Dai et al. (2007) suggested a jet-like break in the XRT lightcurve by considering
the contamination of a nearby source in the field of view. In all our analyses,
we do not try to identify a nearby X-ray contamination source for any GRB, so our
best fit does not reveal this jet-break within the observational error scope.
In view of the analysis of Dai et al. (2007),  we also
cautiously grade this break in the ``Gold'' category.

{\item GRB 060614}: After subtracting the contribution of the host galaxy, the
optical lightcurve of GRB 060614 shows a clear break at $104$ ks (Della Valle et
al. 2006). Mangano et al. (2007) argue that the XRT lightcurve also has a break
at this time. Fitting with our STPL cannot reveal this break. However, we note
that $\alpha_4$ in the X-ray lightcurve of this burst is $\sim 1.9$, consistent
with a post-jet-break decay slope. Although a jet-like break cannot be independently
claimed at the optical break time with the X-ray data alone, the multi-wavelength
data are still consistent with the existence of such a break, with the possibility
that the injection break time and the jet break time are close to each other.
Therefore, we agree with the suggestion
by Mangano et al. (2007) and grade this break as ``Gold''.

\end{itemize}
To be conservative, we do not include GRBs 050730, 060206, and 060605 in our
``Gold'' candidate sample, as discuss below.

\begin{itemize}

{\item GRB050730}: The break happens at $\sim 10$ ks in both the optical and
X-ray bands. The pre-break segment in both the X-ray and optical light curves is
much shallower than the forward shock model predictions. We thus do not grade
this break as a ``Gold'' candidate.

{\item GRB 060605}: A tentative break is observed at $\sim 10$ ks in both the
optical and X-ray afterglow light curves. However, this break time is uncertain
in the optical band because no data point around the break time is available. On the
other hand, the decay slope of the pre-break segment is only $\sim 0.5$.
Similarly to GRB 050730, it is not considered as a ``Gold'' candidate.

{\item GRB 060206}: Curran et al. (2007) fitted the XRT data of GRB 060206 in the
range between $4$ ks and $10^3$  ks after the GRB trigger with the SBPL model,
and reported $t_b= 22^{+2.0}_{-0.8}$ ks, and the decay slopes of pre- and
post-break segments are $1.04\pm 0.1$ and $1.40\pm 0.7$, respectively. The
reduced $\chi^2$ of the fit is 0.79 (63 dof). Fitting with the SPL model, they
got a slope of $1.28\pm 0.02$ with a reduced $\chi^2=1.0$ (65 dof). The fitting
results with the SPL model is more reliable than that of the SBPL model. Also
by checking the consistency with the models, we find that the
the power law spectral index of the WT mode data $1.26\pm 0.06$ after the break
is consistent with the ``normal decay'' phase rather than the post-jet-break phase.
For example, for $\nu_X > {\rm max}(\nu_m,\nu_c)$, the model-predicted temporal
break index in the normal decay phase is $1.39\pm 0.09$, this is well consistent
with the data. We therefore do not consider this break as a ``Gold''
jet break candidate.

\end{itemize}

\subsection{``Platinum'' Jet Break Candidates}
With our definition, a ``Platinum'' jet break should be independently claimed in
at least two bands which should be achromatic. Furthermore, the temporal decay
slopes and spectral indices in both bands should satisfy those required in the
simplest jet break models. Since the optical and the X-ray afterglows could be in
different spectral regimes (Fig. 3), their lightcurve behaviors may be different
(e.g. Sari et al. 1999). However, none of the seven ``Gold'' candidates can be
promoted to the ``Platinum'' sample due to the various issues discussed above.
For some other ``Silver'' candidates in which a prominent ``break'' is observed
in one band, the lightcurve in the other band curiously evolves independently
without showing a signature of break (see \S4.4 for more discussion). It is fair
to conclude that {\em we still have not found a textbook version of jet break
after many years of intense observational campaigns}.

\section{Comparison between the Jet Break Candidates in the X-ray and
Optical Bands}

In this section we compare the statistical characteristics of the jet break
candidates in the X-ray and optical lightcurves. Our final graded jet break
candidates are shown in Table 4. The decay slopes of the pre-break segments of
those ``Bronze'' candidates are much shallower than the prediction of the jet
models. We cannot exclude the possibility that some ``Bronze'' jet break
candidates are due to the energy injection effect in the wind medium (Paper II).
Therefore, for the following analysis, we do not include the ``Bronze'' jet break
candidates.

\subsection{Detection Fraction}
As shown above, within the 103 XRT lightcurves with a good temporal coverage, 27
have ``Silver'' or ``Gold'' jet break candidates. This fraction is 23/57 for
optical lightcurves. The detection fraction of jet break candidates in the XRT
lightcurves is significantly lower than that in the optical
lightcurves\footnote{Note that this effect may be partially due to the
observational effect. Most optical lightcurves with deep and long monitoring
during the pre-{\em Swift} era show a jet-like break. From Table 2, we find that
16 {\em Swift} GRBs have optical monitoring longer than 1 day after the GRB
triggers. Among them 6 have a ``Silver'' or ``Gold'' jet break candidate. This
fraction is smaller than that of the pre-{\em Swift} GRBs. We notice that the
sensitivity of the Swift/BAT is much higher than the pre-{\em Swift} GRB
missions. It can trigger more less energetic GRBs at higher redshifts (Berger et
al. 2005b; Jakobsson et al. 2006e). Considering the suggestion that less
energetic bursts are less beamed (Frail et al. 2001), if the breaks under
discussion are indeed jet breaks, the break times of the {\em Swift} GRBs should
be later than those of the pre-{\em Swift} ones. Due to the time dilation effect,
the observed break time of the {\em Swift} GRBs should be also systematically
later than the pre-{\em Swift} GRBs. In addition, the rate of deep follow up
observations in the optical band drops in the {\em Swift} era, because the number
of bursts is greatly increased. All these effects would contribute to the bias of
detecting jet breaks in the pre-{\em Swift} and the {\em Swift} samples.}.

\subsection{Break Time}
Figure 4 shows the distributions of $t_{j}$ and $\Delta \alpha$ in the X-ray and
optical lightcurves. The distributions of $\log t_{\rm j,X}$/s and $\log t_{\rm
j,O}$/s peak at $4.5 $ and $\sim 5.5$, respectively. The $t_{\rm j,O}$
distribution has a sharp cutoff right at the high edge of the peak, indicating
that the peak is possibly not an intrinsic feature. Since the histogram depends
on the bin size selection, we test the normality of the data set with the
Shapiro-Wilk normality test. It shows that the probability of a normal
distribution for $t_{\rm j,O}$ is $p=11.5\%$ (at 0.05 confidence level), roughly
excluding the normality of the distribution. Therefore, this peak is likely due
to an observational selection bias. By contrast, the $t_{\rm j,X}$ distribution
is log-normal. The Shapiro-Wilk normality test shows $p=79.8\%$ (at confidence
level 0.05). These results suggest that the $t_{\rm j,X}$ is systematically
smaller than $t_{\rm j,O}$ (see also Kocevski \& Butler 2007). This raises the
possibility that X-ray breaks and optical breaks may not be physically of the
same origin.

\subsection{$\Delta \alpha$}
With the closure relations of $p>2$ and assuming sideways expansion, we derive
$\Delta \alpha=(\beta+1)/2$ for the regime-I ISM model and all the wind models,
and $\Delta \alpha=\beta/2+1$ for the regime II ISM model\footnote{As show in
Fig. 3, most of the bursts (25 out of 29 bursts) are consistent with $p>2$.
Therefore we only consider the $p>2$ case.}. The observed $\beta_X$ is $\sim 1$,
hence $\Delta \alpha_X\sim 1$ or $\Delta \alpha_X\sim 1.5$. Figure 4 (right)
shows that the $\Delta \alpha_{X}$ distribution peaks at $\sim 1$, which suggests
that most X-ray afterglows are consistent with the regime II models (i.e. X-ray
is above both $\nu_m$ and $\nu_c$)\footnote{Two GRBs have a $\Delta \alpha_{X}$
greater than 1.5--- GRB 050124 ($\Delta \alpha_{X}=1.91\pm 0.96$) and GRB 051006
($\Delta \alpha_{X}=1.66\pm 0.62$), but they have large errors.}. The
$\Delta\alpha_{O}$ show a tentative bimodal distribution, with two peaks at $\sim
1$ and $\sim 1.7$, roughly corresponding to the regime I ($\nu_{O}
>\rm{max}(\nu_m,\nu_c)$) and regime II ($\nu_m<\nu_{O}<\nu_c$) ISM models,
respectively.

\subsection{Chromaticity}
Being achromatic is the critical criterion to claim a break as
a jet break. As shown above, the distribution of $t_{\rm j,X}$ is systematically
smaller than $t_{\rm j,O}$, which raises the concern of achromaticity of some of
these breaks. Monfardini et al. (2006) have raised the concern that some jet-like
breaks may not be achromatic.
We further check the chromaticity for the jet candidates case by
case. We find 13 bursts that have good temporal coverage in both  X-ray and
optical bands, with a jet break candidate at least in one band. The
results are the following.
\begin{itemize}

{\item The breaks in the X-ray and optical bands are consistent with being achromatic:
GRBs 030329, 050525A, 050820A, 051109A, 060526, and 060614.}

{\item The X-ray and optical breaks are at different epochs}: GRBs 060206 and 060210

{\item A ``Silver'' jet break candidate in the optical band, but no break in the
X-ray band}: GRBs 051111 and 060729.

{\item A ``Silver'' or ``Bronze'' jet break candidate in the X-ray band, but no
break in the optical band}: GRBs 050318 (``Silver''), 050802 (``Bronze''), and
060124 (``Silver'').
\end{itemize}
The ratio of achromatic to chromatic breaks is 6:7, indicating that the
achromaticity is not a common feature of these breaks. It is a great issue to
claim the chromatic breaks as a jet break. If both the X-ray and optical
emissions are from the forward shocks, one can rule out a large fraction (7/13)
of these jet break candidates (many are ``Silver'' candidates) as a jet break! We
indicate the achromaticity of the jet break candidates in Table 4. If the above
achromatic-to-chromatic ratio is a common value, most of the breaks without
multi-wavelength observations (marked with a ``?'' in Table 4) should be also
chromatic. A possible way out to still consider these breaks as jet breaks is to
{\em assume} that the band (either X-ray or optical) in which the break is
detected is from the forward shock, while emission from the other band is either
not from the forward shock or some unknown processes have smeared the jet break
feature from the forward shock in that band. Such a model does not explicitly
exist yet. We therefore suggest that {\em one should be very cautious to claim a
jet break, and further infer the GRB energetics from a jet break candidate}. We
are probably still a long way from understanding GRB collimation and energetics.

\section{Constraints on GRB Jet Collimation and Kinetic Energetics}
As shown above, the observed chromatic feature is not consistent with the forward
shock models, and it is risky to infer GRB collimation and energetics from these
data. On the other hand, it may be still illustrative to perform such a study by
{\em assuming} that ``Silver'' break candidates are jet breaks due to the
following reasons. First, most pre-{\em Swift} works related to jet break and GRB
energetics (Frail et al. 2001; Bloom et al. 2003; Berger et al. 2003) were
carried out with one-band data only. If multi-wavelength data were not available
for the {\em Swift} bursts, one would still confidently take the post-{\em Swift}
``Silver'' breaks as jet breaks. It is therefore valuable to study this expanded
sample and compare the results with the pre-{\em Swift} sample. Second, we notice
that there is no GRB that shows a ``Silver'' jet break candidate in both the
X-ray and optical bands but at different times. For example, although a chromatic
break is observed in both the optical and X-ray lightcurves of GRBs 060206 and
060210, the X-ray break in GRB 060206 ($\alpha_2=0.40\pm 0.05$ and
$\alpha_3=1.26\pm 0.04$) and the optical break in GRB 060210 ($\alpha_2=0.04\pm
0.22$ and $\alpha_3=1.21\pm 0.05$) are not jet break candidates. On the other
hand, the optical afterglow lightcurve of GRB 060729 show a significant jet-like
break, but its XRT lightcurve keeps decaying smoothly without a break. The
lightcurve behaviors in the optical and X-ray bands for most GRBs are also
enormously different (see also Paper II for a discussion of achromaticity of the
shallow-to-normal decay transition in many bursts). These facts suggest that the
jet-break candidates we see may indeed have a genuine origin, but we are probably
far from understanding the lightcurve behaviors of most bursts. In this section,
we {\em assume} that those ``Silver'' or ``Gold'' jet break candidates are jet
breaks, and follow the standard forward shock model to constrain jet collimation
and kinetic energy of the GRB jets.

\subsection{Models}
In the standard afterglow models, the isotropic kinetic energy $(E_{\rm K,iso})$
can be derived from the data in the normal decay phase, and the jet kinetic
energy $E_K$ can be obtained from the jet break information (e.g. Rhoads 1999;
Sari et al. 1999; Frail et al. 2001). The models depend on the power law index
$p$ of the electron distribution, the spectral regime, and the medium
stratification surrounding the bursts (M\'{e}sz\'{a}ros \& Rees 1993; Sari et al.
1998; Dai \& Lu 1998; Chevalier \& Li 2000; Dai \& Cheng 2001). As shown in Fig.
3, most bursts in our sample (25 out of 29) are consistent with $p>2$. We
therefore only consider $p>2$ in this analysis. Essentially all the data are
consistent with the ISM model, although in some bursts the wind model cannot be
confidently ruled out. On the other hand, interpreting the early afterglow
deceleration feature in GRB 060418 (Molinari et al. 2007) requires that the
medium is ISM, even at the very early time (Jin \& Fan 2007).  We therefore
consider only the ISM case in this paper.

We use the X-ray afterglow data to calculate $E_{\rm K,iso}$, following the same
procedure presented in our previous Paper (Zhang et al. 2007a), which gives
\begin{eqnarray}
E_{\rm K,iso,52} & = & \left[\frac{\nu F_\nu (\nu=10^{18}~{\rm Hz})}{5.2\times
10^{-14} ~{\rm ergs~s^{-1} ~cm^{-2}} }\right]^{4/(p+2)}D_{28}^{8/(p+2)}(1+z)^{-1}
t_d^{(3p-2)/(p+2)}\nonumber \\
& \times & (1+Y)^{4/(p+2)} f_p^{-4/(p+2)}\epsilon_{B,-2}^{(2-p)/(p+2)}
\epsilon_{e,-1}^{4(1-p)/(p+2)} \nu_{18}{^{2(p-2)/(p+2)}}
\nonumber \\
& & \  \  \  \ ({\rm Spectral\ regime\ I})\label{Ekiso1}\\
E_{\rm K,iso,52} & = &
\left[\frac{\nu F_\nu (\nu=10^{18}~{\rm Hz})}{6.5\times 10^{-13} ~{\rm
ergs~s^{-1} ~cm^{-2}} }\right]^{4/(p+3)} D_{28}^{8/(p+3)}(1+z)^{-1}
 t_d^{3(p-1)/(p+3)}\nonumber \\
& \times &f_p^{-4/(p+3)} \epsilon_{B,-2}^{-(p+1)/(p+3)}
\epsilon_{e,-1}^{4(1-p)/(p+3)} n^{-2/(p+3)} \nu_{18}{^{2(p-3)/(p+3)}}
\nonumber \\
& & \  \  \  \ ({\rm Spectral\ regime\ II})\label{Ekiso2}
\end{eqnarray}
where $\nu f_\nu(\nu=10^{18}{\rm Hz})$ is the energy flux at $10^{18}$ Hz (in
units of ${\rm ergs~s^{-1} ~cm^{-2}}$) , $z$ the redshift, $D$ the luminosity
distance, $f_p$ a function of the power law distribution index $p$ (Zhang et al.
2007a), $n$ the density of the ambient medium, $t_d$ the time in the observers
frame in days, $Y$ the inverse Compton parameter. The convention $Q_{n}=Q(\rm in\
cgs\ units)/10^{n}$ has been adopted.

If the ejecta are conical, the lightcurve shows a break when the bulk Lorentz
factor declines down to $\sim \theta^{-1}$ at a time (Rhoads 1999; Sari et al.
1999)
\begin{equation}
t_j\sim 0.5 {\rm\ days} (\frac{E_{\rm
K,iso,52}}{n})^{1/3}(\frac{1+z}{2})(\frac{\theta_j}{0.1})^{8/3}.
\end{equation}
The jet opening angle can be derived as
\begin{equation}
\theta_j\sim 0.17\left(\frac{t_j}{1+z}\right)^{3/8}\left(\frac{E_{\rm
K,iso,52}}{n}\right)^{-1/8}.\label{theta}
\end{equation}
The geometrically corrected kinetic energy is then given by
\begin{equation}
E_{\rm K,52}=E_{\rm K,iso,52}(1-\cos\theta_j)~.\label{EKJet}
\end{equation}
\subsection{Results}
Thirty {\em Swift} GRBs in our sample have redshifts available. Among them 14
bursts have a jet break candidate detection  in the optical or X-ray
afterglow lightcurves. For those bursts without jet break detections, we take the
time of the last XRT observation as the lower limit of the jet break time. We
calculate $E_{\rm K,iso}$ and $\theta_j$ (or its lower limit) for these
bursts, then derive their $E_{\rm K}$ (or lower limits). We use the normal decay
phase to identify the spectral regime for each burst using the following method.
We define
\begin{eqnarray}
D=|\alpha^{\rm obs}-\alpha(\beta^{\rm obs})|,\\
\nonumber \\
\delta=\sqrt{(\delta \alpha^{\rm obs})^2+[\delta \alpha(\beta^{\rm obs})]^2},
\end{eqnarray}
where $\alpha^{\rm obs}(\delta \alpha^{\rm obs})$ and $\alpha(\beta^{\rm obs}) $
are the temporal decay slopes (errors) from the observations and that predicted
from the closure relations using the observed $\beta$, respectively, for the
normal decay phase. The ratio $\phi=D/\delta$ reflects the nearness of the data
point to the model lines within errors. In the case of $\phi <1$, the data point
goes across the corresponding closure relation line. We derive $\phi$ from the
data for both the spectral regimes I and II. By comparing the two $\phi$ values,
we then assign each burst to the spectral regime with the smaller $\phi$. We find
that the X-rays of about two-third of the bursts are in the spectral regime I.
Eq.(\ref{Ekiso1}) shows that the calculation of $E_{\rm K,iso}$ is independent of
$n$ and only weakly depends on $\epsilon_B$ and $p$ with the data in this
spectral regime. Therefore, this spectral regime is ideal to measure $E_{\rm K}$.
The X-rays of about one-third of the bursts are in the spectral regime II. The
inferred $E_{\rm K,iso}$ in this spectral regime significantly depends on both
$\epsilon_B$ and $n$. This makes it complicated to derive $E_{\rm K,iso}$. In
this case $\epsilon_B$ or $n$ must be very small (e.g. $\epsilon_B\lesssim
10^{-3}$ or $n\sim 10^{-2}$ cm$^{-3}$) in order to have the cooling frequency
above the observed X-rays at $t\sim 1$ day while retaining a reasonable $E_{\rm
K,iso}$ (Zhang et al. 2007a). Please note that by keeping $\epsilon_e \sim 0.1$,
the $Y$ parameter does not increase significantly for a smaller $\epsilon_B$
since the Klein-Nishina correction factor $\eta_2$ parameter becomes much smaller
(Zhang et al. 2007a).

After identifying the appropriate spectral regime, we derive $p$ from the
relations between $p$ and the spectral index (Table 5). Most derived $p$'s are
greater than 2, except for GRBs 050820A, 060912, and 060926, and we assign
$p=2.01$ for these bursts. In our calculation, we fix $n=0.1$ cm$^{-3}$ (Frail et
al. 2001) and take initial values of $\epsilon_B$ and $Y$ as $10^{-4}$ and $2.7$,
respectively. We iteratively search for the maximum value of $\epsilon_B$ that
ensures the X-rays are in the proper spectral regime. Previous broadband fits and
statistical analyses suggest that $\epsilon_e$ is typically around 0.1 (Wijers \&
Galama 1999; Panaitescu \& Kumar 2002; Yost et al. 2003; Liang et al. 2004; Wu et
al. 2004)\footnote{With the observations of {\em Swift}, some authors suggested
that the microphysical parameters possibly evolve with time (Panaitescu et al.
2006; Ioka et al. 2006)}. Therefore we take $\epsilon_e=0.1$ for all the bursts.
The $E_{\rm K,iso}$ is calculated with the observed energy flux at a given time.
After the energy injection is over, $E_{\rm K,iso}$ is a constant in the scenario
of an adiabatic decelerating fireball. In principle, one can derive $E_{\rm
K,iso}$ at any time $t_d$ with Eqs.(\ref{Ekiso1}) and (\ref{Ekiso2}). We take the
flux at a time $\log t=(\log t_{b,2}+\log t_{b,1})/2$. Our results are reported
in Table 4.

We calculate $\theta_j$ and $E_{\rm K}$ for the pre-{\em Swift} GRBs with the
same method. We collect the X-ray afterglow data of the pre-{\em Swift} GRBs
from the literature. The results are shown in Table 5. Eight bursts in Table 5
are included in the sample presented by Frail et al. (2001). Assuming $n=0.1$
cm$^{-3}$ and GRB efficiency $\eta=0.2$, Frail et al. (2001) derived the jet
opening angles $\theta_j$ of these bursts with the observed gamma-ray energy.
We compare our results with theirs ($\theta_j^{'}$) in Fig. 5. They are
generally consistent with each other.
Since our calculations derive $E_{\rm K,iso}$ directly rather than assuming an
$\eta$ value, this result
indicates that the derivation of $\theta_j$ is insensitive to $\eta$, as
suggested by the ${-1/8}$ dependence of $E_{\rm K,iso}$ in Eq.(\ref{theta}).

The distributions of $E_{\rm K,iso}$ and $p$ are displayed in Fig. 6. No
significant differences between the pre-{\em Swift} and the {\em Swift} samples
are found for these parameters. The Kolmogorov-Smirnov test shows that
$p_{K-S}=0.61$ for the $E_{\rm K,iso}$ distribution and $p_{K-S}=0.81$ for the
$p$ distribution. As mentioned above, since we only consider $p>2$, the sharp
cutoff at $p=2$ is an artifact. A small fraction of bursts might have $p<2$ (such
as GRBs 050820A, 060912, and 060926), which would extend the $p$-distribution to
smaller values. No evidence for $p$-clustering among bursts is found (see also
Shen et al. 2006; Paper II). The $E_{\rm K,iso}$ distribution spans almost 3
orders of magnitude, ranging from $2\times 10^{52}$ to $1\times 10^{55}$ ergs
with a log-normal peak at $7\times 10^{53}$ ergs. The probability of the
normality is $73\%$ at 0.05 confidence level.

The $\theta_j$ and $E_K$ distributions are shown in Fig.7. A sharp cutoff at
$\theta_j\sim 1.5^{\rm o}$ is observed. The $\theta_j$ of the {\em Swift} GRBs
derived from XRT observations tends to be smaller than that of the pre-{\em
Swift} GRBs. The $E_K$ of the pre-{\em Swift} GRBs log-normally distribute around
$1.5\times 10^{51}$ with a dispersion of 0.44 dex (at $1\sigma$ confidence
level). However, the $E_K$ of the {\em Swift} GRBs randomly distribute in the
range of $10^{50}\sim 10^{52}$ ergs (see also Kocevski \& Butler 2007). We examine
the correlation between $E_{\rm K,iso}$ and $\theta_j$ in Fig. 8. A tentative
anti-correlation is found, but it has a large scatter. The best fit yields
$E_{\rm K,iso}\propto \theta_j^{-2.35\pm 0.52}$, with a linear correlation
coefficient $r=-0.66$ and a chance probability of $p\sim 10^{-4}$ (N=28). This
suggests that although $E_K$ has a much larger scatter than the
pre-{\em Swift} sample, it is still quasi-universal among bursts.

\section{Conclusions and Discussion}
We have presented a systematic analysis on the {\em Swift}/XRT data of 179 GRBs
observed between Jan., 2005 and Jan., 2007 and the optical afterglow lightcurves
of 57 GRBs detected before Jan. 2007, in order to systematically investigate the
jet-like breaks in the X-ray and optical afterglow lightcurves. Among the 179 XRT
lightcurves, 103 have good temporal coverage and have no significant flares in
the afterglow phase. The 103 XRT lightcurves are fitted with the STPL, SBPL, or
SPL model, and the spectral index of each segment of the lightcurves is derived
by fitting the spectrum with a simple absorbed power law model. The same fitting
is also made for the 57 optical light curves. We grade the jet break candidates
through examining the data with the forward shock models with ``Bronze'', ``Silver'',
``Gold'', or ``Platinum''. We show that among the 103 well-sampled XRT
lightcurves with a break, 42 are ``Bronze'', and 27
are ``Silver''. Twenty-seven out of 57 optical breaks are
``Bronze'', and 23 ``Silver''. Thirteen bursts have well-sampled lightcurves of
both the X-ray and optical bands, but only 6 cases are consistent with being
achromatic. Together with the GRB 990510 (in which an achromatic break in optical
and radio bands can be claimed, Harrison et al. 1999), we have 7 ``Gold''
jet break candidates. However, none of them
can be classified as ``Platinum'', i.e. a textbook
version of a jet break. Curiously, 7 out of the 13 jet-break candidates with
multi-wavelength data suggest a chromatic break at the ``jet break'', in contrary
to the expectation of the jet models. The detection fraction of a jet break
candidate in the XRT lightcurves is lower than that of the optical lightcurves,
and the break time is also statistically earlier. These facts suggest that one
should be very cautious in claiming a jet break and using the break information
to infer GRB collimation and energetics.

On the other hand, the possibility that some of these breaks are jet breaks is
not ruled out. The ``Silver'' and ``Gold'' jet break candidates have both the pre-
and post-break temporal decay segments satisfying the simplest jet models,
suggesting that these break are likely indeed jet breaks. In order to compare with
the previous work on jet breaks, we then cautiously assume that the breaks in
discussion are indeed jet breaks and proceed to constrain the $\theta_j$ and
$E_K$ by using the X-ray afterglow data using the conventional jet models. We
show that the geometrically corrected afterglow kinetic energy $E_K$ has a
broader distribution than the pre-{\em Swift} sample, disfavoring the standard
energy reservoir argument. On the other hand, a tentative anti-correlation
between $\theta_j$ and $E_{\rm K,iso}$ is found for both the pre-{\em Swift} and
{\em Swift} GRBs, indicating that the $E_K$ could still be quasi-universal.

The GRB jet models had been extensively studied in the pre-{\em Swift} era (e.g.,
Rhoads 1999; Sari et al. 1999; Panaitescu \& M\'{e}sz\'{a}ros 1999; Moderski et
al. 2000; Huang et al. 2000; Wei \& Lu 2000; see reviews by M\'{e}sz\'{a}ros
2002; Zhang \& M\'{e}sz\'{a}ros 2004; Piran 2005). The results of this paper
suggest that for most bursts the X-ray and optical afterglows cannot be
simultaneously explained within the simplest jet models. Data suggest that we may
be missing some basic ingredients to understand GRB afterglows. There have been
skepticism about the jet break interpretations before (e.g. Dai \& Lu 1999; Wei
\& Lu 2002a,b). The current data call for more open-minded thoughts on the origin
of lightcurve breaks (Zhang 2007). Observationally, at the epoch when the
jet-like breaks show up the flux level is typically low. Source contaminations
(e.g. GRB 060526; Dai et al. 2007) would complicate the picture. Careful analyses
are needed to claim the breaks. On the other hand, most of the curious late
afterglow break behaviors are likely not caused by these observational
uncertainties. For example, even if the contamination source is removed, the
broad band afterglow lightcurves of GRB 060526 (Dai et al. 2007) cannot be
incorporated within any simplest jet models.

As cosmic beacons extending to high redshift universe (e.g. Lamb 2000; Bromm \&
Loeb 2002; Gou et al. 2004; Lin et al. 2004), GRBs have the potential to probe
the high-$z$ universe. Using the pre-{\em Swift} jet break sample, Ghirlanda et
al. (2004a) discovered a tight correlation between the cosmic rest-frame peak
energy ($E_p$) of the GRB $\nu f_\nu$ spectrum  and the geometrically-corrected
GRB jet energy ($E_{\gamma}$). This correlation was taken as a potential standard
candle to perform cosmography studies (e.g. Dai et al. 2004; Ghirlanda et al.
2004b). Liang \& Zhang (2005) proceed with a model-independent approach, and
derived a tight correlation among three observables, $E_{iso}$, $E_p^{'}$, and
$t^{'}_{b, O}$, with the later being the cosmic rest frame optical break time
only. This correlation was also used to constrain cosmological parameters (Liang
\& Zhang 2005;  Wang \& Dai 2006). As shown in this paper, it is difficult to
accommodate both the X-ray and optical afterglow data within a unified jet model,
so that the Ghirlanda relation is not longer supported by the {\em Swift} data.
In fact, even with the optical data only, the {\em Swift} bursts make the
Ghirlanda relation more dispersed than the pre-{\em Swift} sample (Campana et al.
2007). As shown in Figs. 4 and 6, the break times in the XRT lightcurves are
significantly smaller than that in the optical lightcurves (most are pre-{Swift}
bursts), but no significant difference is observed in the $E_{iso}$ distributions
of the pre-{\em Swift} and {\em Swift} GRBs. These results tend to suggest that
the jet break candidates in the XRT lightcurves do not share the same Liang-Zhang
relation derived from the pre-{\em Swift} optical data. Since the energy band of
{\em Swift} BAT is too narrow to reliably derive $E_p$ and $E_{iso}$ for most
GRBs, it is non-trivial to test the Liang-Zhang relation rigorously. We plan to
explore this interesting question in the future.

GRBs fall into short-hard and long-soft categories (Kouveliotou et al. 1993) or
more generally Type I and Type II categories (Zhang et al. 2007b; Zhang 2006).
The progenitors of the two classes are distinctly different: Type II GRBs are
related to deaths of massive stars (Woosley \& Bloom 2006 and references
therein), and Type I GRBs are likely related to mergers of compact objects
(Gehrels et al. 2005; Fox et al. 2005; Barthelmy et al. 2005; Berger et al.
2005a). Inspecting our sample of GRBs with known redshifts, there are two Type I
GRBs: 051221A and 060614\footnote{The classification of GRB 060614 is not
conclusive. Based on the fact that no supernovae associated nearby bursts and the
similarity of the temporal and spectral behaviors with short GRB 050724, it was
proposed that it would be from the merger of compact objects (Zhang et al. 2007b;
Gehrels et al. 2006; Zhang 2006). However, the apparently long duration suggests
it may be from a new type of collapsar with a small amount of $^{56}$Ni ejection
(Fynbo et al.2006; Della Valle et al. 2006; Gal-Yam et al. 2006; King et al.
2007; Tominaga et al. 2007).}. Their X-ray afterglows are very bright, and the
derived $E_K$ from the XRT data are $\sim 6\times 10^{49}$ ergs and $\sim 2\times
10^{50}$ ergs, respectively, roughly about 1 order of magnitude smaller than that
of the typical Type II GRBs. The $\theta_j$ of the two bursts are $\sim 12^{\rm
o}$ and $\sim 7^{\rm o}$, respectively. They are wider than those of the other
(Type II) {\em Swift} GRBs in our sample. Combining our results with the fact
that the $\theta_j$ of another short GRB 050724 is $>25^{\rm o}$ (Grupe et
al.2006; Malesani et al. 2007), we cautiously suggest that the short GRBs might
be less collimation, if the breaks are explained as a jet break.

\acknowledgments

We thank the referee for helpful suggestions, and Z. G. Dai, Dirk Grupe and
Goro Sato for valuable comments.
We acknowledge the use of the public data from the Swift data archive. This
work is supported by NASA under grants NNG06GH62G, NNG05GB67G, NNX07AJ64G,
NNX07AJ66G, and the National Natural Science Foundation of China under grant No.
10463001 (EWL) and 10640420144.

\begin{deluxetable}{lllllllllllllllllllll}


\rotate
\tablewidth{550pt} \tabletypesize{\tiny}

\tablecaption{XRT observations and the Fitting results}

\tablenum{1}


\tablehead{\colhead{GRB}&\colhead{$t_1$(ks)\tablenotemark{a}}&\colhead{$t_2$(ks)\tablenotemark{a}}&\colhead{$t_{b,1}(\delta
t_{b,1})$(ks)\tablenotemark{b}}&\colhead{$t_{b,2}(\delta
t_{b,2})$(ks)\tablenotemark{b}}&\colhead{$\alpha_2(\delta
\alpha_2)$\tablenotemark{b}}&\colhead{$\alpha_3(\delta
\alpha_3)$\tablenotemark{b}}&\colhead{$\alpha_4(\delta
\alpha_4$\tablenotemark{b})}&\colhead{$\chi^2$(dof)}&\colhead{$\Gamma_2(\delta\Gamma_2)$}&\colhead{$\Gamma_3(\delta\Gamma_3$)}&\colhead{$\Gamma_4(\delta
\Gamma_4$)}}

\startdata
&STPL\\
\hline
050128  & 0.25 & 70.72 & 1.13(0.74) & 30.67(14.19) & 0.34(0.15) & 1.00(0.13) & 1.98(0.39) & 27(46) & 1.76(0.07) & 2.05(0.08) & 1.95(0.15)\\
060210 & 3.90 & 861.94 & 5.51(0.86) & 186.65(76.48) & -0.20(0.39) & 1.00(0.05) & 1.85(0.27) & 134(131) & -- & 2.12(0.08) & 2.11(0.33)\\
060510A & 0.16 & 343.41 & 2.89(1.87) & 47.65(16.75) & 0.01(0.09) & 0.87(0.17) & 1.74(0.12) & 84(140) & 1.91(0.07) & 2.04(0.14) & 2.06(0.14)\\
060807  & 0.28 & 166.22 & 3.80(1.15) & 14.89(5.88) & -0.22(0.13) & 0.96(0.24) & 1.92(0.12) & 42(34) & 2.19(0.16) & 2.18(0.09) & 2.40(0.20)\\
060813  & 0.09 & 74.25 & 0.19(0.04) & 15.24(3.88) & -0.01(0.19) & 0.87(0.03) & 1.63(0.13) & 56(73) & 2.05(0.09) & 1.99(0.05) & 2.10(0.07)\\
060814  & 0.87 & 203.31 & 5.92(2.88) & 68.58(23.27) & 0.32(0.13) & 1.06(0.12) & 2.38(0.40) & 44(48) & 2.21(0.05) & -- & 2.30(0.05)\\
\hline
&SBPL\\
\hline

050124 & 11.37 & 58.66 &-- &29.37(12.61)  & -- & 0.62(0.56) & 2.53(0.78) & 6(11) & --& 2.05(0.29) & 1.93(0.21) \\
050315 & 5.40 & 450.87 &-- &224.64(38.68) & -- & 0.66(0.03) & 1.90(0.28) & 42(52) & --& 2.31(0.12) & 2.17(0.07) \\
050318 & 3.34 & 45.19 &--& 10.64(4.97) & -- & 0.90(0.23) & 1.84(0.19) & 27(20) & --& 2.01(0.08) & 2.02(0.06) \\
050319 & 6.11 & 84.79 & 11.20(13.26) &--  & 0.23(0.59) & 0.99(0.25) &--& 9(9) & 2.00(0.06) & 2.04(0.07) &--\\
050401 & 0.14 & 801.04 & 5.86(0.78) &  & 0.58(0.02) & 1.39(0.06)&-- & 107(92) & 2.06(0.06) & 2.03(0.04) &--\\
050416A & 0.25 & 261.69 & 1.74(1.12) &  &  0.43(0.12) & 0.90(0.04)&-- & 36(38) & 2.19(0.20) & 2.15(0.10)& -- \\
050505 & 3.07 & 97.19 & 7.87(1.57) &   & 0.15(0.19) & 1.30(0.06) & --& 26(45) & 2.00(0.07) & 2.03(0.04) & --\\
050713A & 4.61 & 1600.08 & 5.86(1.24) &  & -0.27(1.05) & 1.16(0.03) & -- & 28(17) & 2.25(0.05) & 2.21(0.17)& -- \\
050713B & 0.79 & 478.50 & 10.80(1.59) &   & -0.00(0.07) & 0.94(0.04) & --& 40(63) & 1.83(0.11) & 1.94(0.09) & --\\
050716 & 0.64 & 74.40 & 7.53(9.02) &   & 0.76(0.16) & 1.35(0.24) & --& 31(36) & 1.60(0.08) & 2.01(0.13)& -- \\
050717 & 0.32 & 11.23 & -- & 1.84(0.95) & -- & 0.57(0.21) & 1.65(0.12) & 28(56) & --& 1.61(0.08) & 1.89(0.12) \\
050726 & 0.42 & 17.05 & --&1.17(0.33)& -- & 0.80(0.03) & 2.32(0.22) & 27(34) & --& 2.06(0.08) & 2.14(0.09) \\
050730 & 3.93 & 108.75 & --& 6.66(0.29)& -- & -0.37(0.25) & 2.49(0.04) & 203(215) & --& 1.65(0.03) & 1.70(0.03) \\
050801 & 0.07 & 46.10 & 0.25(fixed) &   & 0(fixed) & 1.10(0.03) & --& 44(45) & -- & 1.91(0.12)& -- \\
050802 & 0.51 & 83.83 &-- &4.09(0.61) & -- & 0.32(0.10) & 1.61(0.04) & 58(72) & --& 1.92(0.05) & 1.89(0.07) \\
050803 & 0.50 & 368.89 & --&13.71(0.90) &  -- & 0.25(0.03) & 2.01(0.07) & 94(57) & --& 1.78(0.10) & 2.00(0.08) \\
050820A & 4.92 & 1510.14 &--& 420.78(179.33) &  -- & 1.11(0.02) & 1.68(0.21) & 246(292) & --& 1.63(0.05) & 1.87(0.04) \\
050822 & 6.41 & 523.32 & 66.99(44.38) &  & 0.60(0.10) & 1.25(0.19) & -- & 29(44) & 2.29(0.23) & 2.36(0.11)& -- \\
050824 & 6.31 & 330.49 & 11.52(4.25) &  & -0.40(0.52) & 0.61(0.06) & -- & 45(41) & 2.00(0.16) & 2.01(0.09) & --\\
050908 & 3.97 & 33.36 & --&7.81(5.33) & -- & 0.13(0.96) & 1.58(0.46) & 0(1) & --& - & 2.09(0.25) \\
050915A & 0.32 & 88.77 & 1.94(1.11) & & 0.39(0.27) & 1.24(0.09) & -- & 7(6) & 2.32(0.17) & 2.42(0.20)& -- \\
051006 & 0.23 & 13.13 & &0.93(0.71) & -- & 0.57(0.26) & 2.23(0.56) & 15(19) & --& 1.61(0.14) & 1.84(0.20) \\
051008 & 3.09 & 43.77 & 14.67(3.82) &  & -- & 0.86(0.09) & 2.01(0.19) & 52(49) & --& 2.15(0.32) & 2.11(0.10) \\
051016A & 0.37 & 37.41 & 0.63(0.40) &   & -0.41(1.18) & 0.91(0.12) & --& 0(7) & 2.40(0.26) & -- & --\\
051016B & 4.78 & 150.47 & --&66.40(23.09) & -- & 0.71(0.08) & 1.84(0.46) & 15(16) & --& -& 2.19(0.13) \\
051109A & 3.73 & 639.16 & --&27.28(7.90)& -- & 0.79(0.07) & 1.53(0.08) & 39(48) & --& 1.91(0.07) & 1.90(0.07) \\
051109B & 0.39 & 87.63 & 5.11(4.73) &  & 0.56(0.17) & 1.22(0.17) & -- & 15(17) & 2.73(0.44) & 2.35(0.24)& -- \\
051117A & 18.19 & 970.14 & 104.23(151.17) &   & 0.51(0.25) & 1.07(0.24) & --& 21(19) & 2.25(0.04) & 2.39(0.15) & --\\
051221A & 6.87 & 118.64 & --&40.74(15.89) & -- & 0.46(0.16) & 1.75(0.41) & 11(14) & --& 2.08(0.09) & 2.02(0.19) \\
060105 & 0.10 & 360.83 & --&2.31(0.14)& -- & 0.84(0.01) & 1.72(0.02) & 653(754) & --& 2.23(0.05) & 2.15(0.03) \\
060108 & 0.77 & 165.26 & --&22.08(7.38)& -- & 0.26(0.09) & 1.43(0.17) & 7(7) & --& 2.17(0.32) & 1.75(0.15) \\
060109 & 0.74 & 48.01 & 4.89(1.10) &   & -0.17(0.14) & 1.32(0.09) & --& 19(13) & 2.32(0.15) & 2.34(0.14)& -- \\
060124 & 13.30 & 664.01 & --&52.65(10.33)& -- & 0.78(0.10) & 1.65(0.05) & 165(132) & --& 2.10(0.06) & 2.06(0.08) \\
060202 & 1.03 & 96.23 & 3.50(6.95) &   & 0.68(0.37) & 1.14(0.13) & --& 51(31) & 2.96(0.19) & 3.41(0.14)& --\\
060203 & 3.80 & 32.95 & --&12.95(6.69) & -- & 0.40(0.30) & 1.65(0.47) & 4(7) & --& 2.08(0.19) & 2.25(0.13) \\
060204B & 4.06 & 98.80 & --&5.55(0.66) & -- & -0.49(0.65) & 1.47(0.07) & 21(34) & --& 2.54(0.14) & 2.64(0.16) \\
060206 & 0.11 & 621.77 & 8.06(1.46) &   &  0.40(0.05) & 1.26(0.04) &--&43(44) & 2.31(0.12) & 2.33(0.32)&--\\
060211A & 5.40 & 527.10 & --&267.24(165.67) & -- & 0.38(0.08) & 1.63(1.27) & 10(9) & --& 2.15(0.06) & 2.11(0.26) \\
060306 & 0.25 & 124.39 & 4.67(2.91) &   & 0.40(0.11) & 1.05(0.07) & --& 30(32) & 2.10(0.11) & 2.21(0.10) & --\\
060313 & 0.09 & 93.22 & --&11.18(2.89) & -- & 0.82(0.03) & 1.76(0.18) & 95(128) & --& 1.84(0.34) & 1.78(0.09) \\
060319 & 0.33 & 304.52 & --&99.70(26.78) & -- & 0.84(0.02) & 1.92(0.30) & 72(93) & --& 1.93(0.22) & 2.25(0.11) \\
060323 & 0.33 & 16.28 & --&1.29(0.32)  & -- & -0.11(0.23) & 1.55(0.16) & 4(7) & --& 1.99(0.16) & 2.02(0.13) \\
060428A & 0.23 & 271.10 & --&125.31(47.19)& -- & 0.48(0.03) & 1.46(0.37) & 26(21) & --& 2.11(0.24) & 1.97(0.10) \\
060428B & 0.96 & 200.36 & 3.95(5.55) &   & 0.53(0.41) & 1.16(0.13) & --& 19(21) & 2.41(0.24) & 2.10(0.33)& -- \\
060502A & 0.24 & 593.06 & --&72.57(15.05) & -- & 0.53(0.03) & 1.68(0.15) & 11(26) & --& 2.11(0.29) & 2.15(0.13) \\
060507 & 3.00 & 86.09 & 6.95(1.68) &   & -0.06(0.55) & 1.12(0.07) && -- 13(24) & 2.06(0.23) & 2.15(0.14)& -- \\
060510B & 4.40 & 77.71 & --&67.90(29.88) & -- & 0.44(0.18) & 2.40(0.00) & 4(8) & --& 1.71(0.04) & -- \\
060526 & 1.09 & 45.20 & --&11.60(6.39) & -- & 0.42(0.12) & 1.58(0.34) & 5(9) & --& 2.07(0.09) & 2.08(0.16) \\
060604 & 4.14 & 403.81 & 11.51(9.81) &   & 0.20(0.77) & 1.17(0.09) & --& 32(36) & 2.44(0.15) & 2.43(0.17)& -- \\
060605 & 0.25 & 39.85 & --& 7.14(0.93) & -- & 0.45(0.04) & 1.80(0.13) & 22(34) & --& 1.62(0.17) & 1.83(0.09) \\
060614 & 5.03 & 451.71 & --&49.84(3.62)  & -- & 0.18(0.06) & 1.90(0.07) & 70(54) & --& 2.02(0.02) & 1.93(0.06) \\
060707 & 5.32 & 813.53 & 22.21(54.08) &  & 0.37(0.96) & 1.09(0.17) & -- & 8(11) & 1.88(0.08) & 2.06(0.20)& -- \\
060708 & 0.25 & 439.09 & 7.28(2.34) &   & 0.57(0.08) & 1.32(0.07) & --& 39(35) & 2.30(0.20) & 2.36(0.11) & --\\
060712 & 0.56 & 317.56 & 7.89(2.67) &  & 0.12(0.16) & 1.15(0.10) & -- & 15(14) & 3.21(0.38) & 2.94(0.28) & --\\
060714 & 0.32 & 331.97 & 3.70(0.97) &   & 0.34(0.10) & 1.27(0.05) & --& 53(73) & 2.15(0.08) & 2.04(0.11)& -- \\
060719 & 0.28 & 182.15 & 9.57(2.70) &   & 0.40(0.06) & 1.31(0.10) & --& 19(26) & 2.35(0.13) & 2.28(0.26) & --\\
060729 & 0.42 & 2221.24 & 72.97(3.02) &   & 0.21(0.01) & 1.42(0.02) & --& 459(459) & 2.33(0.08) & 2.29(0.07)& -- \\
060804 & 0.18 & 122.07 & 0.86(0.22) &   & -0.09(0.15) & 1.12(0.07) & --& 18(24) & 2.04(0.23) & 2.14(0.15) & --\\
060805A & 0.23 & 75.91 & 1.30(0.70) &   & -0.17(0.41) & 0.97(0.13) & --& 11(17) & -- & 1.97(0.37) & --\\
060906 & 1.32 & 36.69 & --&13.66(3.29) & -- & 0.35(0.10) & 1.97(0.36) & 3(7) & --& 2.28(0.37) & 2.12(0.17) \\
060908 & 0.08 & 363.07 & --& 0.95(0.34)  & -- & 0.70(0.07) & 1.49(0.09) & 98(59) & --& 2.01(0.22) & 2.00(0.08) \\
060912 & 0.12 & 86.80 & 2.92(2.77) &   & 0.65(0.12) & 1.24(0.11) & --& 31(56) & -- & 2.03(0.12)& -- \\
060923A & 0.22 & 280.62 & 3.33(1.03) &   & -0.16(0.22) & 1.30(0.06) & --& 34(21) & 2.05(0.25) & 1.86(0.18) & --\\
060923B & 0.16 & 6.03 & 0.42(0.64) &   & -0.73(0.99) & 1.08(0.82) & --& 2(10) & 2.47(0.53) & 2.25(0.31)& -- \\
060926 & 0.09 & 5.96 & 1.13(0.92) &   & 0.04(0.14) & 1.23(0.52) & --& 11(9) & 1.93(0.16) & 1.88(0.14)& -- \\
060927 & 0.11 & 5.64 & --&4.24(8.22) & -- & 0.73(0.32) & 1.82(2.60) & 4(7) & --& 1.65(0.19) & 1.92(0.15) \\
061004 & 0.39 & 69.99 & 1.50(0.52) &   & -0.08(0.29) & 1.04(0.09) & --& 13(17) & 1.84(0.34) & 3.04(0.34) & --\\
061019 & 9.07 & 287.03 & 10.84(2.15) &   & -1.38(2.88) & 1.15(0.08) & --& 6(10) & 2.32(0.20) & 1.93(0.28)& -- \\
061021 & 0.30 & 594.16 & 9.59(2.17) &   & 0.52(0.03) & 1.08(0.03) & --& 94(87) & 1.90(0.06) & 1.72(0.05)& -- \\
061121 & 4.89 & 353.10 & --&24.32(4.38) & -- & 0.75(0.06) & 1.63(0.05) & 121(147) & --& 1.71(0.03) & 1.96(0.07) \\
061201 & 0.10 & 15.42 & --&2.09(0.75) & -- & 0.57(0.07) & 1.61(0.23) & 20(29) & --& 1.30(0.09)&-- \\
061222A & 10.94 & 724.64 & --&60.51(8.89) & -- & 0.81(0.07) & 1.86(0.06) & 144(95) & --& 2.45(0.06) & 2.22(0.12) \\
070103 & 0.11 & 143.98 & --&2.88(0.48) & -- & 0.20(0.10) & 1.63(0.08) & 43(30) & --& 2.32(0.25) & 2.52(0.21) \\
070129 & 1.32 & 546.36 & 20.12(3.14) &  & 0.15(0.07) & 1.31(0.06) & --&42(70) & 2.25(0.07) & 2.30(0.10) & --\\
\hline
&SPL\\
\hline

050219B & 3.21 & 85.26 &  &  & -- & 1.14(0.03) & -- & 24(32) & -- & 2.27(0.14) & --\\
050326 & 3.34 & 142.24 &  &  & -- & --&1.63(0.04) & 45(34) & -- & --&2.15(0.14) \\
050408 & 2.60 & 3223.36 &  &  & -- & 0.78(0.01) & -- & 52(44) & -- & 2.01(0.18) & --\\
050525A & 5.94 & 157.85 &  &  & -- & 1.40(0.05) & -- & 11(11) & -- & 2.17(0.18) & --\\
050603 & 39.72 & 166.22 &  &  & -- & --&1.71(0.10) & 8(10) & -- &--& 1.84(0.09) \\
050721 & 0.30 & 257.24 &  &  & -- & 1.18(0.02) & -- & 80(98) & -- & 1.77(0.10) & --\\
050814 & 2.17 & 87.85 &  &  & -- & 0.65(0.05) & -- & 21(16) & -- & 1.91(0.07) & --\\
050826 & 0.13 & 61.93 &  &  & -- & 1.02(0.03) & -- & 23(21) & -- & 2.19(0.19) & --\\
050827 & 65.95 & 246.35 &  &  & -- & 1.24(0.15) & -- & 12(15) & -- & 1.88(0.15) & --\\
051001 & 6.71 & 273.86 &  &  & -- & 0.70(0.06) & -- & 30(25) & -- & 1.93(0.19) & --\\
051111 & 10.98 & 34.24 &  &  & -- & 1.09(0.17) & -- & 1(6) & -- & -- & --\\
051117B & 0.22 & 0.62 &  &  & -- & --&1.68(0.27) & 0(2) & -- & -- & --\\
060115 & 5.44 & 326.04 &  &  & -- & 0.88(0.04) & -- & 12(12) & -- & 2.50(0.38) & --\\
060116 & 0.21 & 6.87 &  &  & -- & 0.88(0.06) & -- & 3(6) & -- & 2.33(0.39) & --\\
060403 & 0.05 & 79.82 &  &  & -- & --&1.67(0.07)  & 70(57) & -- & --&1.58(0.13) \\
060418 & 0.20 & 201.65 &  &  & -- & 1.45(0.02) & -- & 272(283) & -- & 2.24(0.05) & --\\
060421 & 0.12 & 6.52 &  &  & -- & 0.93(0.05) & -- & 11(7) & -- & 1.60(0.35) & --\\
060512 & 0.11 & 104.01 &  &  & -- & 1.39(0.02) & -- & 76(58) & -- & 3.60(0.19) & --\\
060522 & 5.50 & 432.75 &  &  & -- & 1.07(0.10) & -- & 7(13) & -- & -- & --\\
060825 & 0.23 & 63.15 &  &  & -- & 1.08(0.04) & -- & 4(6) & -- & 1.64(0.29) & --\\
061007 & 0.09 & 97.82 &  &  & -- & &1.68(0.01) & 2153(1880) & -- & -- &2.08(0.05)\\
061019 & 2.90 & 287.03 &  &  & -- & 0.95(0.03) & -- & 28(20) & -- & 2.12(0.21) & --\\
070110 & 43.70 & 439.51 &  &  & -- & 1.05(0.14) & -- & 9(5) & -- & 2.36(0.24) & --\\

\enddata

\tablenotetext{a} {The time interval of our fitting.}

\tablenotetext{b}{The fitting results of the two-segment lightcurves with the
SBPL model are reported in columns for the jet break candidate (Columns
$t_{b,2}$, $\alpha_3$, $\alpha_4$, $\Gamma_3$, and $\Gamma_4$) if their
post-break segments are steeper than $\gtrsim 1.5$; otherwise, the results are
reported in the columns of the energy injection break(Columns $t_{b,1}$,
$\alpha_2$, $\alpha_3$, $\Gamma_2$, and $\Gamma_3$). The results of the fitting
results of the one-segment XRT lightcurves with the SPL model are similarly
reported in the columns of the energy injection break or of the jet break
candidate depending on their temporal decay slopes.}





\end{deluxetable}


\begin{deluxetable}{lllllllll}

\tablewidth{400pt} \tabletypesize{\tiny}

\tablecaption{Optical Data and the Fitting results}

\tablenum{2}

\tablehead{ \colhead{GRB\tablenotemark{a}}& \colhead{$t_1$(ks)\tablenotemark{b}}&
\colhead{$t_2$(ks)\tablenotemark{b}}& \colhead{$t_{b,O}(\delta t_{b,O})$(ks)}&
\colhead{$\alpha_{O,3}(\delta \alpha_{O,3})$}& \colhead{$\alpha_{O,4}(\delta
\alpha_{O,4})$}& \colhead{$\chi^2$(dof)\tablenotemark{c}} 
}

\startdata
970508 & 30.00 & 7421.93 & 139.67(3.16) & -2.73 & 1.21(0.02) & 29(21) & \\
980703 & 81.26 & 343.92 & 214.92(10.15) & 1.11 & 2.83 & 7(7) & \\
990123 & 13.31 & 1907.45 & 155.13(78.79) & 0.98(0.10) & 1.71(0.10) & 12(8) & \\
990510 & 12.44 & 340.24 & 101.91(12.48) & 0.86(0.03) & 1.95(0.14) & 17(17) & \\
990712 & 15.25 & 2991.47 & 2000.00(fixed) & 0.97 & 2.32 & 15(11) & \\
991216 & 41.17 & 1100.60 & 248.71(67.63) & 1.22(0.04) & 2.17 & 27(13) & \\
000301 & 134.00 & 4198.10 & 562.87(18.70) & 1.04 & 2.97 & 25(24) & \\
000926 & 74.48 & 591.61 & 175.18(4.62) & 1.48 & 2.49 & 35(24) & \\
010222 & 13.09 & 2124.75 & 32.12(3.62) & 0.43(0.08) & 1.29(0.02) & 29(48) & \\
011211 & 34.40 & 2755.47 & 198.66(16.68) & 0.85(0.05) & 2.36 & 26(33) & \\
020124 & 5.77 & 2787.67 & 8.47(7.39) & 0.76(1.19) & 1.85(0.11) & 8(9) & \\
020405 & 85.04 & 882.60 & 236.88(15.90) & 1.21 & 2.48 & 6(10) & \\
020813 & 14.18 & 362.83 & 40.03(0.21) & 0.63 & 1.42 & 69(43) & \\
021004 & 21.12 & 2030.14 & 300.30(fixed) & 0.82(0.02) & 1.39(0.05) & 82(90) & \\
030226 & 17.34 & 609.12 & 88.83(16.30) & 0.88(0.12) & 2.41(0.12) & 10(12) & \\
030323 & 34.68 & 895.74 & 400.00(fixed) & 1.29 & 2.11 & 10(10) & \\
030328 & 4.90 & 227.46 & 18.50(4.32) & 0.52(0.09) & 1.25(0.05) & 52(70) & \\
030329 & 4.60 & 100.00 & 41.00(0.42) & 0.84 & 1.89(0.01) & 870(956) & \\
030429 & 12.53 & 574.04 & 158.73(fixed) & 0.72(0.03) & 2.72 & 30(10) & \\
030723 & 15.00 & 800.00 & 103.22(5.02) & 0.05(0.06) & 2.01(0.05) & 20(15) & \\
040924 & 0.95 & 134.12 & 1.49(0.96) & 0.34(0.64) & 1.11(0.06) & 19(10) & \\
041006 & 0.23 & 550.00 & 14.24(1.15) & 0.44(0.02) & 1.27(0.01) & 97(69) & \\
050319 & 0.03 & 3.00 & 0.61(0.25) & 0.38(0.06) & 1.02(0.12) & 29(29) & \\
050525 & 2.83 & 91.80 & 40.72(8.18) & 1.02(0.12) & 3.00(0.57) & 28(5) & \\
050730 & 0.07 & 358.90 & 11.61(1.95) & 0.26(0.08) & 1.67(0.09) & 58(16) & \\
050801 & 0.02 & 9.49 & 0.20(0.01) & 0.00(0.02) & 1.11(0.01) & 140(42) & \\
050820A & 0.12 & 663.30 & 344.98(32.78) & 0.88(0.01) & 1.48 & 439(25) & \\
050922C & 0.25 & 69.60 & 3.13(2.75) & 0.63(0.13) & 1.14(0.10) & 14(17) & \\
051109A & 0.04 & 265.20 & 36.02(8.28) & 0.68(0.01) & 1.42(0.12) & 116(40) & \\
051111 & 0.03 & 20.00 & 2.61(0.25) & 0.79(0.01) & 1.70(0.14) & 107(84) & \\
060206 & 20.00 & 201.58 & 71.21(3.65) & 1.07(0.02) & 1.96 & 25(50) & \\
060210 & 0.09 & 7.19 & 0.72(0.17) & 0.04(0.22) & 1.21(0.05) & 13(12) & \\
060526 & 0.06 & 893.55 & 84.45(5.88) & 0.67(0.02) & 1.80(0.04) & 116(56) & \\
060605A & 0.43 & 111.96 & 8.83(1.21) & 0.41 & 2.33(0.16) & 2(1) & \\
060607A & 0.07 & 13.73 & 0.16(fixed) & -3.07(0.25) & 1.18(0.02) & 92(35) & \\
060614 & 20.00 & 934.36 & 112.35(8.53) & 0.77(0.10) & 2.70(0.07) & 16(16) & \\
060714 & 3.86 & 285.87 & 10.00(fixed) & 0.01 & 1.41(0.03) & 35(11) & \\
060729 & 70.00 & 662.39 & 297.49(69.62) & 1.09(0.10) & 2.13(0.44) & 18(19) & \\
061121 & 0.26 & 334.65 & 1.70(0.73) & 0.17 & 0.99(0.05) & 18(23) & \\
980326 & 36.46 & 117.68 &  & 2.14(0.09) &  & 15(6) & \\
991208 & 179.52 & 613.24 & - & 2.30(0.12) &  & 17(9) & \\
000131 & 357.44 & 699.06 & - & 2.55(0.29) & - & 0(1) & \\
000418 & 214.27 & 2000.00 & - & 0.81(0.03) & - & 13(9) & \\
000911 & 123.35 & 1466.26 & - & 1.36(0.06) & - & 9(2) & \\
011121 & 33.36 & 1000.00 & - & 1.98(0.06) & - & 7(5) & \\
021211 & 0.13 & 1865.64 & - & 1.18(0.01) & - & 78(50) & \\
050318 & 3.23 & 22.83 & - & 0.84(0.22) & - & 0(1) & \\
050401 & 0.06 & 1231.18 & - & 0.80(0.01) & - & 43(12) & \\
050408 & 8.64 & 434.81 & - & 0.72(0.04) & - & 9(15) & \\
050502 & 6.12 & 29.22 & - & 1.42(0.02) & - & 31(19) & \\
050603 & 34.09 & 219.71 & - & 1.75(0.20) & - & 16(7) & \\
050802 & 0.34 & 127.68 & - & 0.85(0.02) & - & 50(10) & \\
050908 & 1.32 & 57.81 & - & 0.71(0.09) & - & 11(10) & \\
060124 & 3.34 & 1979.30 & - & 0.85(0.02) & - & 11(19) & \\
060418 & 3.92 & 69.53 & - & 1.36(0.04) & - & 8(11) & \\
060904B & 0.50 & 163.13 & - & 0.86(0.02) & - & 60(19) & \\
070110 & 0.66 & 34.76 & - & 0.43(0.08) & - & 1(4) & \\
\enddata
\tablenotetext{a}{Taken from Liang \& Zhang (2006) and Paper II and the
references therein.}

\tablenotetext{b}{Time interval for temporal analysis.}

\tablenotetext{c}{The fitting $\chi^2$ and degree of freedom. Please note that we
take the observed uncertainty as $\sigma_{\log F_O}=0.05$ for those detection
without observed error or with $\sigma_{\log F_O}<0.05$, in order to properly fit
the data. The uncertainties of the fitting parameters of these bursts thus cannot
be properly constrained.}




\end{deluxetable}

\newpage

\begin{deluxetable}{cccccc}

\tablewidth{480pt} \tabletypesize{\tiny}

\tablecaption{Definition of Jet Break Candidate Grades}

\tablenum{3}

\tablehead{\colhead{Grade}& \colhead{No Spectral Evolution}&
\colhead{$\alpha_4>1.5$}& \colhead{Closure Relations}&
\colhead{Achromaticity}&\colhead{Number}}
\startdata
``Bronze''&Y&Y&&&42(XRT)+27(Opt.)\\
``Silver''&Y&Y&Y&&27(XRT)+23(Opt.)\\
``Gold''&Y&Y&Y(1 band)&Y&7\\
``Platinum''&Y&Y&Y (at least 2 bands)&Y&0\\
\enddata
\end{deluxetable}

\begin{deluxetable}{llllllllll}

\tablewidth{480pt} \tabletypesize{\tiny}

\tablecaption{Jet Break Candidates and Their Grades}

\tablenum{4}

\tablehead{ \colhead{GRB}& \colhead{$\beta_2(\delta \beta_2)$}&
\colhead{$\beta_4(\delta \beta_4)$}&\colhead{$\alpha_3 (\delta \alpha_3)$}&
\colhead{$\alpha_4(\delta \alpha_4)$}& \colhead{$t_j(\delta t_j)$(ks)}&
\colhead{$\Delta \alpha(\delta \Delta
\alpha)$}&\colhead{Grade}&Achromaticity\tablenotemark{*}}

\startdata
Radio&&&&\\
\hline
970508\tablenotemark{a}&&&&&$\sim 25$ (days)&&Bronze&?\\
000418\tablenotemark{b}&&&&&$\sim 26$ (days)&&Bronze&?\\
\hline
Optical&&&&\\
\hline
980703 & 1.01(0.02) & -- & 1.11 & 2.83 & 214.92(10.15) & 1.71 & Silver&?\\
990123 & 0.80(0.10) & -- & 0.98(0.10) & 1.71(0.10) & 155.13(78.79) & 0.73(0.14) & Silver&?\\
990510 & 0.75(0.07) & -- & 0.86(0.03) & 1.95(0.14) & 101.91(12.48) & 1.09(0.14) & Gold&$\surd$\\
990712 & 0.99(0.02) & -- & 0.97 & 2.32 & 2000 & 1.35 & Silver&?\\
991216 & 0.74(0.05) & -- & 1.22(0.04) & 2.17 & 248.71(67.63) & 0.95(0.04) & Silver&?\\
000301C& 0.90(0.02) & -- & 1.04 & 2.82 & 562.87(18.70) & 1.78 & Silver&?\\
000926 & 1.00(0.20) & -- & 1.48 & 2.49 & 175.18(4.62) & 1.01 & Silver&?\\
011211 & 0.74(0.05) & -- & 0.85(0.05) & 2.36 & 198.66(16.68) & 1.52(0.05) & Silver&?\\
020124 & 0.91(0.14) & -- & 0.76(1.19) & 1.85(0.11) & 8.47(7.39) & 1.09(1.19) & Silver&?\\
020405 & 1.23(0.12) & -- & 1.21 & 2.48 & 236.88(15.90) & 1.27 & Silver&?\\
020813 & 0.85(0.07) & --   & 0.63 & 1.42 & 40.03(0.21)   &0.79  & Silver&?\\
021004 & 0.39(0.12) & -- & 0.65(0.02) & 1.57(0.05) & 300.30 & 0.92(0.05) & Silver&?\\
030226 & 0.70(0.03) & -- & 0.88(0.12) & 2.41(0.12) & 88.83(16.30) & 1.53(0.17) & Silver&?\\
030323 & 0.89(0.04) & -- & 1.29 & 2.11 & 400 & 0.82 & Silver&?\\
030329 & 0.66 & --  & 0.84 & 1.89(0.01) & 41.00(0.42) & 1.05(0.01) & Gold&$\surd$\\
030429 & 1.22(0.04) & -- & 0.72(0.03) & 2.72 & 158.73 & 2.00(0.03) & Silver&?\\
030723 & 1          & --  & 0.05(0.06)  &2.01(0.05)& 103.22(5.02)&1.96(0.08)& Bronze&?\\
050525 & 0.97(0.10) & -- & 1.02(0.12) & 3.00(0.57) & 40.72(8.18) & 1.98(0.58) & Gold&$\surd$\\
050730 & 0.75 & -- & 0.26(0.08) & 3.00(0.57) & 1.67(0.09) & 2.74(0.58) & Bronze&?\\
050820A & 0.57(0.06) & -- & 0.88(0.01) & 1.48 & 344.98(32.78) & 0.60 & Gold&$\surd$\\
051109A & 0.65(0.15) & -- & 0.68(0.01) & 1.42(0.12) & 36.02(8.28) & 0.74(0.12) & Gold&$\surd$\\
051111 & 0.84(0.02) & -- & 0.79(0.01) & 1.70(0.14) & 2.61(0.25) & 0.91(0.14) & Silver&X\\
060206 & 0.70 & -- & 1.07(0.02) & 2.00(0.26) & 71.21(3.65) & 0.93(0.26) & Silver&X\\
060605 & 0.8  & --  & 0.41       & 2.33(0.16) &    8.83(1.21) &1.92& Bronze &$\surd$\\
060526 & 1.69(0.53) & -- & 0.67(0.02) & 1.80(0.04) & 84.45(5.88) & 1.13(0.04) & Gold&$\surd$\\
060614 & 0.94(0.08) & -- & 0.77(0.10) & 2.70(0.07) & 112.35(8.53) & 1.93(0.12) & Gold&$\surd$\\
060729 & 0.74(0.07) & -- & 1.09(0.10) & 2.13(0.44) & 297.49(69.62) & 1.03(0.45) & Silver&X\\
\hline
X-Ray&&&\\
\hline 980828 &  $\sim 1$  & &1.44 &2.6 & 190 &1.16 &Silver& ?\\
030329 & 1.17 & 0.8(0.3)&0.87(0.05) &1.84(0.07)&44.93(4.32) &0.97(0.09)&Gold&$\surd$\\

050124 & 1.05(0.29) & 0.93(0.21) & 0.62(0.56) & 2.53(0.78) & 29.37(12.61) & 1.91(0.96) & Silver&?\\
050128 & 1.05(0.08) & 0.95(0.15) & 1.00(0.13) & 1.98(0.39) & 30.70(14.20) & 0.98(0.41) & Silver&?\\
050315 & 1.31(0.12) & 1.17(0.07) & 0.66(0.03) & 1.90(0.23) & 224.64(38.68) & 1.24(0.23) & Silver&?\\
050318 & 1.01(0.08) & 1.02(0.06) & 0.90(0.23) & 1.84(0.19) & 10.60(4.97) & 0.94(0.30) & Silver&X\\
050525A\tablenotemark{c} & 1.17(0.18) & 1.17(0.18) & 1.20(0.03) & 1.62(0.16) & 13.73(7.47) & 0.42(0.16) & Gold&$\surd$\\
050717 & 0.61(0.08) & 0.89(0.12) & 0.57(0.21) & 1.65(0.12) & 1.84(0.95) & 1.08(0.24) & Silver&?\\
050726 & 1.06(0.08) & 1.14(0.09) & 0.79(0.03) & 2.32(0.22) & 8.78(1.11) & 1.53(0.22) & Silver&?\\
050730 & 0.65(0.03) & 0.70(0.03) & -0.37(0.25) & 2.49(0.04) & 6.66(0.29) & 2.86(0.25) & Bronze&$\surd$\\
050802 & 0.92(0.05) & 0.89(0.07) & 0.32(0.10) & 1.61(0.04) & 4.09(0.61) & 1.29(0.11) & Bronze&X\\
050803 & 0.78(0.10) & 1.00(0.08) & 0.25(0.03) & 2.01(0.07) & 13.71(0.90) & 1.76(0.08) & Bronze&?\\
050820A & 0.63(0.05) & 0.87(0.04) & 1.11(0.02) & 1.68(0.21) & 421.00(179.00) & 0.57(0.21) & Gold&$\surd$\\
050908 & 2.09(0.25) & 1.09(0.25) & 0.13(0.96) & 1.58(0.46) & 7.81(5.33) & 1.45(1.06) & Bronze&X\\
051006 & 0.61(0.14) & 0.84(0.20) & 0.57(0.26) & 2.23(0.56) & 0.93(0.71) & 1.66(0.62) & Silver&?\\
051008 & 1.15(0.32) & 1.11(0.10) & 0.86(0.09) & 2.01(0.19) & 14.67(3.82) & 1.15(0.21) & Silver&?\\
051016B & 1.19(0.13) & 1.19(0.13) & 0.71(0.08) & 1.84(0.46) & 66.40(23.09) & 1.13(0.47) & Silver&?\\
051109A & 0.91(0.07) & 0.90(0.07) & 0.79(0.07) & 1.53(0.08) & 27.28(7.90) & 0.74(0.11) & Gold&$\surd$\\
051221A\tablenotemark{d} & 1.07(0.36) & 1.02(0.19) & 1.20(0.06) & 1.92(0.52) & 354.00(103.00) & 0.72(0.52) & Silver&$\surd$\\
060105 & 1.23(0.05) & 1.15(0.03) & 0.84(0.01) & 1.72(0.02) & 2.31(0.14) & 0.88(0.02) & Silver&?\\
060108 & 1.17(0.32) & 0.75(0.15) & 0.26(0.09) & 1.43(0.17) & 22.08(7.38) & 1.17(0.19) & Bronze&?\\
060124 & 1.10(0.06) & 1.06(0.08) & 0.81(0.09) & 1.66(0.05) & 52.60(10.30) & 0.85(0.10) & Silver&X\\
060203 & 1.08(0.19) & 1.25(0.13) & 0.40(0.30) & 1.65(0.47) & 12.95(6.69) & 1.25(0.56) & Bronze&?\\
060204B & 1.54(0.14) & 1.64(0.16) & -0.49(0.65) & 1.47(0.07) & 5.55(0.66) & 1.96(0.65) & Bronze&?\\
060210 & 1.12(0.08) & 1.11(0.33) & 1.00(0.05) & 1.85(0.27) & 187.00(76.50) & 0.85(0.27) & Silver&X\\
060211A & 1.15(0.06) & 1.11(0.26) & 0.38(0.08) & 1.63(1.27) & 267.24(165.67) & 1.25(1.27) & Bronze&?\\
060313 & 0.84(0.34) & 0.78(0.09) & 0.82(0.03) & 1.76(0.18) & 11.18(2.89) & 0.94(0.18) & Silver&?\\
060319 & 0.93(0.22) & 1.25(0.11) & 0.84(0.02) & 1.92(0.30) & 99.70(26.78) & 1.08(0.30) & Silver&?\\
060323 & 0.99(0.16) & 1.02(0.13) & -0.11(0.23) & 1.55(0.16) & 1.29(0.32) & 1.66(0.28) & Bronze&?\\
060428A & 1.11(0.24) & 0.97(0.10) & 0.48(0.03) & 1.46(0.37) & 125.31(47.19) & 0.98(0.37) & Bronze&?\\
060502A & 1.11(0.29) & 1.15(0.13) & 0.53(0.03) & 1.68(0.15) & 72.57(15.05) & 1.15(0.15) & Bronze&?\\
060510A & 1.04(0.05) & 1.06(0.14) & 0.93(0.14) & 1.77(0.10) & 47.70(16.70) & 0.84(0.17) & Silver&?\\
060526 \tablenotemark{e}& 1.07(0.09) & 1.08(0.16) & 0.42(0.12) & 1.58(0.34) & 11.60(6.39) & 1.16(0.36) & Gold&$\surd$\\
060605 & 0.62(0.17) & 0.83(0.09) & 0.45(0.04) & 1.80(0.13) & 7.14(0.93) & 1.35(0.14) & Bronze&$\surd$\\
060614\tablenotemark{f} & 0.96(0.16) & 0.93(0.06) & 1.03(0.02) & 2.13(0.07) & 36.60(2.40) & 1.10(0.07) & Gold&$\surd$\\
060807 & 1.18(0.09) & 1.40(0.20) & 0.96(0.24) & 1.92(0.12) & 14.90(5.88) & 0.96(0.27) & Silver&?\\
060813 & 0.99(0.05) & 1.10(0.07) & 0.87(0.03) & 1.63(0.13) & 15.20(3.88) & 0.76(0.13) & Silver&?\\
060814 & 1.30(0.05) & 1.30(0.05) & 1.06(0.12) & 2.38(0.40) & 68.60(23.30) & 1.32(0.42) & Silver&?\\
060906 & 1.28(0.37) & 1.12(0.17) & 0.35(0.10) & 1.97(0.36) & 13.66(3.29) & 1.62(0.37) & Bronze&?\\
060908 & 1.01(0.22) & 1.00(0.08) & 0.70(0.07) & 1.49(0.09) & 0.95(0.34) & 0.79(0.11) & Bronze&?\\
060927 & 0.65(0.19) & 0.92(0.15) & 0.73(0.32) & 1.82(2.60) & 4.24(8.22) & 1.09(2.62) & Silver&?\\
061121 & 0.71(0.03) & 0.96(0.07) & 0.75(0.06) & 1.63(0.05) & 24.32(4.38) & 0.88(0.08) & Silver&?\\
061201 & 0.30(0.15) & 0.30(0.15) & 0.57(0.07) & 1.61(0.23) & 2.09(0.75) & 1.04(0.24) & Bronze&?\\
061222A & 1.45(0.06) & 1.22(0.12) & 0.81(0.07) & 1.86(0.06) & 60.50(8.89) & 1.05(0.09) & Silver&?\\
070103 & 1.32(0.25) & 1.52(0.21) & 0.20(0.10) & 1.63(0.08) & 2.88(0.48) & 1.43(0.13) & Bronze&?\\
\enddata
\tablenotetext{*}{If a break is confirmed to be achromatic, we mark the break
with a ``$\surd$''. If a break is clearly chromatic, we mark it with ``X''. For
most of breaks without multi-wavelength observations, we have no information to
access to the chromaticity of these breaks, so we mark them with a ``?''
sign.}

\tablerefs{a:Frail et al.(2000); b:Berger et al.(2001);c: Blustin et al. (2006);
d: Burrows et al. (2006); e: Dai et al. (2007); f: Mangano et al. (2007) }

%
%

%
%
%
%






\end{deluxetable}


\begin{deluxetable}{llllllllllllll}

\tablewidth{450pt}

\tabletypesize{\footnotesize}

\tablecaption{Derivation of Jet Opening Angles and Kinetic Energies}

\tablenum{5}
\tablehead{\colhead{GRB}&\colhead{$z$}&\colhead{Reg.\tablenotemark{a}}&\colhead{$p$}&\colhead{$\epsilon_{B,-4}$}&\colhead{$Y$}&\colhead{$\theta_j(^{o})$}&
\colhead{$\log E_{\rm K,iso}$\tablenotemark{b}}&\colhead{$\log E_{\rm
K}$\tablenotemark{b}}&\colhead{$\log \nu_m$\tablenotemark{c}}&\colhead{$\log
\nu_c$\tablenotemark{c}}&\colhead{ref.\tablenotemark{d}}}

\startdata
050315&1.95&I&2.76&1.00&2.45&2.9&55.06&52.17(0.05)&11.71&16.61&1\\
050318&1.44&I&2.08&1.01&6.91&1.6&53.30&49.91(0.16)&11.86&18.00&2\\
050319&3.24&I&2.16&1.00&4.93&$>$2.5&53.76&$>$50.75&11.69&17.70&3\\
050401&2.9&II&2.98&0.20&2.59&$>$4.2&54.97&$>$52.40&12.51&$>$18.00&4\\
050416A&0.65&I&2.32&10.78&0.72&$>$9.3&51.94&$>$50.06&11.80&$>$18.00&5\\
050505&4.27&I&2.1&1.00&6.28&$>$2.5&54.03&$>$51.00&11.56&17.53&6\\
050525A&0.606&II&3.34&0.99&0.10&2.6&53.98&50.99(0.06)&12.79&$>$18.69&7\\
050820A&2.61&I&2.01&1.00&11.05&3.6&54.88&52.17(0.14)&9.65&17.08&8\\
050922C&2.2&II&2.44&1.49&3.16&$>$2.9&53.24&$>$50.35&13.15&$>$18.00&9\\
051016B&0.94&I&2.18&3.30&1.31&4.82&52.24&49.79(0.12)&10.31&18.00&10\\
051221A&0.5465&I&2.14&9.76&0.79&12.06&51.53&49.87(0.10)&10.19&18 & 11\\
060124&2.3&II&3.12&0.23&1.06&1.5&55.41&51.91(0.06)&12.11&$>$18.00&12\\
060206&4.05&I&2.62&1.00&2.65&1.8&54.48&51.18(0.02)&11.74&16.90&13\\
060210&3.91&I&2.24&1.00&5.75&2.7&54.33&51.39(0.14)&12.25&17.37&14\\
060502A&1.51&II&3.3&0.79&0.15&2.1&54.88&51.71(0.07)&12.04&18.00&15\\
060512&0.44&II&3.36&1.00&0.12&$>$6.1&52.38&$>$50.14&13.41&$>$19.09&16\\
060522&5.11&I&2.26&1.00&2.91&$>$4.8&53.17&$>$50.73&11.50&17.81&17\\
060526&3.21&I&2.14&1.01&5.26&2.82&53.41&50.49(0.02)&11.58&17.65&18\\
060604&2.68&I&2.54&1.00&2.38&$>$4.6&53.75&$>$51.26&12.02&17.59&19\\
060605&3.7&II&2.98&0.60&2.38&$>$1.6&54.21&$>$50.81&13.32&$>$18.00&20\\
060614&0.13&II&2.72&1.00&0.43&6.8&52.45&50.30(0.02)&10.74&$>$18.32&21\\
060714&2.71&I&2.12&1.00&4.26&$>$5.1&53.32&$>$50.91&11.17&17.94&22\\
060729&0.54&I&2.26&1.00&2.42&6.6&53.39&51.21(0.08)&9.93&17.54&23\\
060814&0.84&I&2.60&2.33&1.41&3.63&53.34&50.64(0.12)&12.6&18 0&24\\
060908&2.43&II&2.5&1.00&1.62&$>$8.4&52.78&$>$50.81&11.99&$>$18.26&25\\
060912&0.94&I&2.01&1.02&7.06&$>$4.4&52.88&$>$50.35&9.46&17.98&26\\
060926&3.2&I&2.01&1.01&15.71&$>$0.9&54.21&$>$50.26&11.21&17.21&27\\
061007&1.26&II&3.16&1.00&1.93&$>$7.6&53.99&$>$51.94&14.35&$>$18.22&28\\
061121&1.31&II&2.7&0.95&1.01&1.93&53.88&50.63(0.06)&11.57&18.01&29\\
070110&2.35&I&2.72&1.00&1.70&$>$7.8&54.31&$>$52.27&11.38&16.98&30\\
\enddata
\tablenotetext{a}{The spectral regime of the X-rays:
I---$\nu_X>\max(\nu_m,\nu_c)$; II---$\nu_m<\nu_X<\nu_c$.}

\tablenotetext{b}{The kinetic energies are in units of ergs. The calculation of
the error of $E_{\rm K}$ for those bursts with detection of a jet break  takes
only the uncertainty of the jet break time into account. }

\tablenotetext{c}{The frequencies are in units of Hz. The $\nu_c$ for those
X-rays in the spectral regime II is a lower limit.}

\tablenotetext{d}{The reference of redshift.}

\tablerefs{1: Kelson \& Berger(2005); 2: Berger \& Mulchaey(2005); 3: Fynbo et
al.(2005a); 4: Fynbo et al.(2005b); 5: Cenko et al.(2005; 6: Berger et
al.(2005c); 7: Fynbo et al.(2005c); 8: Ledoux et al.(2005); 9: D'Elia et
al.(2005); 10: Soderberg et al.(2005); 11: Berger \& Soderberg(2005); 12: Cenko
et al.(2006a); 13: Aoki et al.(2006); 14: Cucchiara et al.(2006a); 15: Cucchiara
et al.(2006b); 16: Bloom et al.(2006a); 17: Cenko et al.(2006b); 18: Berger \&
Gladders(2006); 19: Castro-Tirado et al.(2006); 20: Still et al.(2006); 21:
Fugazza et al.(2006); 22: Jakobsson et al.(2006a); 23: Thoene et al.(2006) ;
24:Thoene (2007); 25:Rol et al.(2006); 26: Jakobsson et al.(2006b); 27: D'Elia et
al.(2006); 28: Jakobsson et al.(2006c); 29: Bloom et al.(2006b); 30: Jaunsen et
al.(2006) }

\end{deluxetable}

\begin{deluxetable}{lllllllllllllllllll}


\rotate
\tablewidth{620pt} \tabletypesize{\tiny}

\tablecaption{Observations of pre-{\em Swift} GRBs derived parameters}

\tablenum{6}
\tablehead{\colhead{GRB}&\colhead{$z$\tablenotemark{a}}&\colhead{Reg.}&\colhead{time
(s)\tablenotemark{a}}&\colhead{$F_x(\delta
 F_X$)\tablenotemark{a}}&\colhead{$\alpha(\delta \alpha)$\tablenotemark{a}}&\colhead{$\beta$\tablenotemark{b}}&\colhead{$t_j$(ks)}&\colhead{$\theta^{o}$}&\colhead{$\theta_j$ (rad)\tablenotemark{a}}&\colhead{$p$}&\colhead{$\epsilon_{B,-4}$}&\colhead{$Y$}&\colhead{$E_{\rm K,iso}$}&\colhead{$E_{\rm K}$}&\colhead{$\log \nu_m$}&\colhead{$\log \nu_c$}}
\startdata
970508&0.835&I&47160&7.13&1.1&$1.14^{+0.51}_{-0.36}$& 2160.00(432.00)& 16.7&0.391&2.28&3.4&1.34&52.53&51.15(0.07)&11.26&18.00\\
970828&0.958&II&14400&118&1.44(0.07)&$1.1^{+0.3}_{-0.3}$&190.08(34.56)&3.9&0.128&3.2&0.99&0.80&54.31&51.68(0.06)&13.46&18.20\\
980703&0.966&I&122400&4(1)&1.24(0.18)&$1.77^{+0.6}_{_0.47}$&214.92(10.15)&6.1&0.2&2.1&1.01&3.52&52.91&50.67(0.02)&9.81&17.87\\
990123&1.6&I&84240&19.11(2.2)&1.41(0.05)&$0.99^{0.07}_{-0.08}$&155.13(787.86) &2.9&0.089&2.98&0.40&1.51&54.77&51.87(0.18)&12.32&18.00\\
990510&1.619&I&42120&32.8(1.4)&1.41(0.18)&$1.19^{+0.14}_{-0.14}$&101.91(124.81)&3.1&0.054&2.38&1.00&3.63&53.98&51.14(0.04)&12.07&17.62\\
990705&0.84&II&52200&1.9(0.6)&--&1.05&86.40(17.28)&3.8&0.096&3.1&0.99&0.22&53.52&50.85(0.06)&12.18&18.67\\
991216&1.02&II&39240&250(10)&1.61(0.07)&$0.7^{+0.1}_{-0.1}$&248.71(67.63)&3.7&0.051&2.01&1.0&11/11&54.79&52.12(0.11)&9.57&17.29\\
000926&2.307&I&197640&2.23(0.77)&--&$0.9^{+0.3}_{-0.2}$&175.18(4.62)&3.3&0.14&2.01&1.00&7.11&54.12&51.34(0.01)&8.29&17.16\\
010222&1.477&I&117720&1.87(0.18)&1.33(0.04)&$1^{+0.1}_{-0.1}$&80.35(12.96)&2.7&0.08&2.02&1.00&7.59&54.21&51.25(0.05)&9.21&17.29\\
011211&2.14&II&29600&0.248&0.95(0.02)&$1.16^{+0.03}_{0.03}$&198.66(16.68)&3.5&-&3.32&0.99&0.21&54.11&51.39(0.02)&12.84&18.31\\
020405&0.689&I&147600&13.6(2.5)&1.15(0.95)&$1^{+0.2}_{-0.1}$&236.88(15.90)&5.7&0.285&2.02&1.00&5.55&53.53&51.21(0.02)&8.63&17.67\\
020813&1.254&II&114840&22&1.42(0.05)&$0.8^{+0.1}_{-0.1}$&397.44(0.864)&2.2&0.066&2.6&0.55&1.86&54.08&50.95(0.01)&11.57&18.00\\
021004&2.323&I&113040&4.3(0.7)&1(0.2)&$1.1^{+0.1}_{-0.1}$&300.30(8.64)&4.7&0.24&2.2&1.00&3.69&53.61&51.13(0.01)&10.85&17.54\\
030329\tablenotemark{c}&0.1678&I&22377&157.0(8.7) &1.2(0.1)&$1.17\pm 0.04$ & 40.95(0.43) &3.8&0.052&2.34&2.15&2.23&53.09&50.43(0.01)&11.95&18.00\\

\enddata

\tablenotetext{a}{Taken from Berger et al. (2003) and Bloom et al. (2003).}

\tablenotetext{b}{Taken from Sako et al. 2005.}

\tablenotetext{c}{Taken from Willingale et al. (2004).}


\end{deluxetable}


\clearpage

\begin{figure}
\epsscale{0.8} \plotone{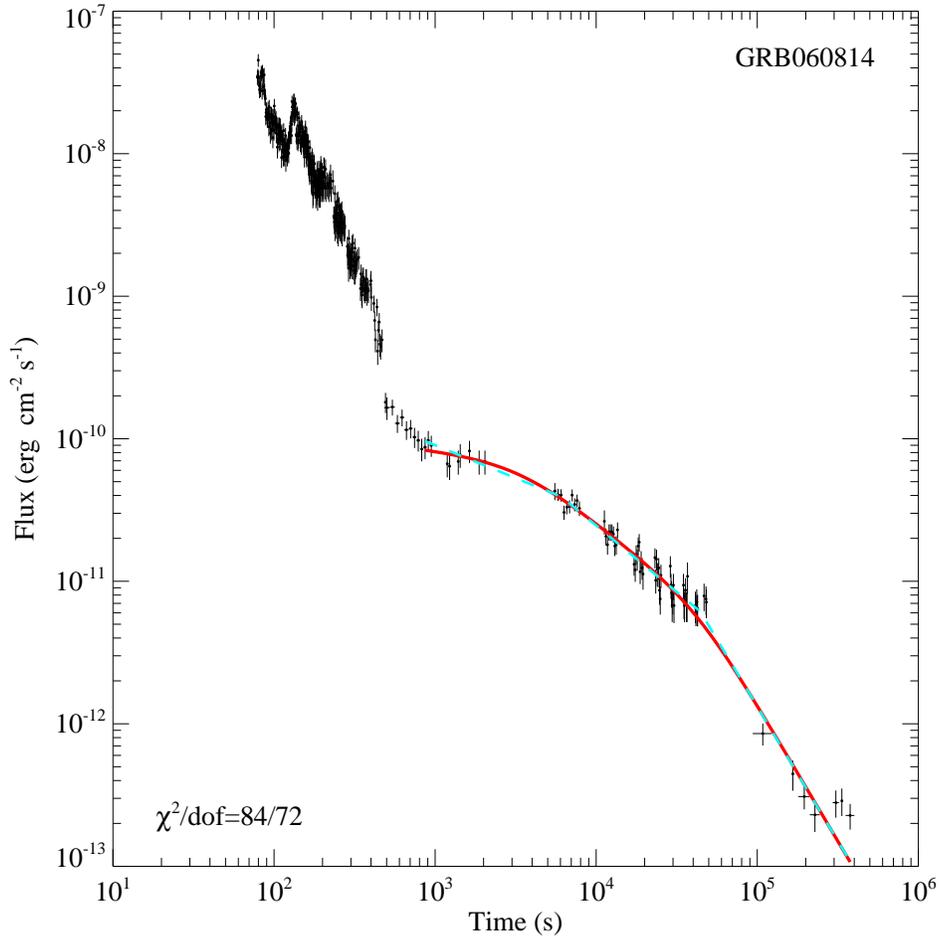} \caption{Comparison of the fitting results with
the STPL (solid line) and the JTPL(dashed) models. The last three data points are
excluded in the fits.} \label{Fig_fitcom}
\end{figure}

\clearpage
\thispagestyle{empty}
\setlength{\voffset}{-18mm}

\begin{figure*}
\includegraphics[angle=0,scale=0.35]{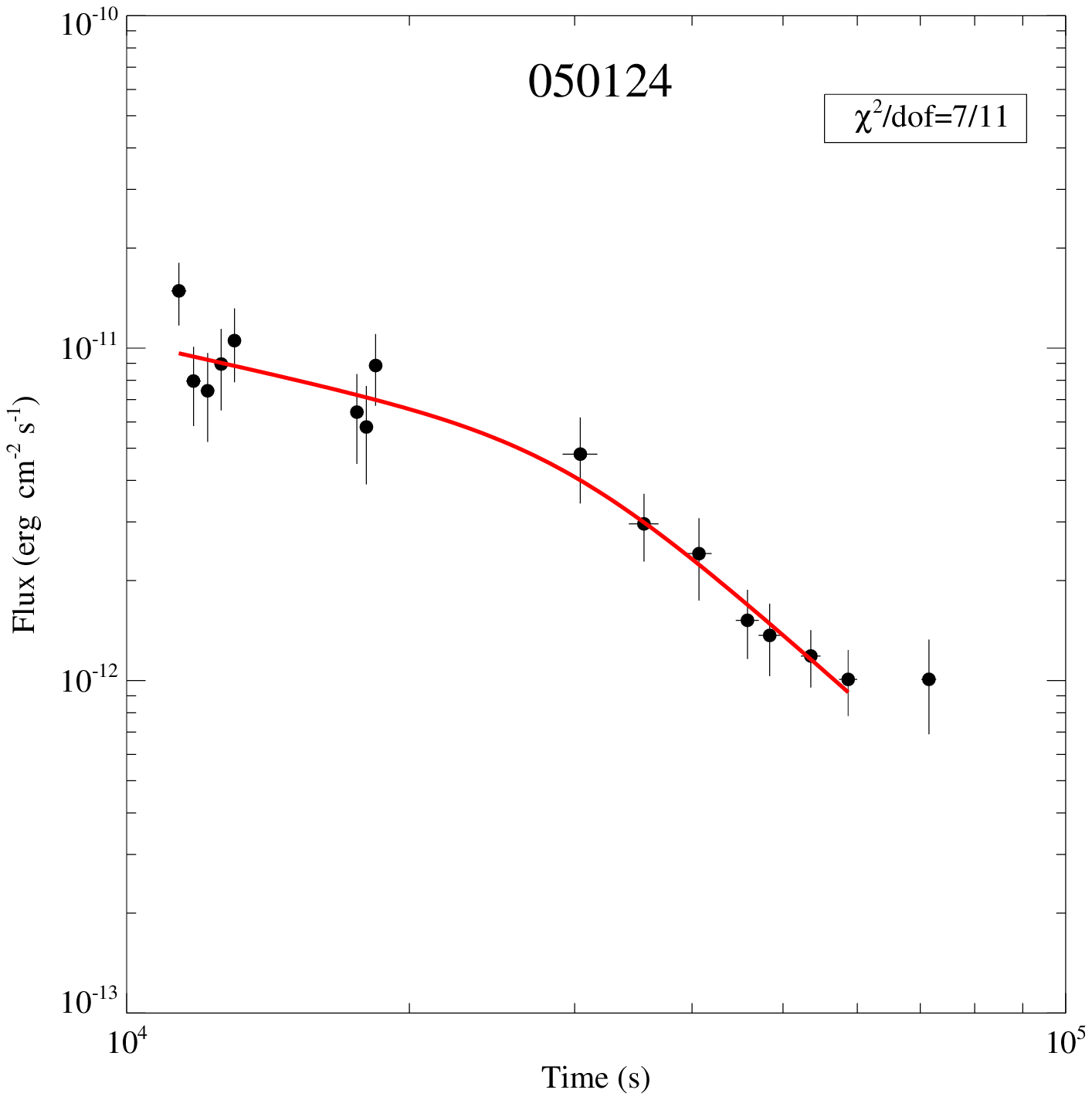}
\includegraphics[angle=0,scale=0.35]{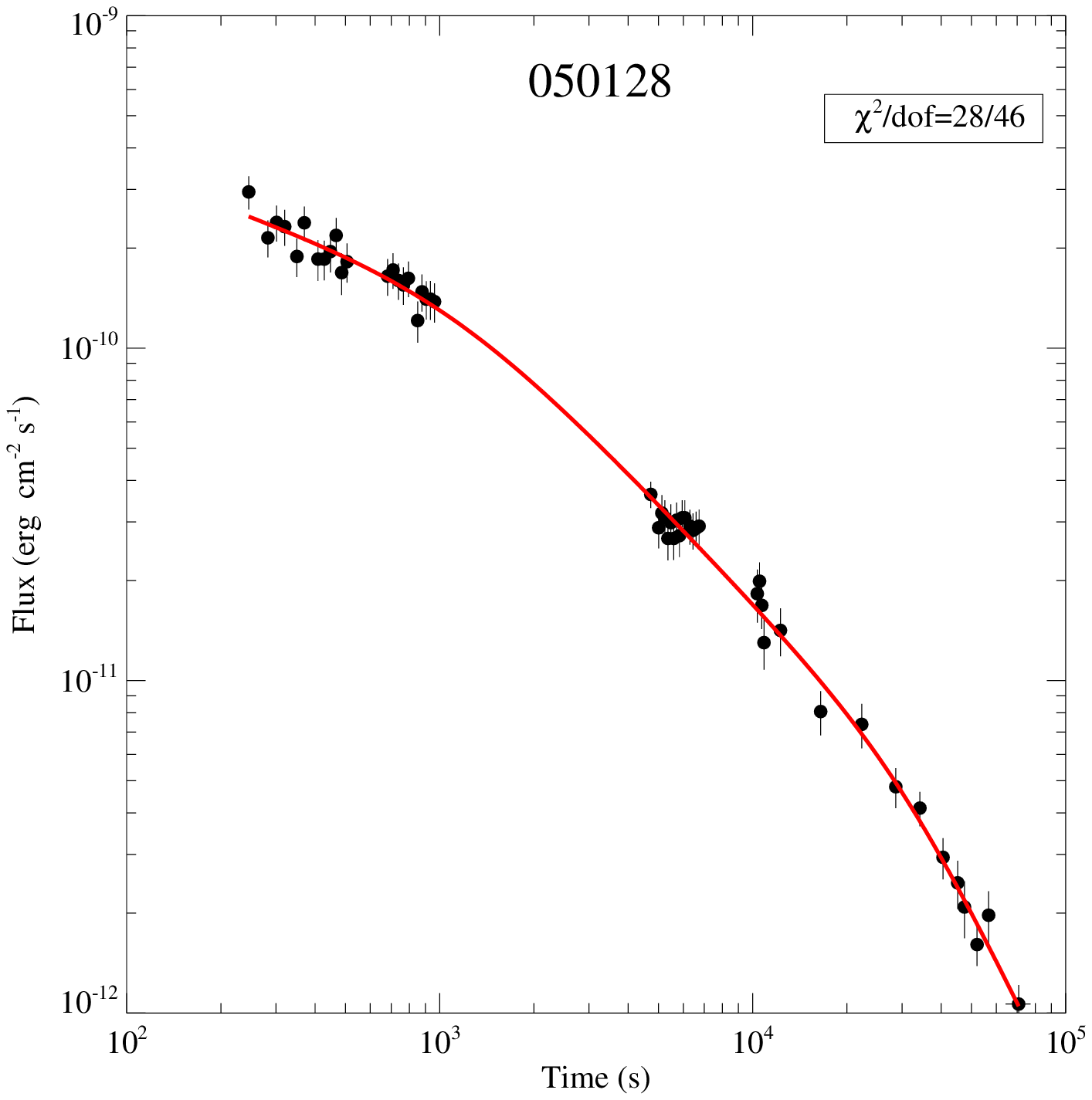}
\includegraphics[angle=0,scale=0.35]{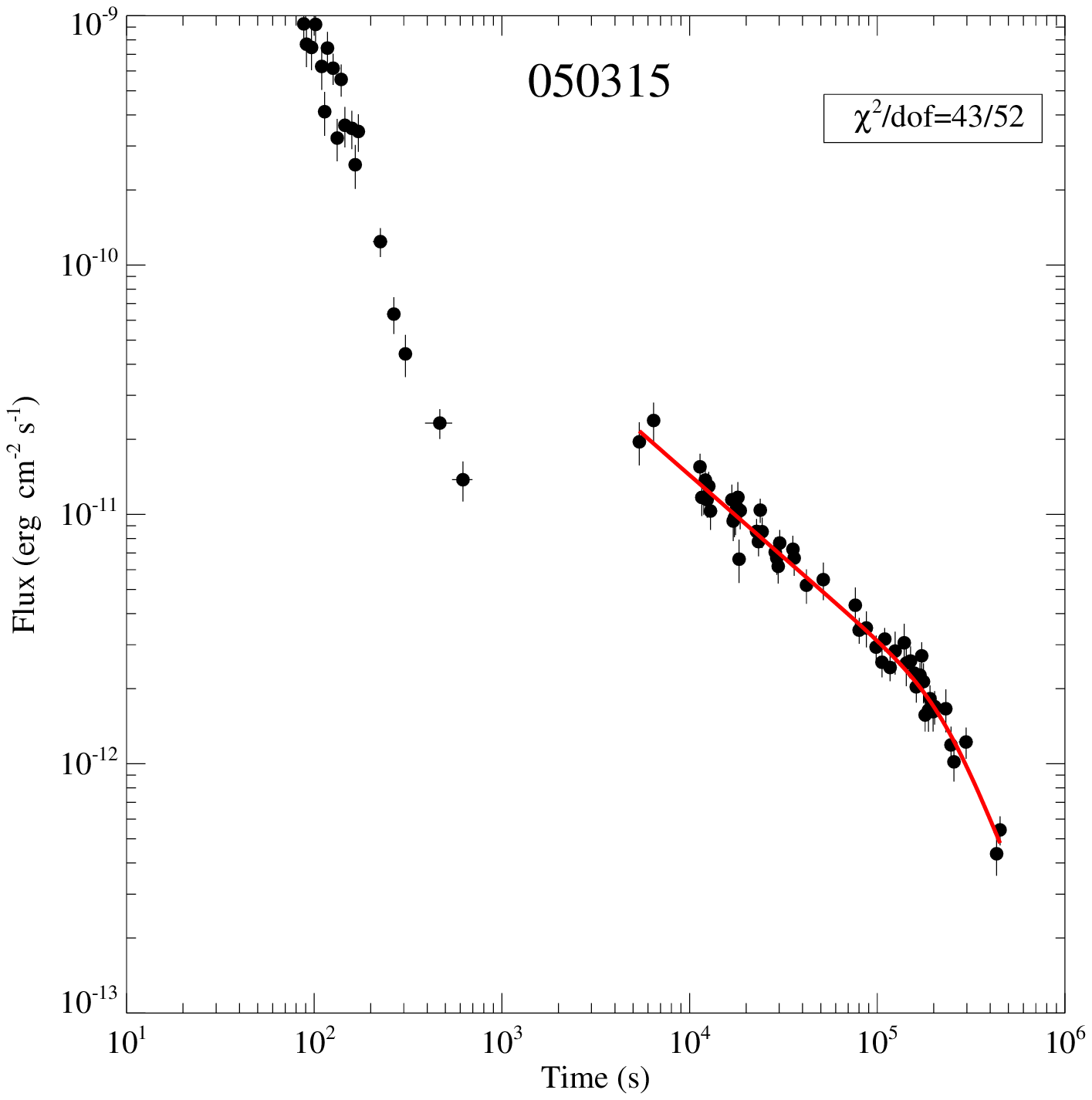}
\includegraphics[angle=0,scale=0.35]{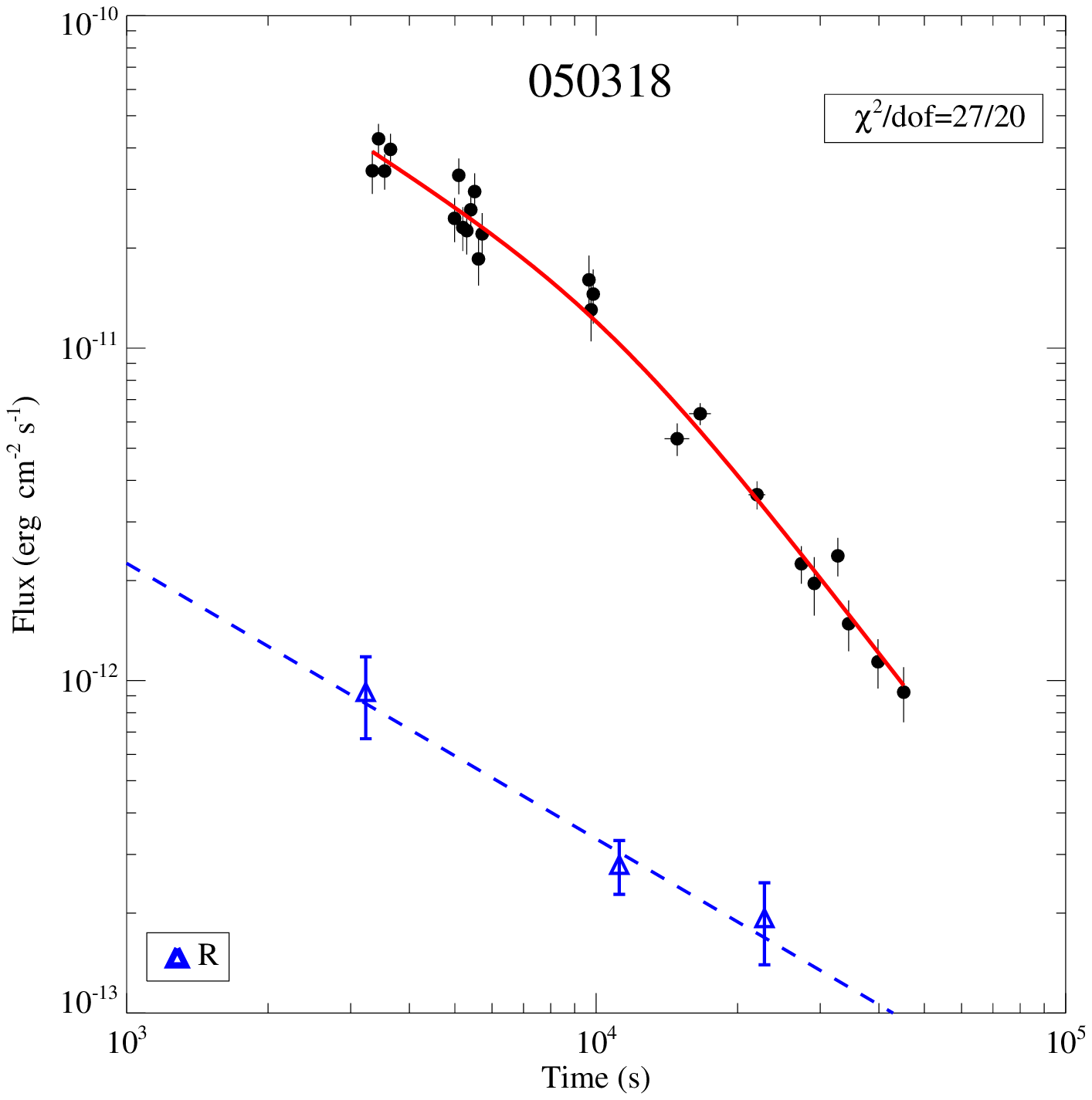}
\includegraphics[angle=0,scale=0.35]{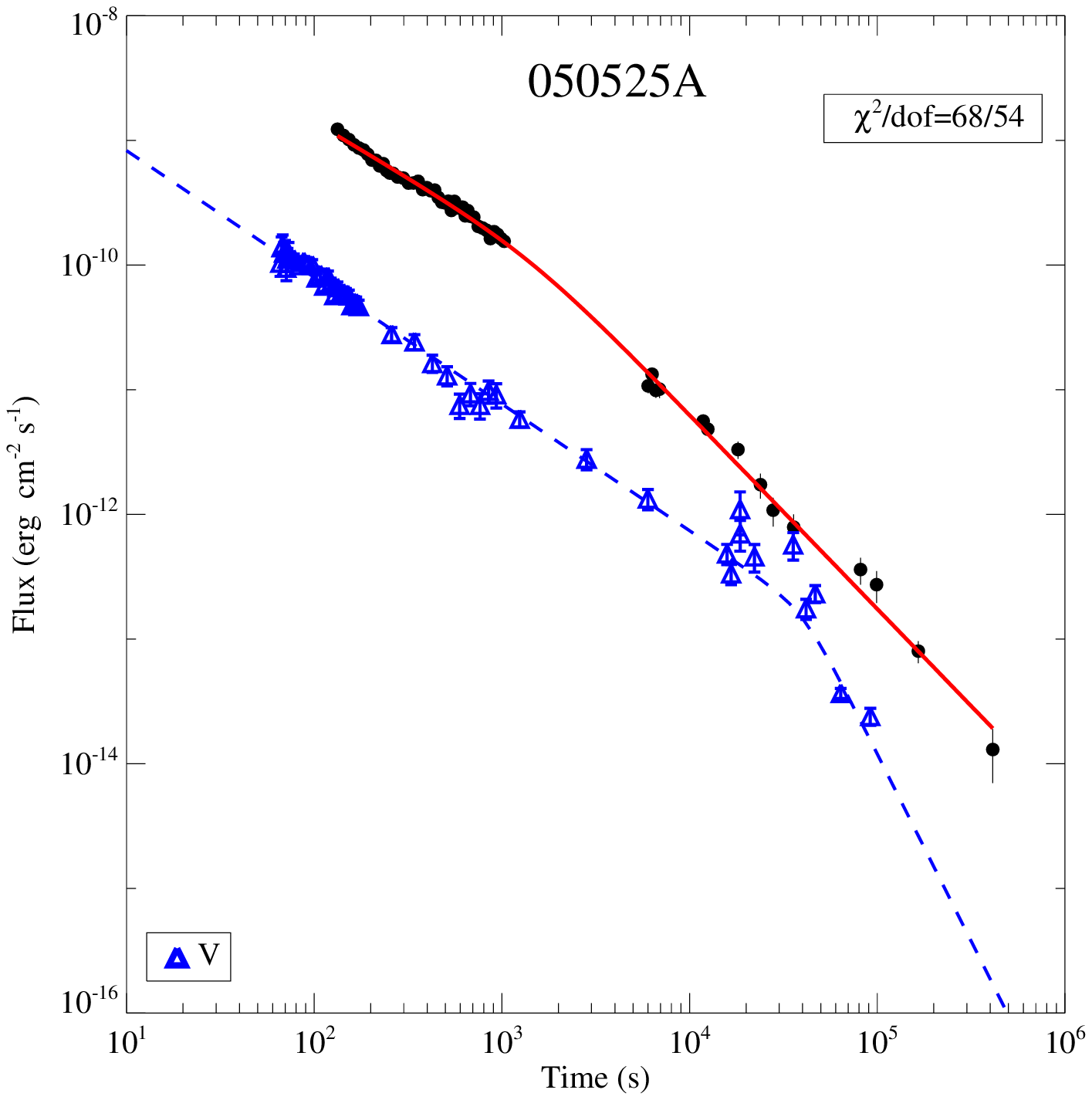}
\includegraphics[angle=0,scale=0.35]{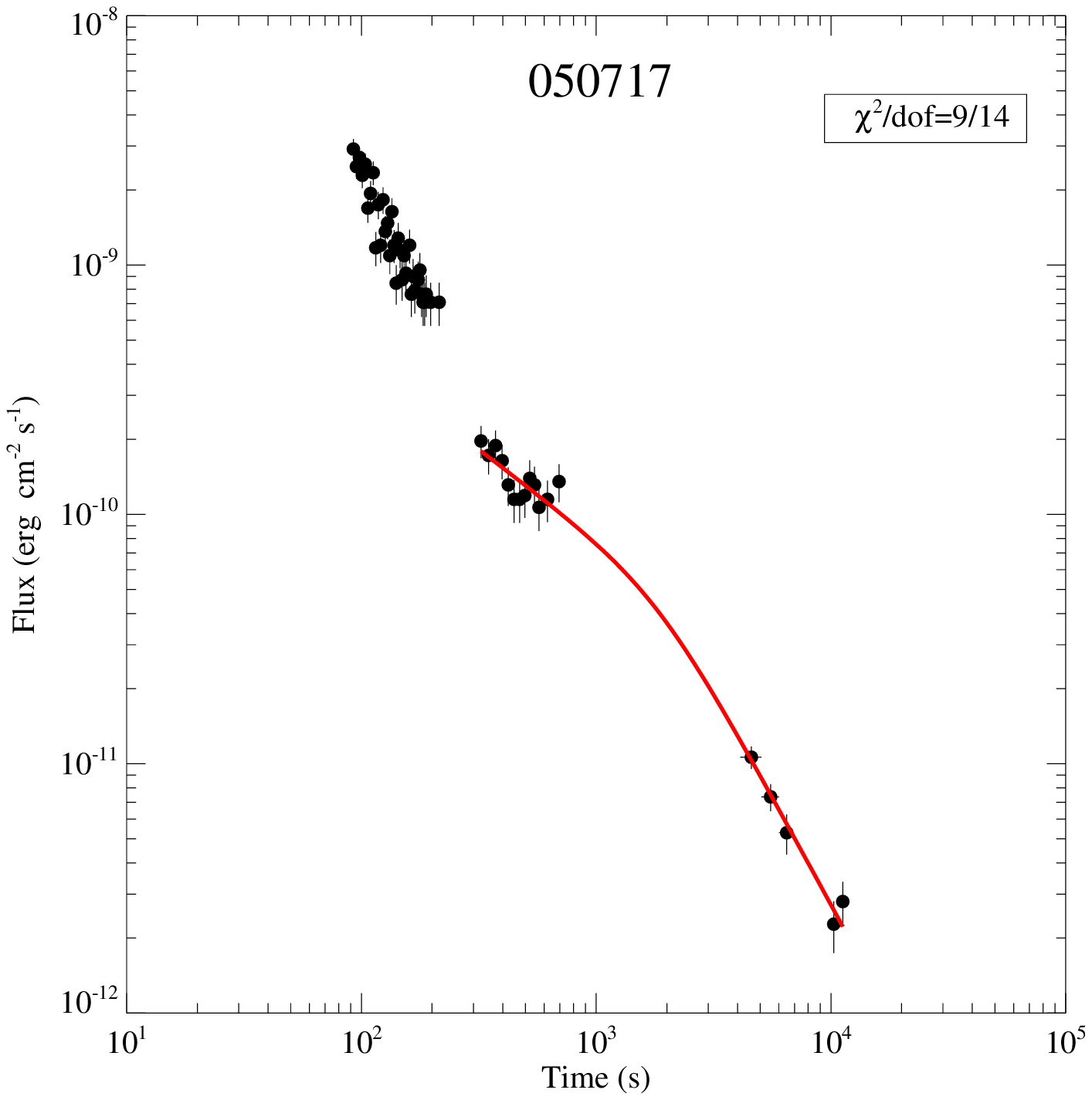}
\includegraphics[angle=0,scale=0.35]{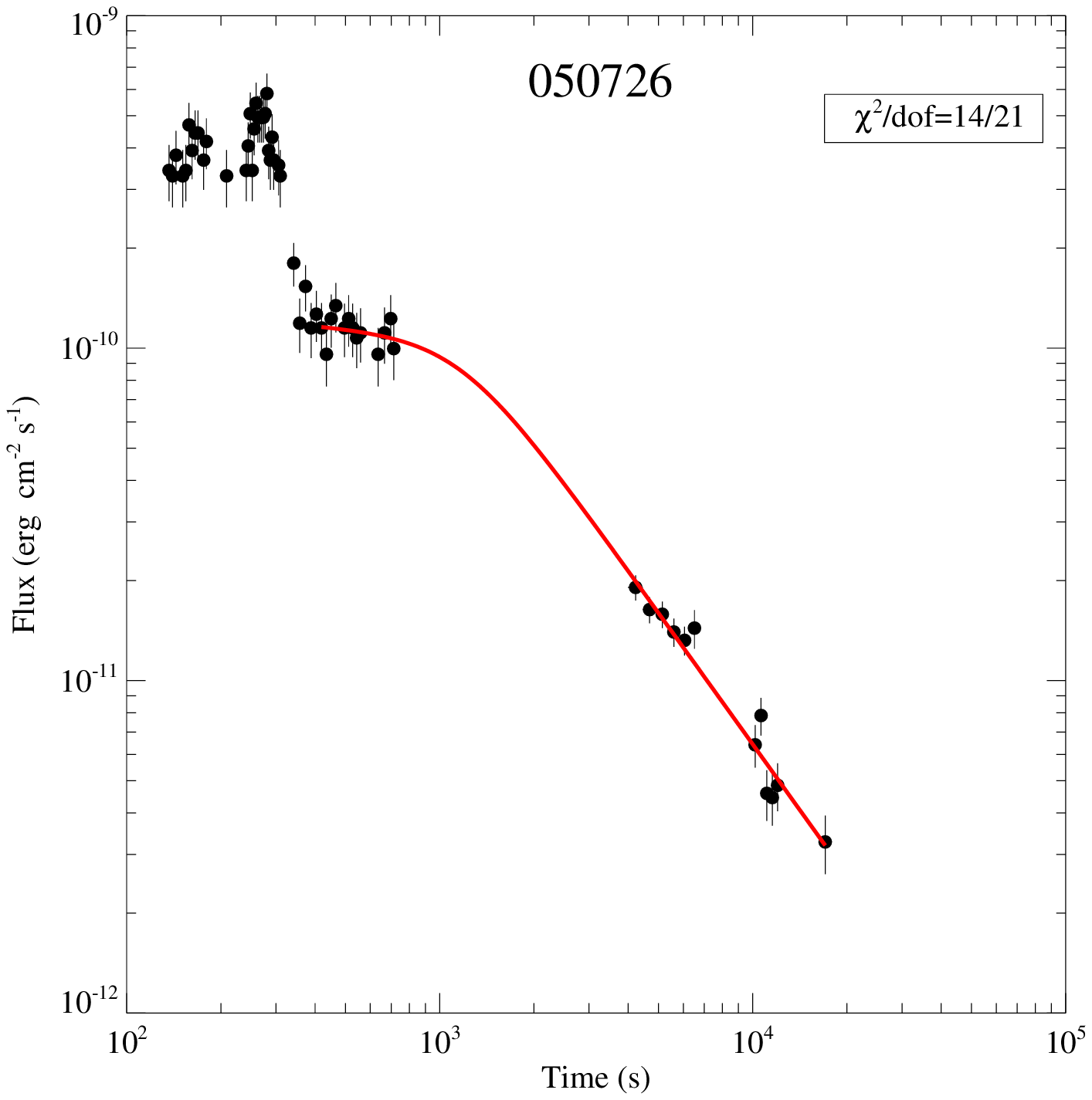}
\includegraphics[angle=0,scale=0.35]{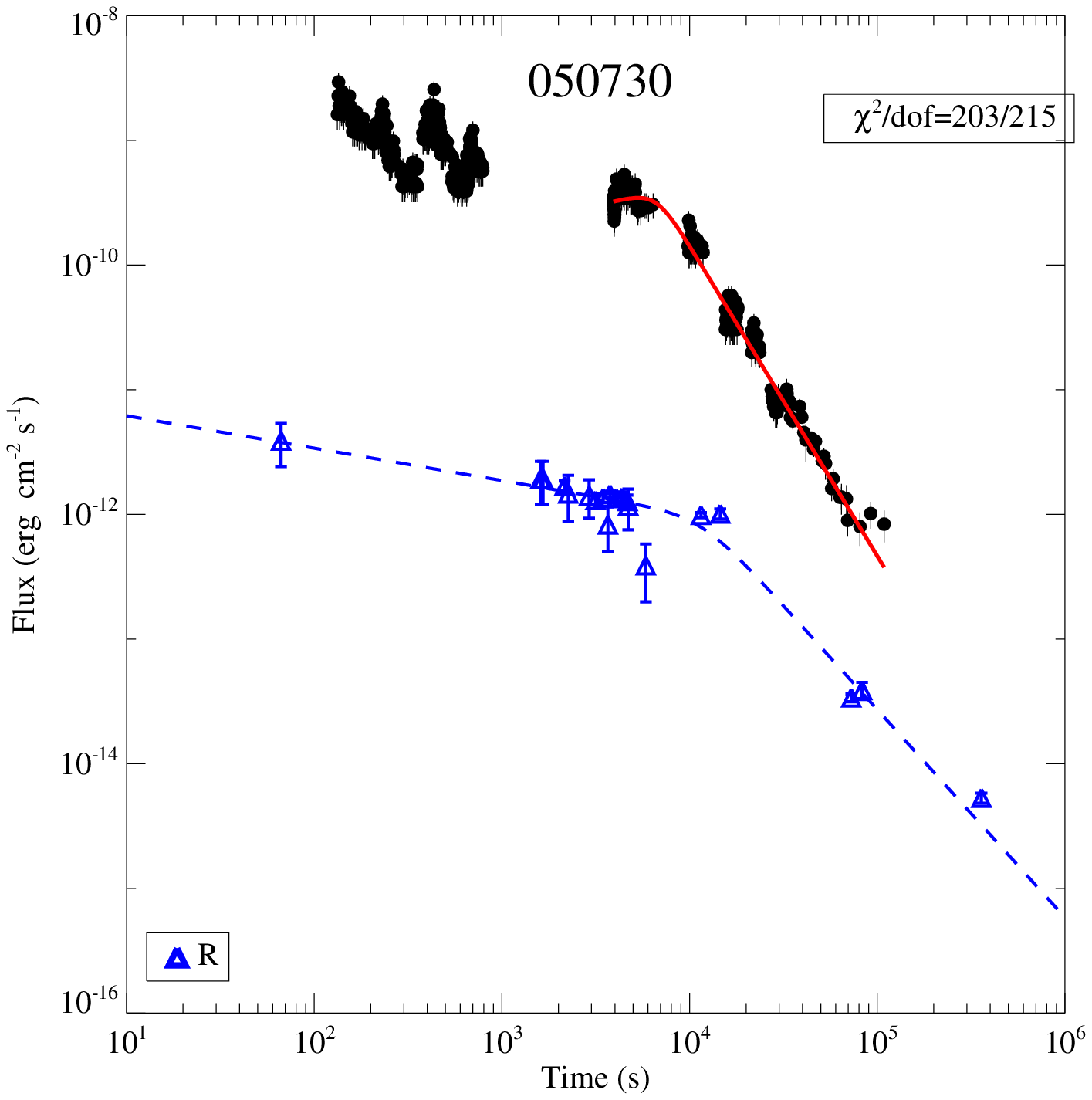}
\includegraphics[angle=0,scale=0.35]{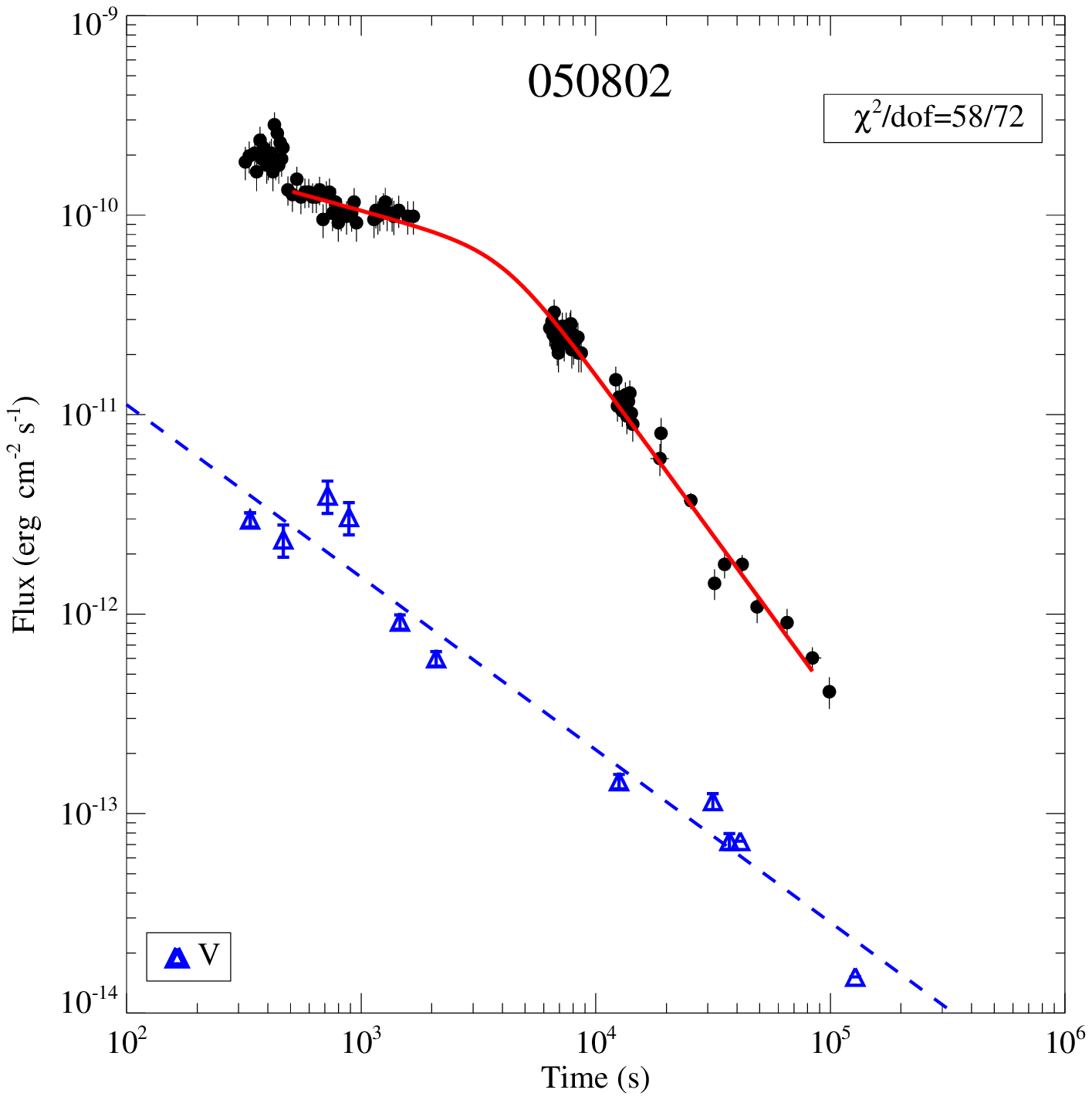}
\includegraphics[angle=0,scale=0.35]{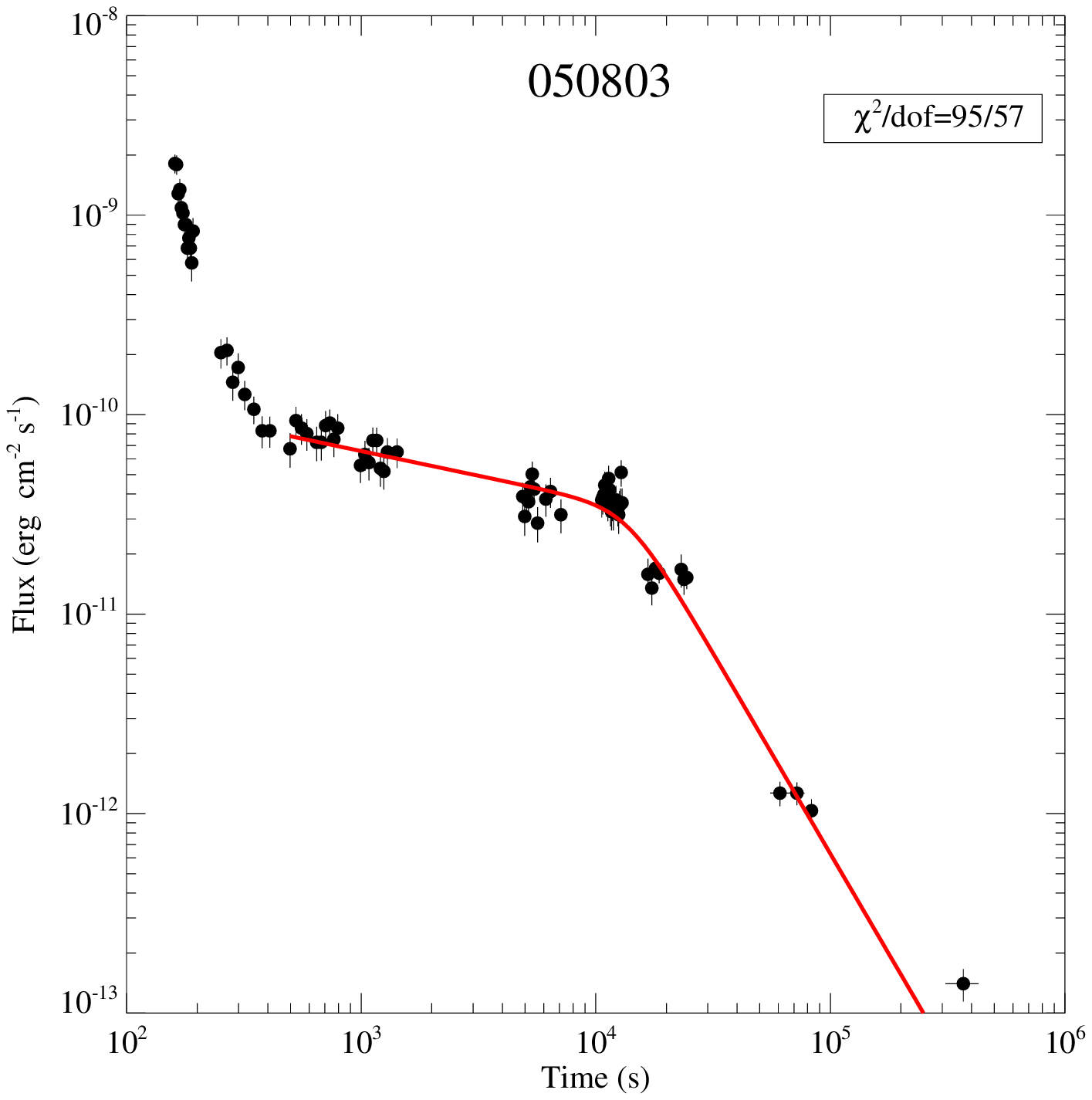}
\includegraphics[angle=0,scale=0.35]{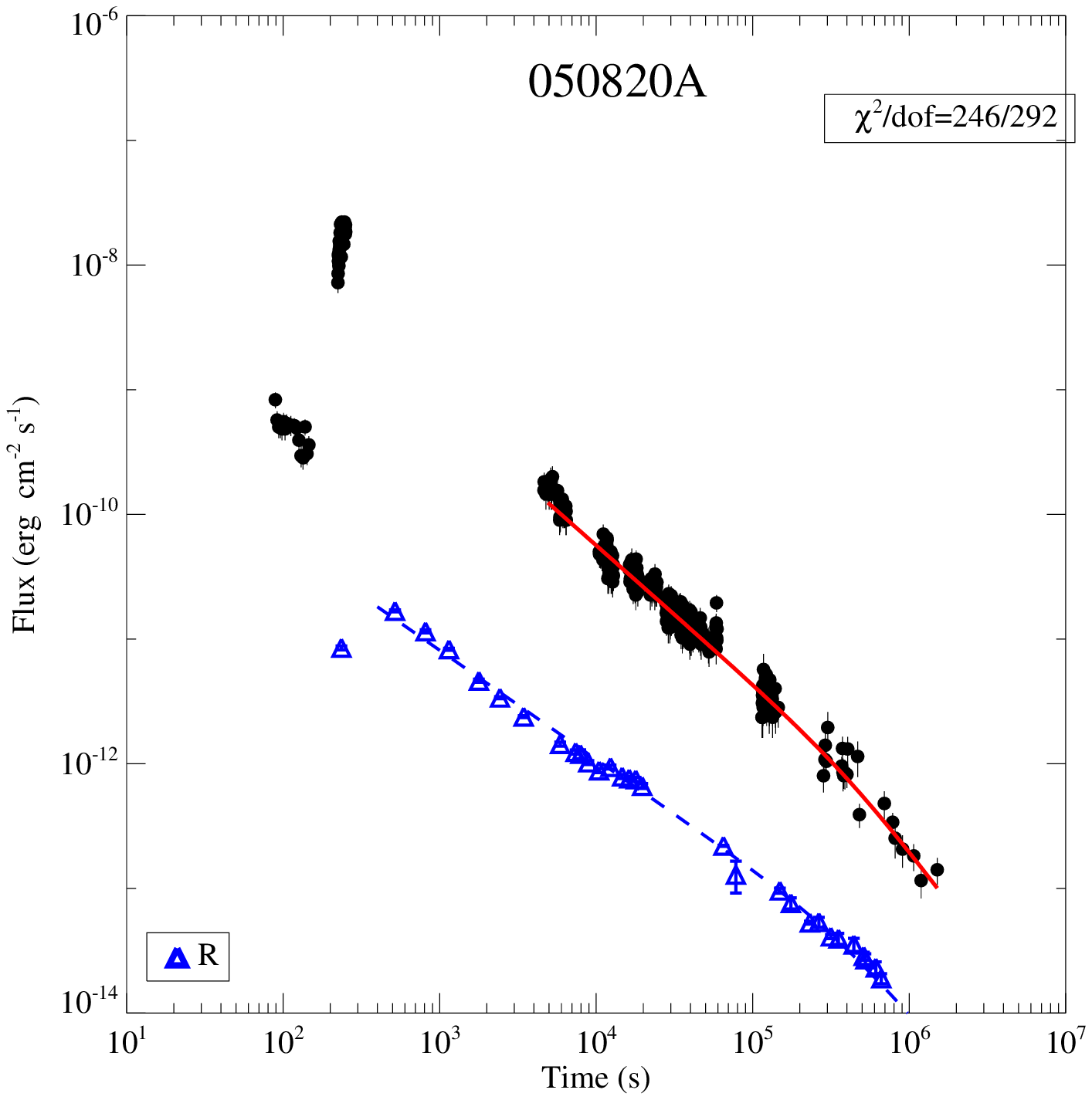}
\hfill
\includegraphics[angle=0,scale=0.35]{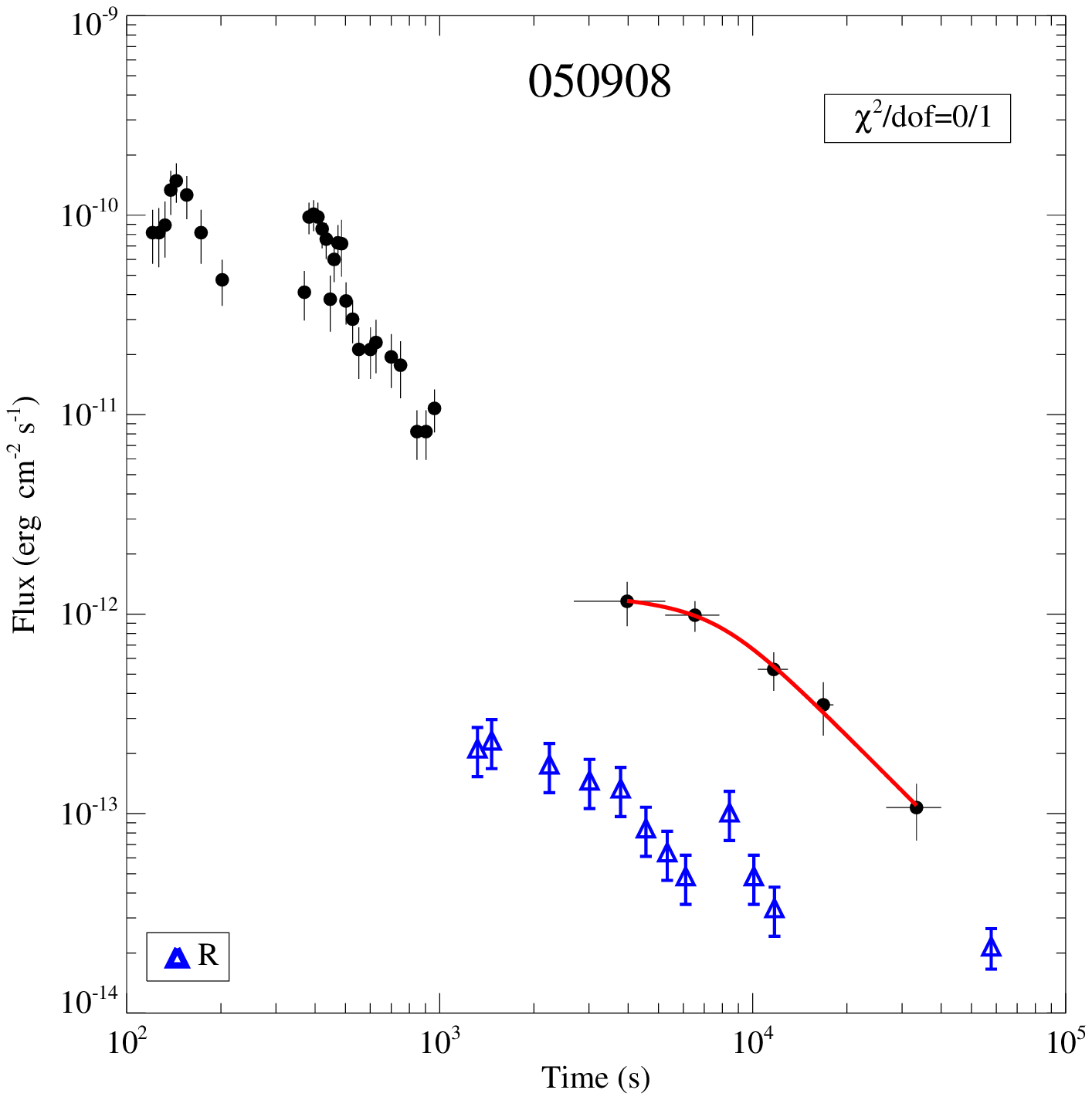}
\caption{The X-ray (solid dots) and optical (open triangles) lightcurves and
their fitting results as well for derived the jet break candidates
.}\label{XRT_LC}
\end{figure*}
\clearpage
\setlength{\voffset}{0mm}

\begin{figure*}
\includegraphics[angle=0,scale=0.35]{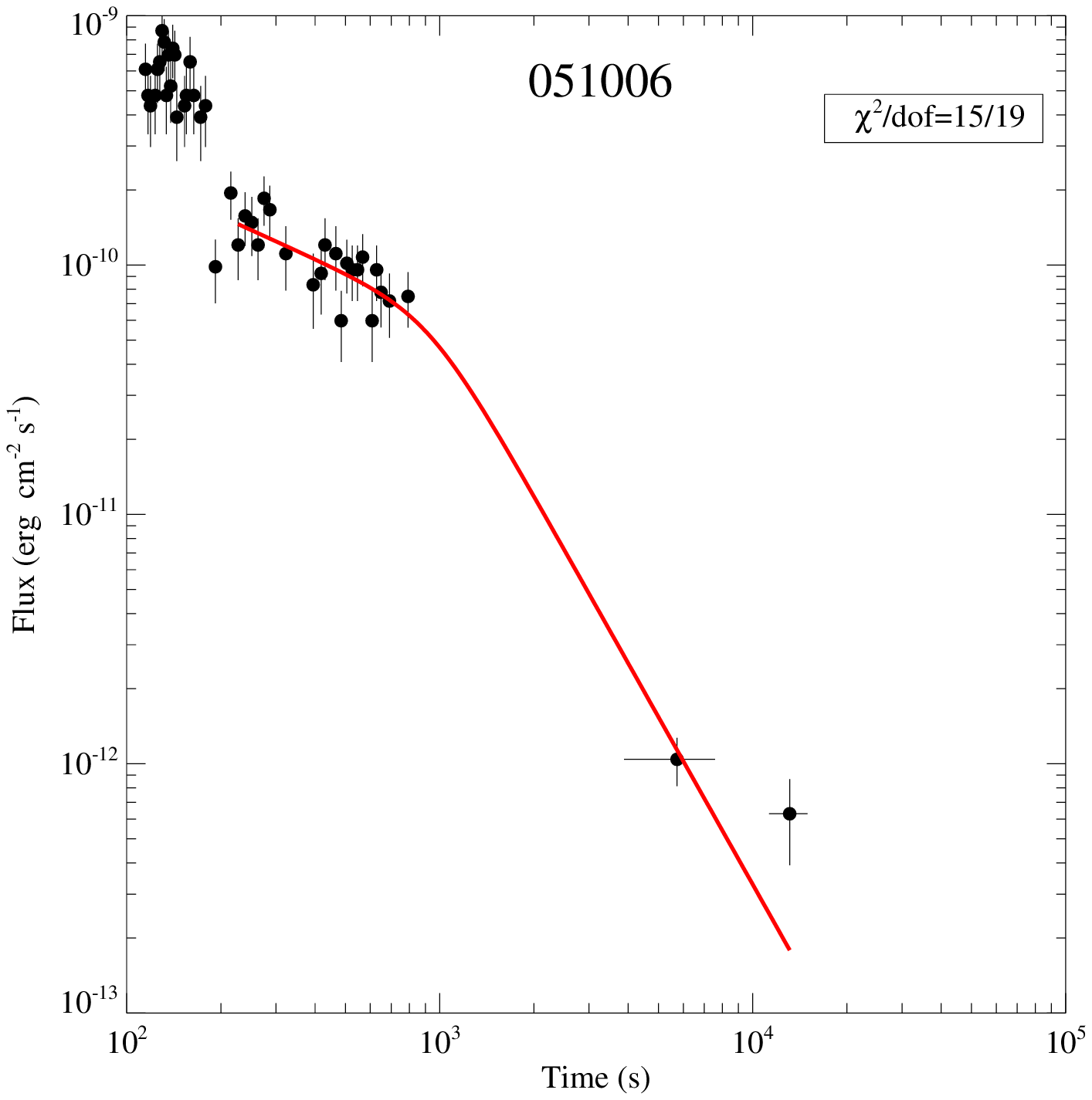}
\includegraphics[angle=0,scale=0.35]{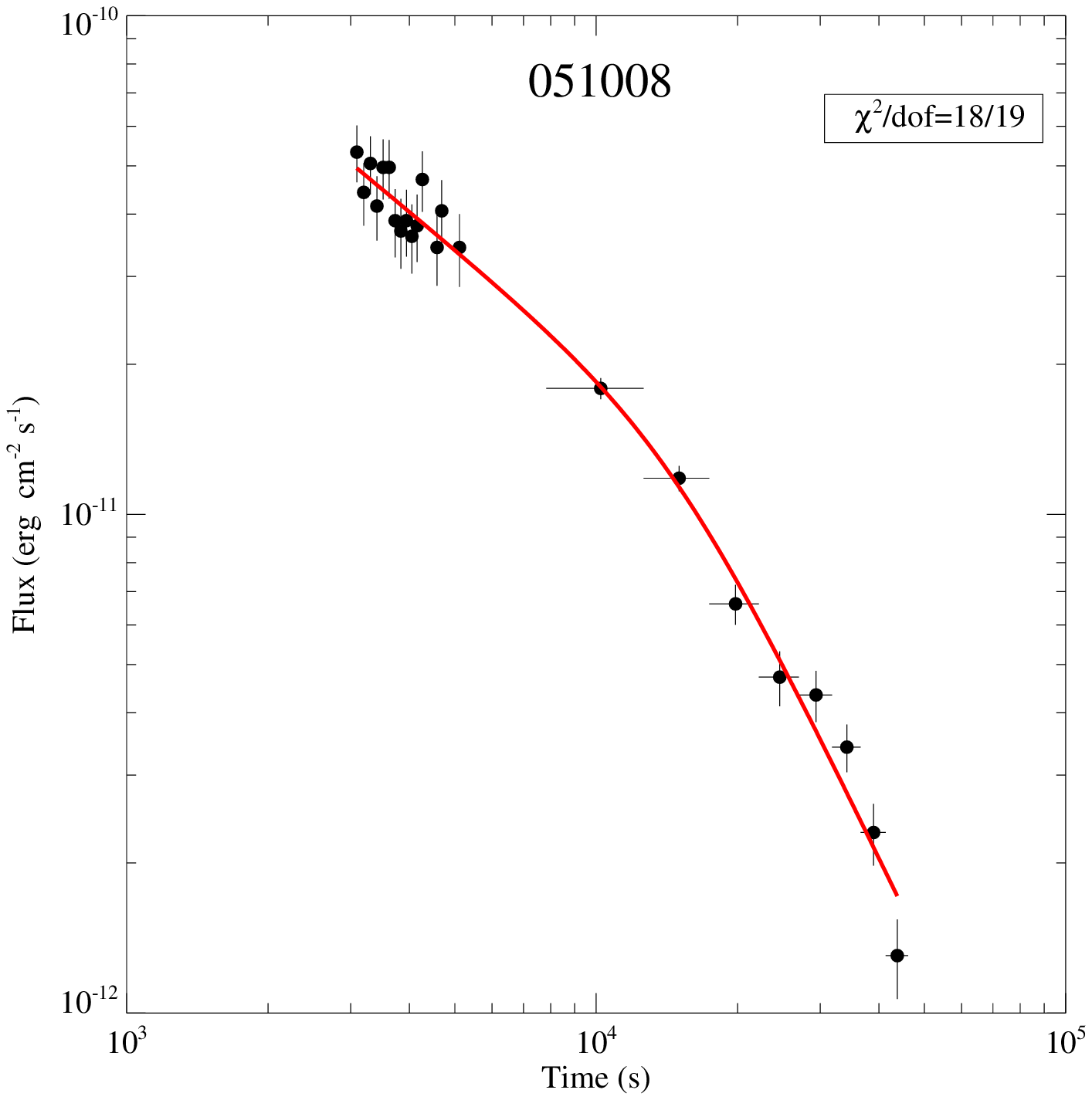}
\includegraphics[angle=0,scale=0.35]{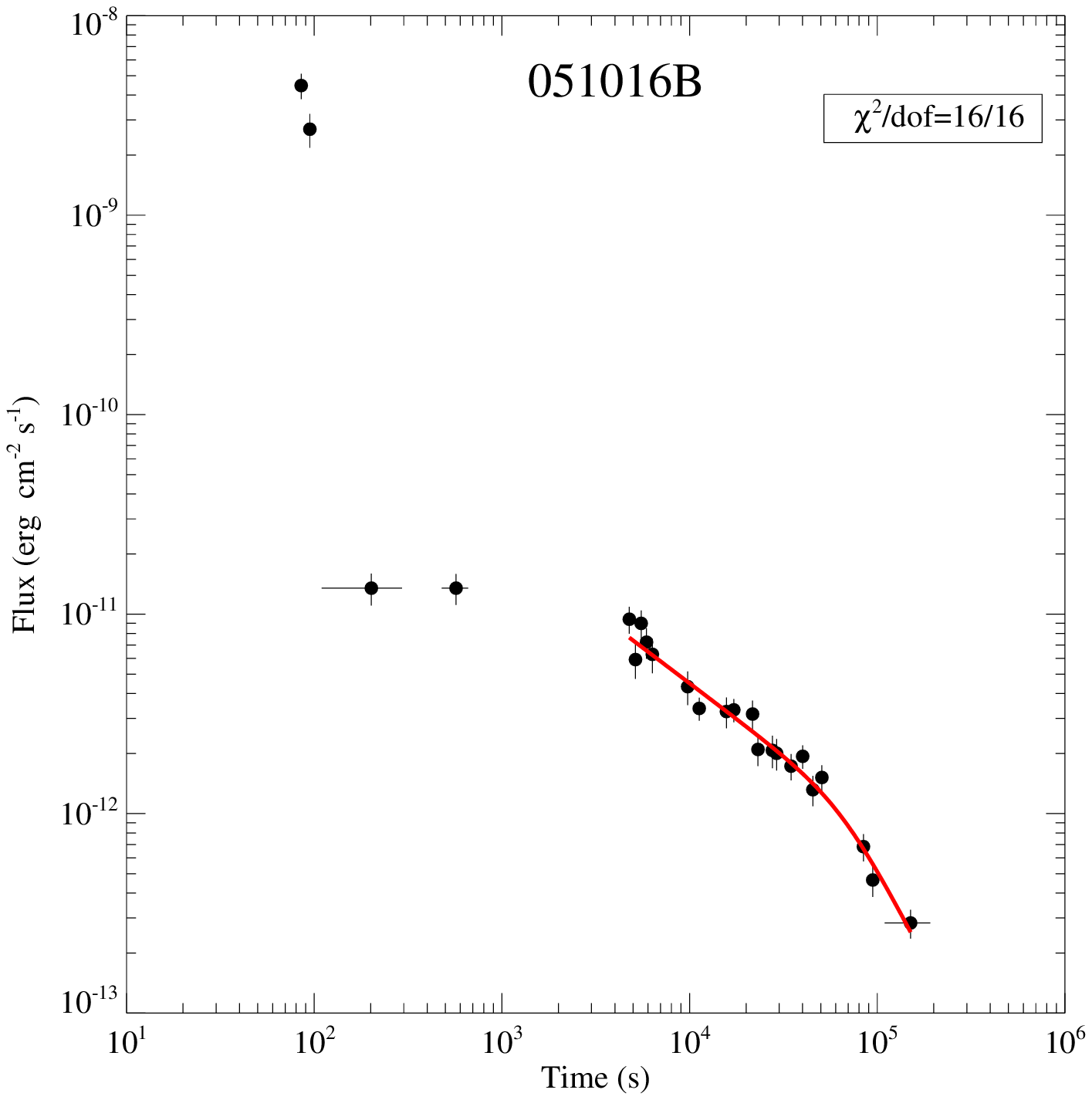}
\includegraphics[angle=0,scale=0.35]{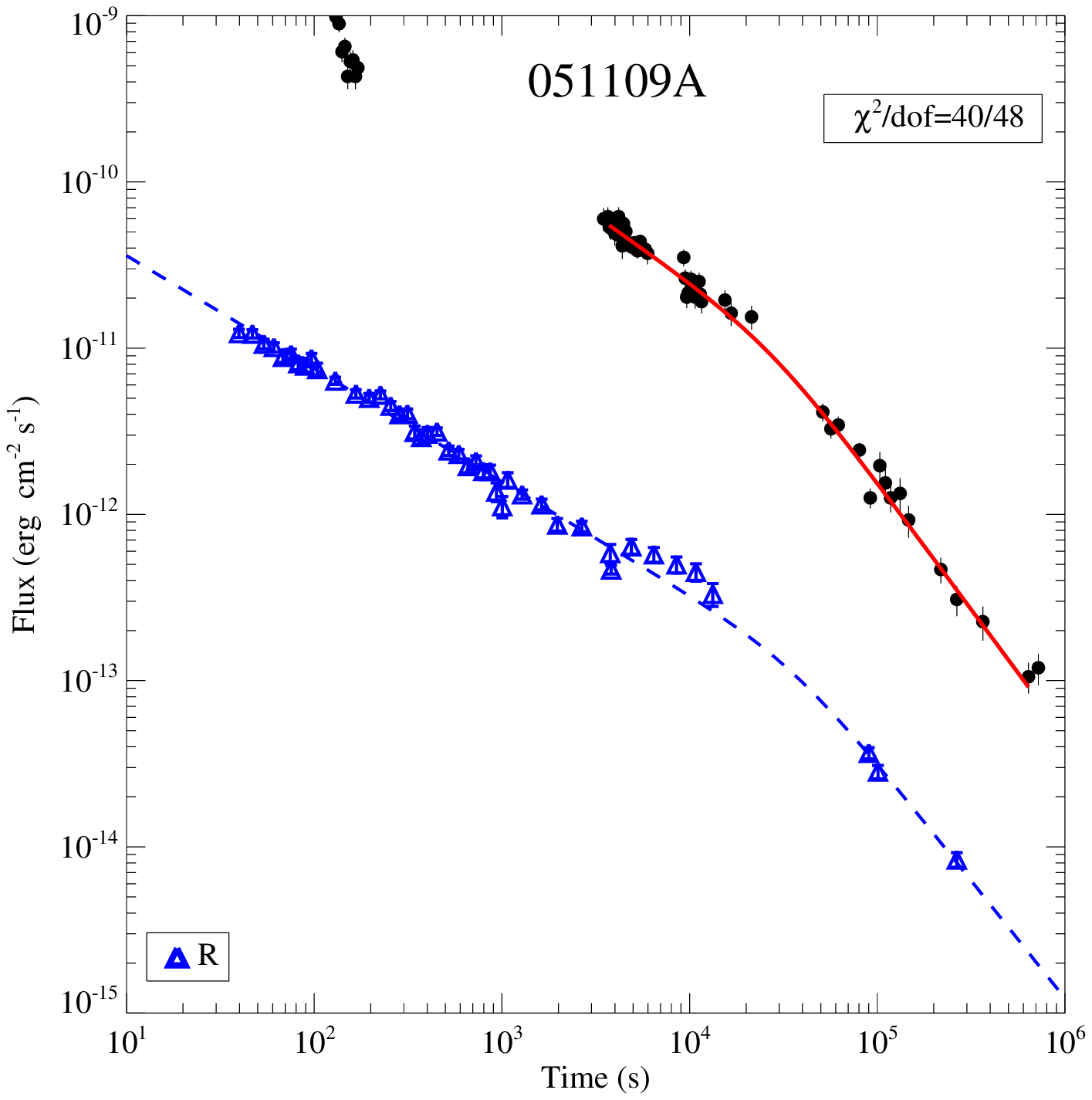}
\includegraphics[angle=0,scale=0.35]{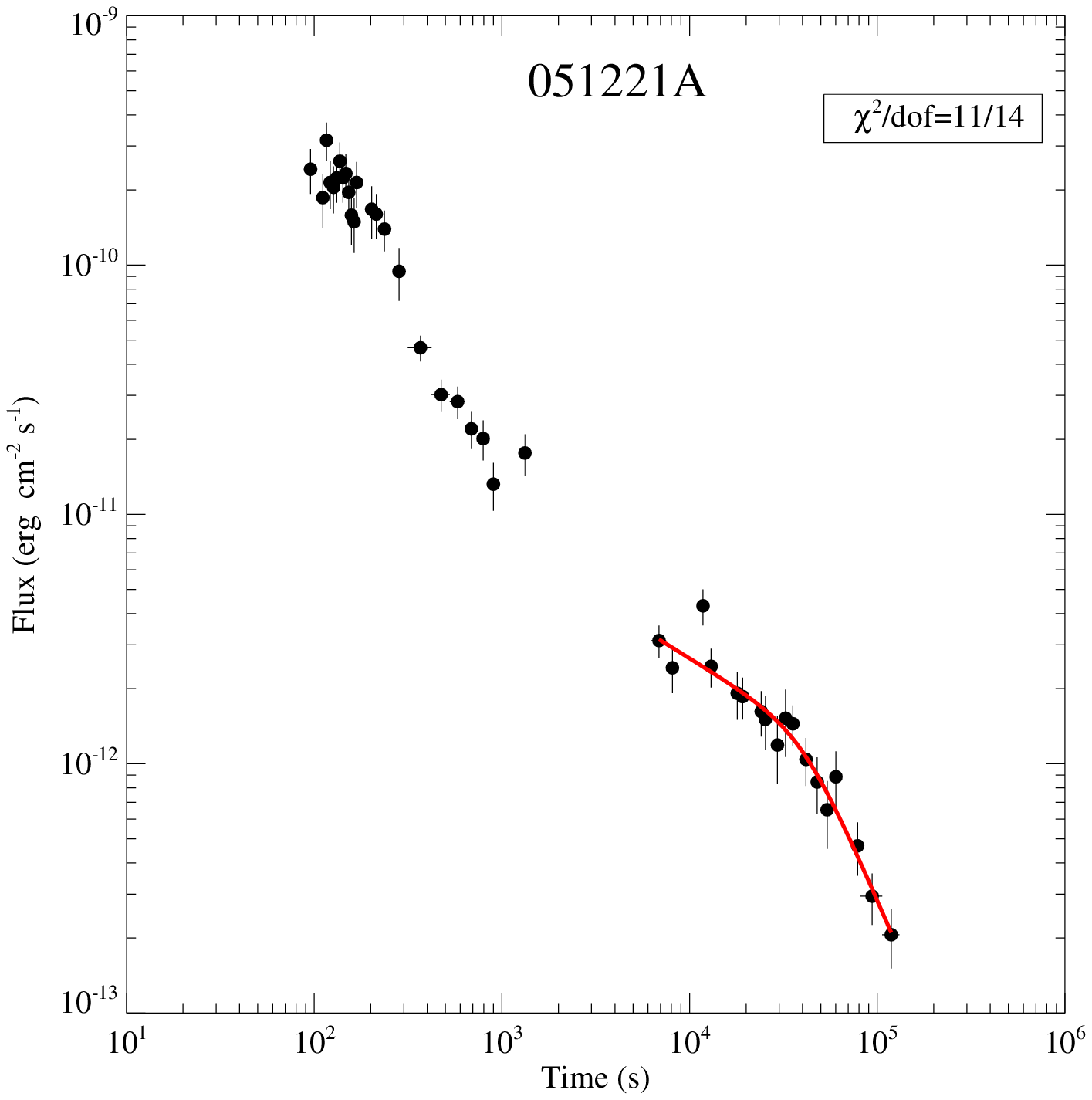}
\includegraphics[angle=0,scale=0.35]{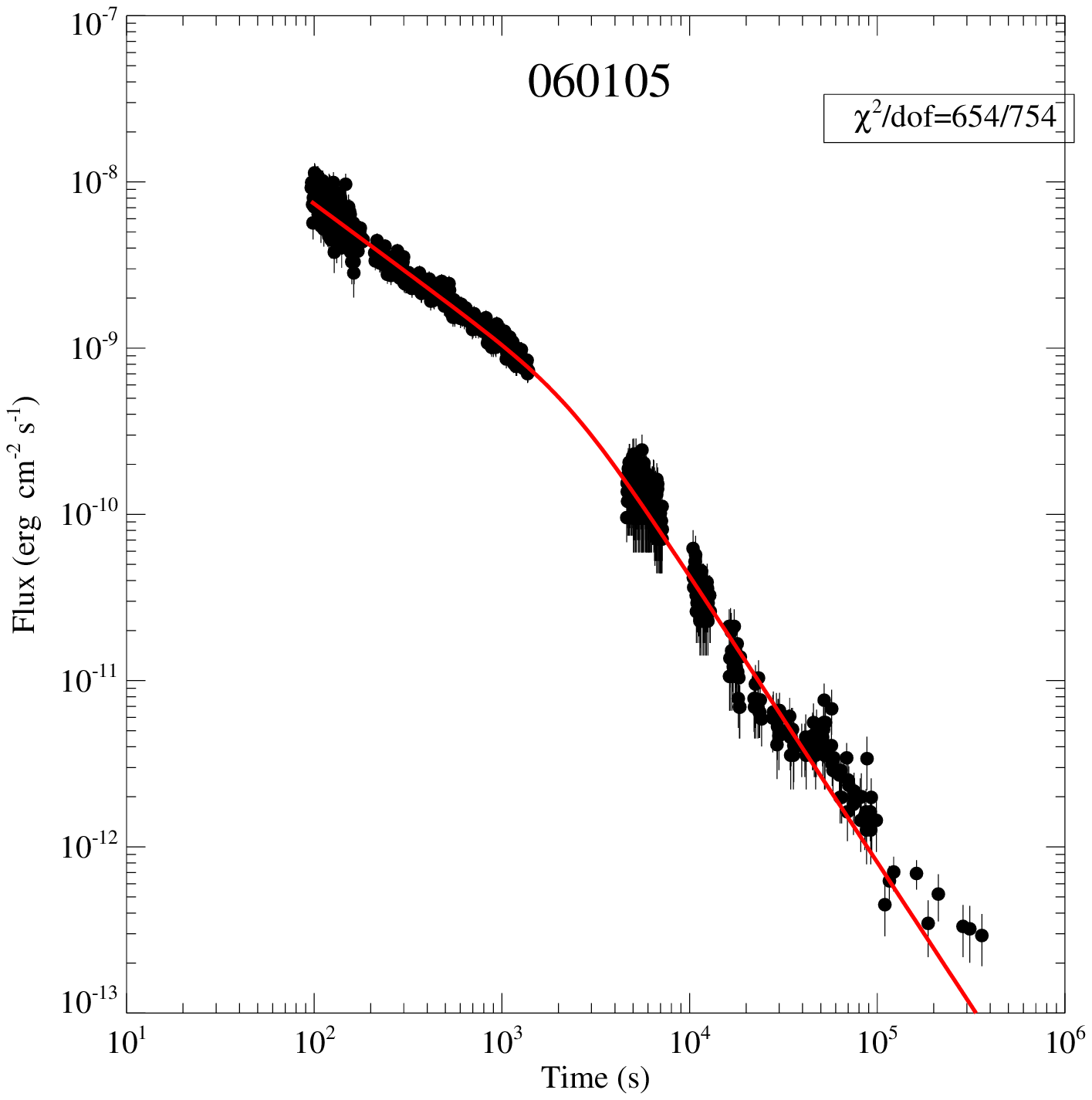}
\includegraphics[angle=0,scale=0.35]{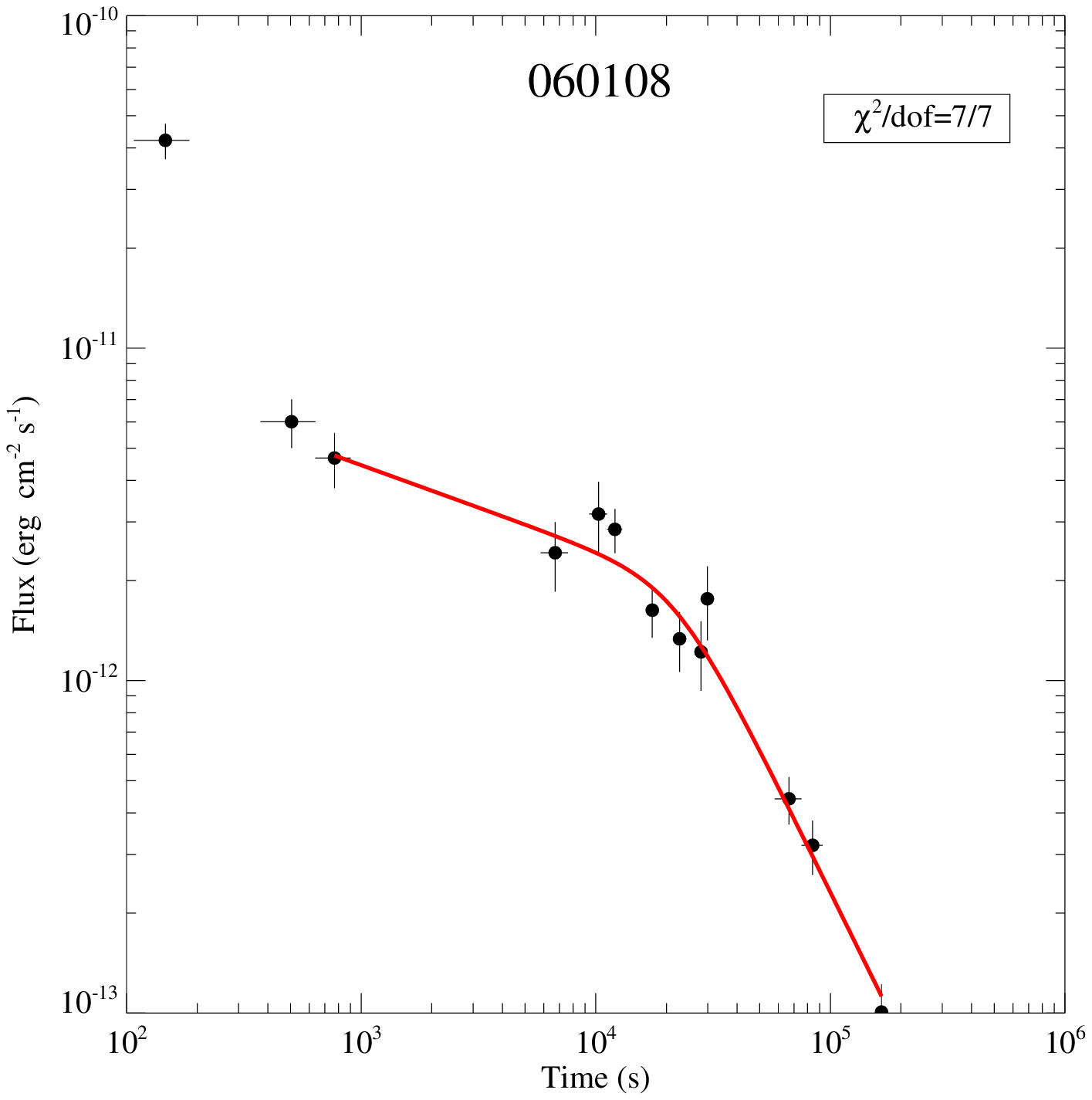}
\includegraphics[angle=0,scale=0.35]{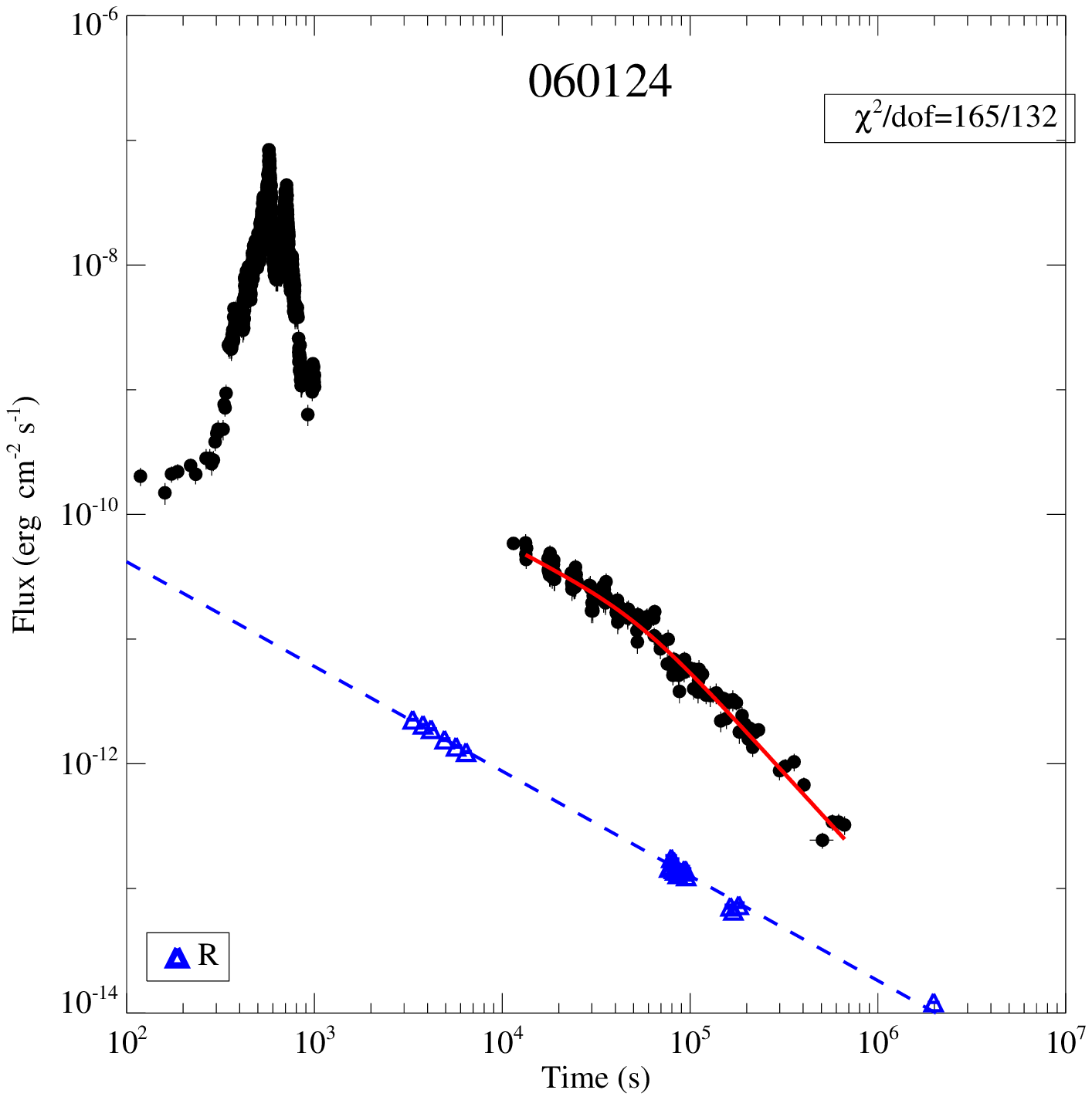}
\includegraphics[angle=0,scale=0.35]{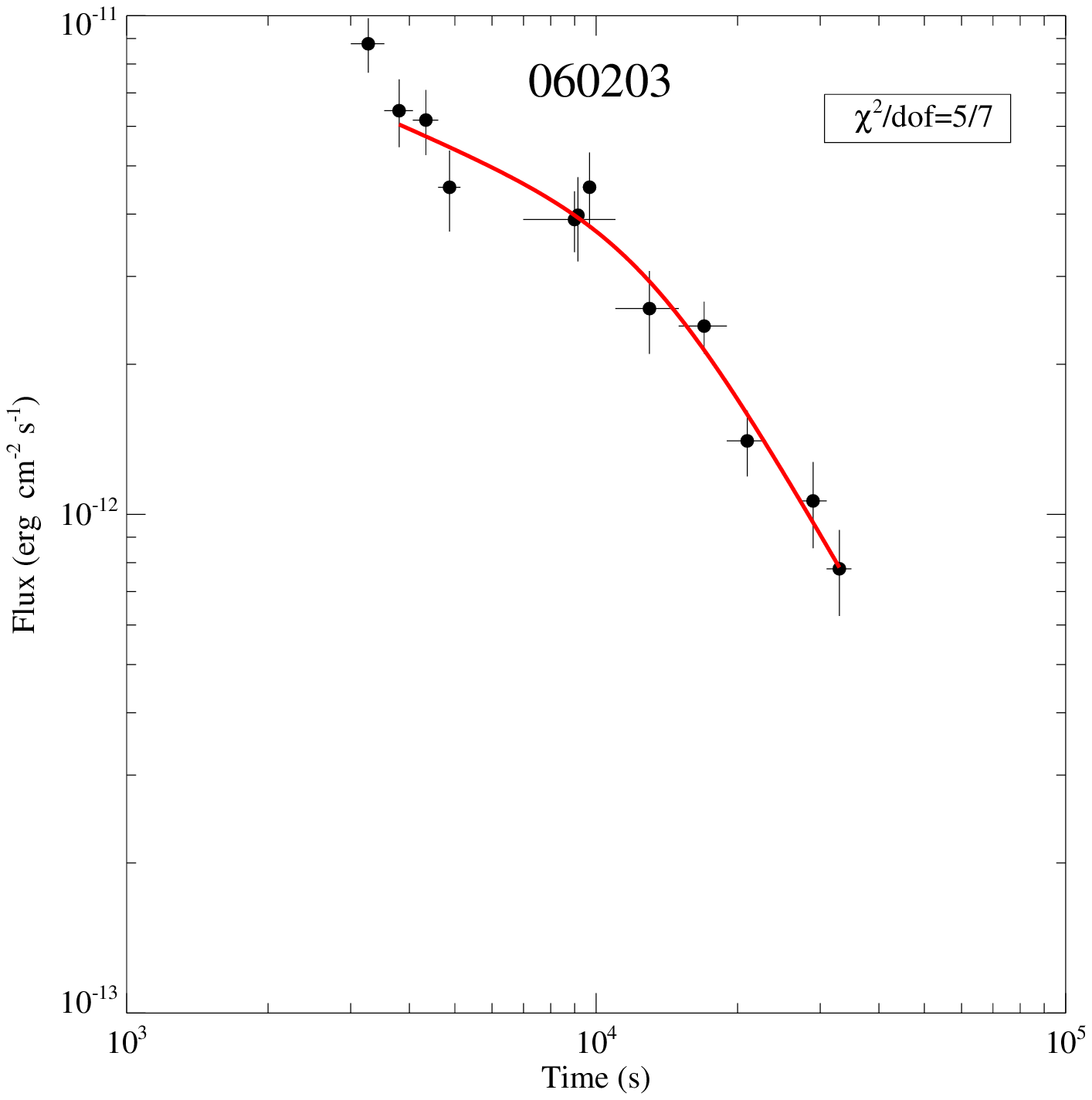}
\includegraphics[angle=0,scale=0.35]{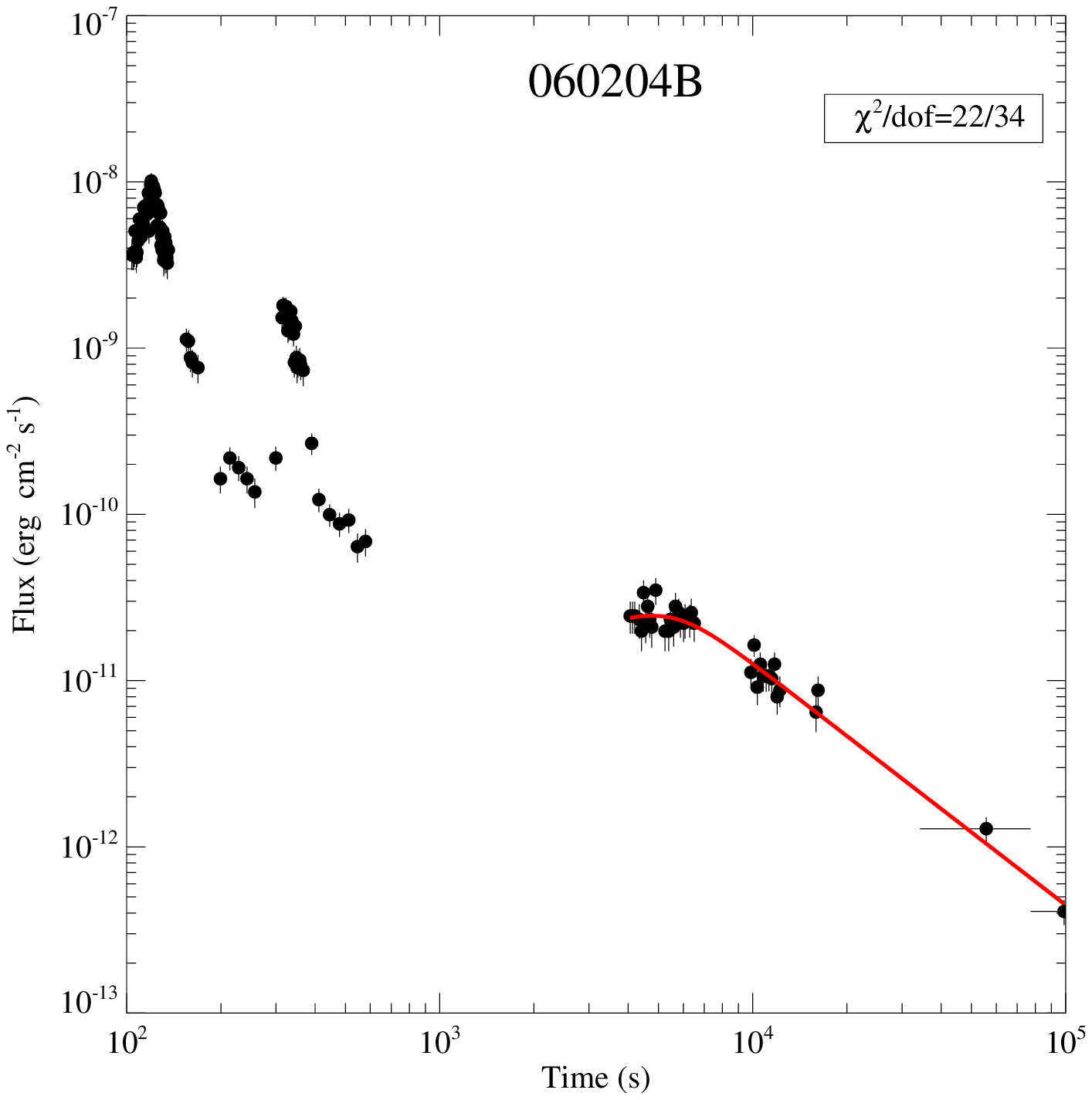}
\includegraphics[angle=0,scale=0.35]{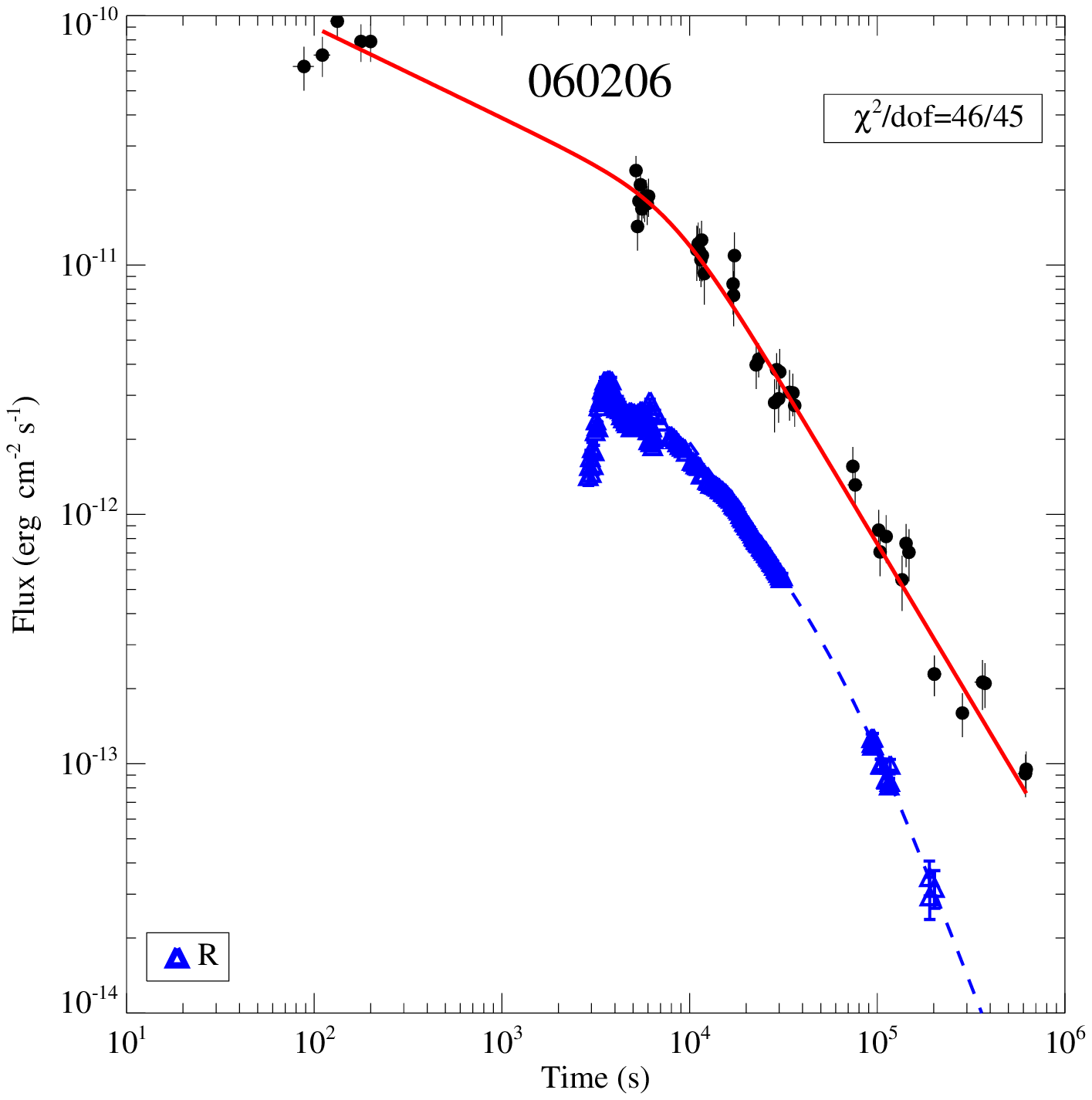}
\hfill
\includegraphics[angle=0,scale=0.35]{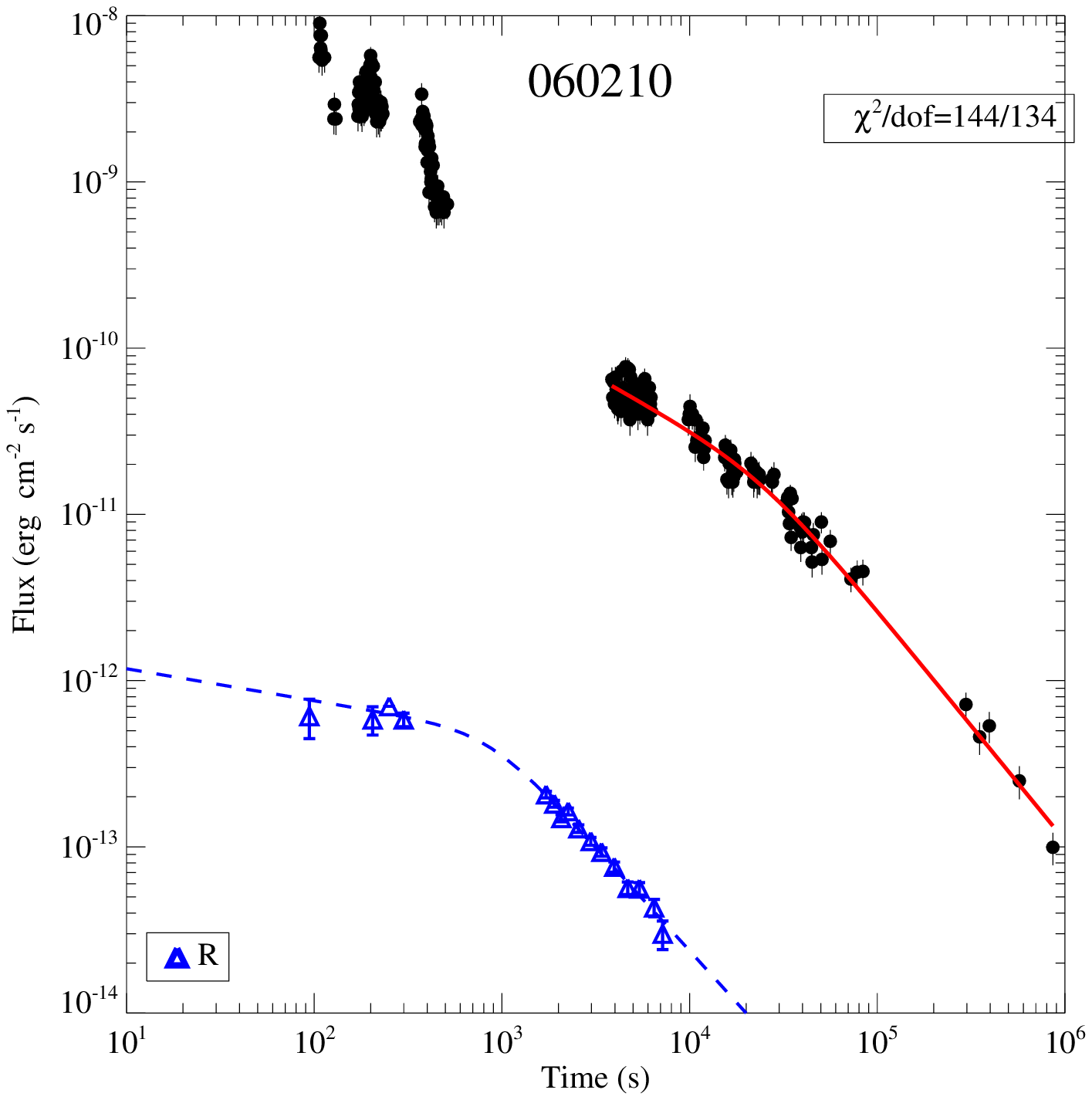}
\center{Fig.2--- continued.}\nonumber
\end{figure*}

\begin{figure*}
\includegraphics[angle=0,scale=0.35]{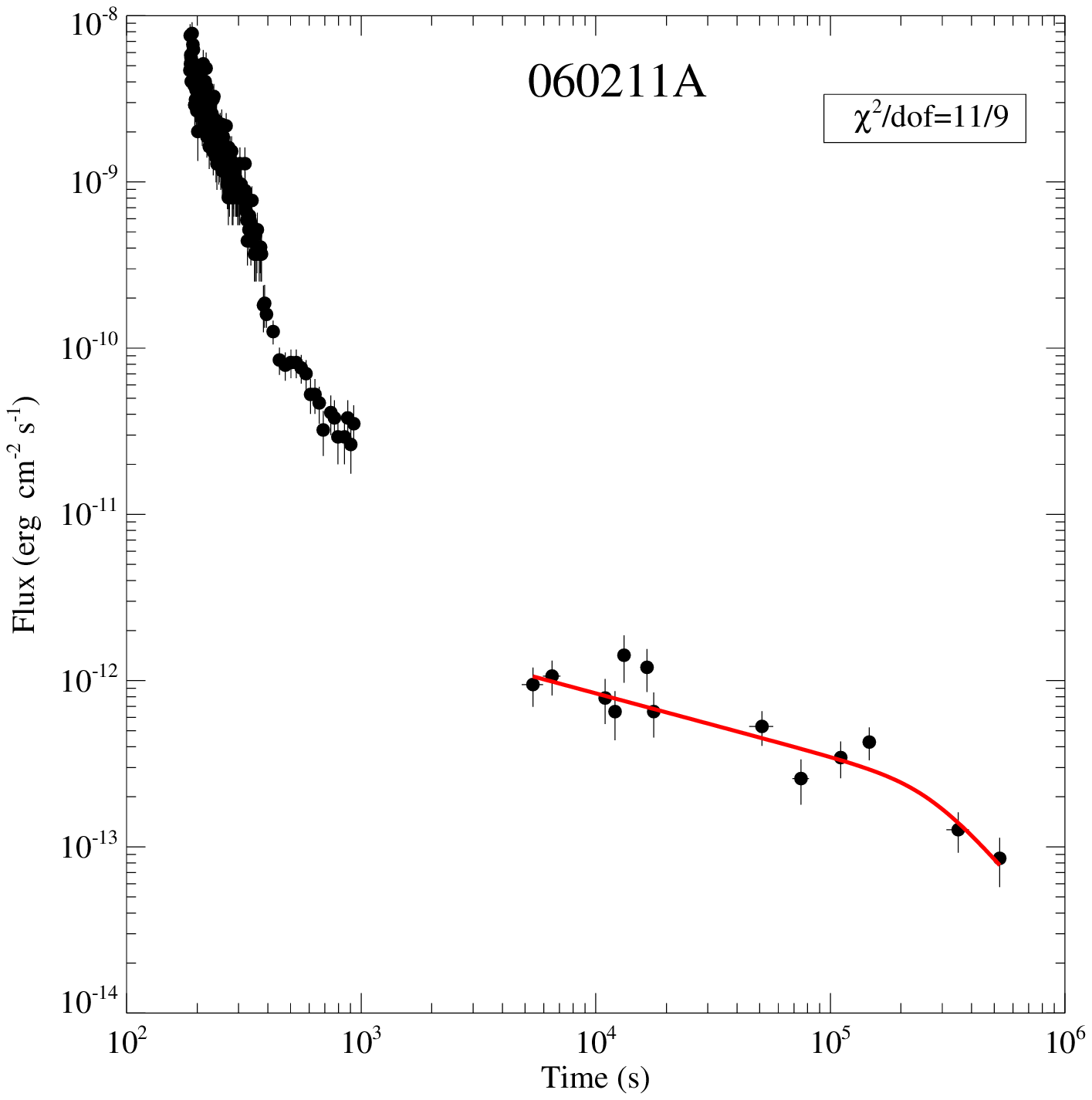}
\includegraphics[angle=0,scale=0.35]{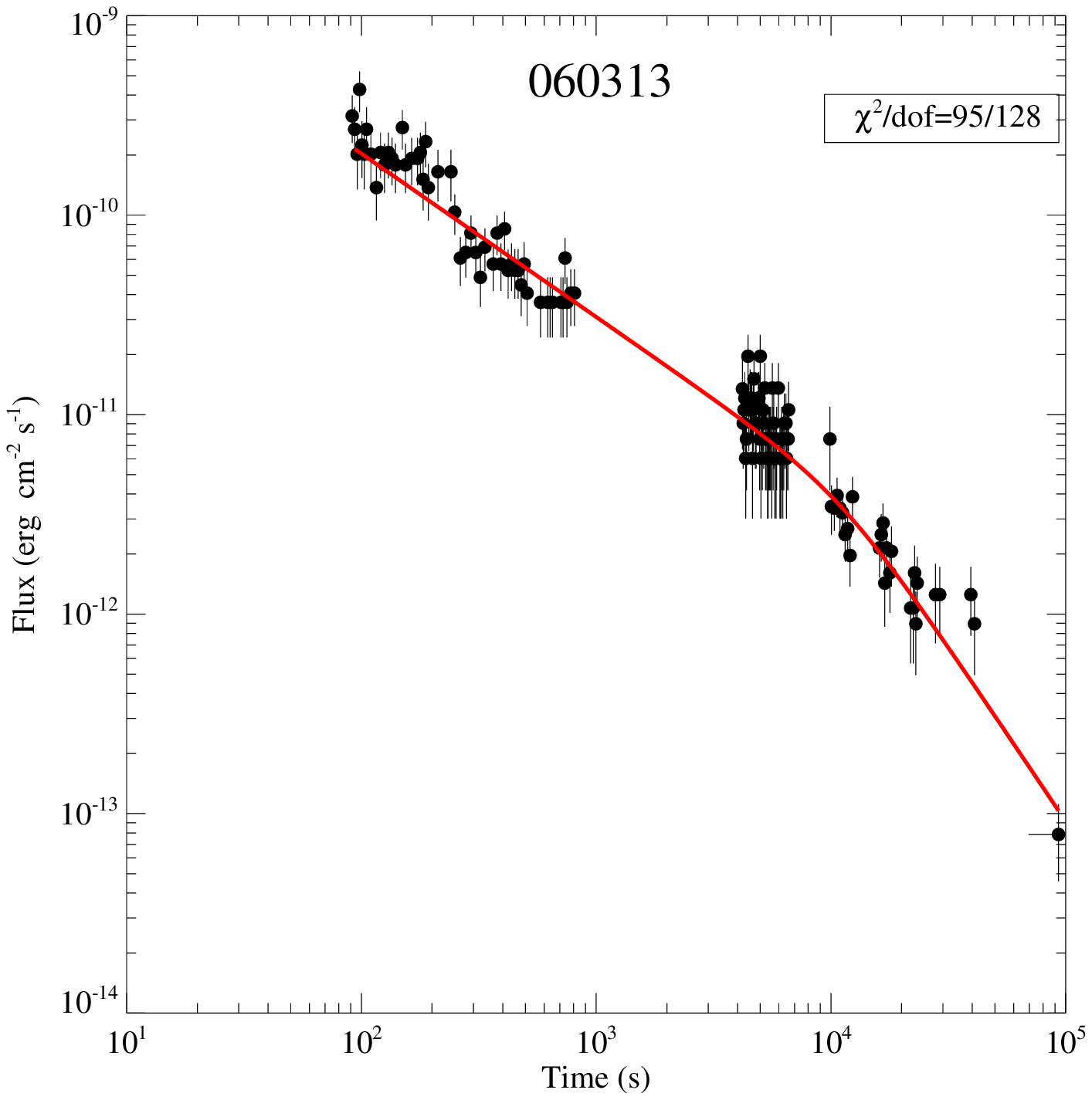}
\includegraphics[angle=0,scale=0.35]{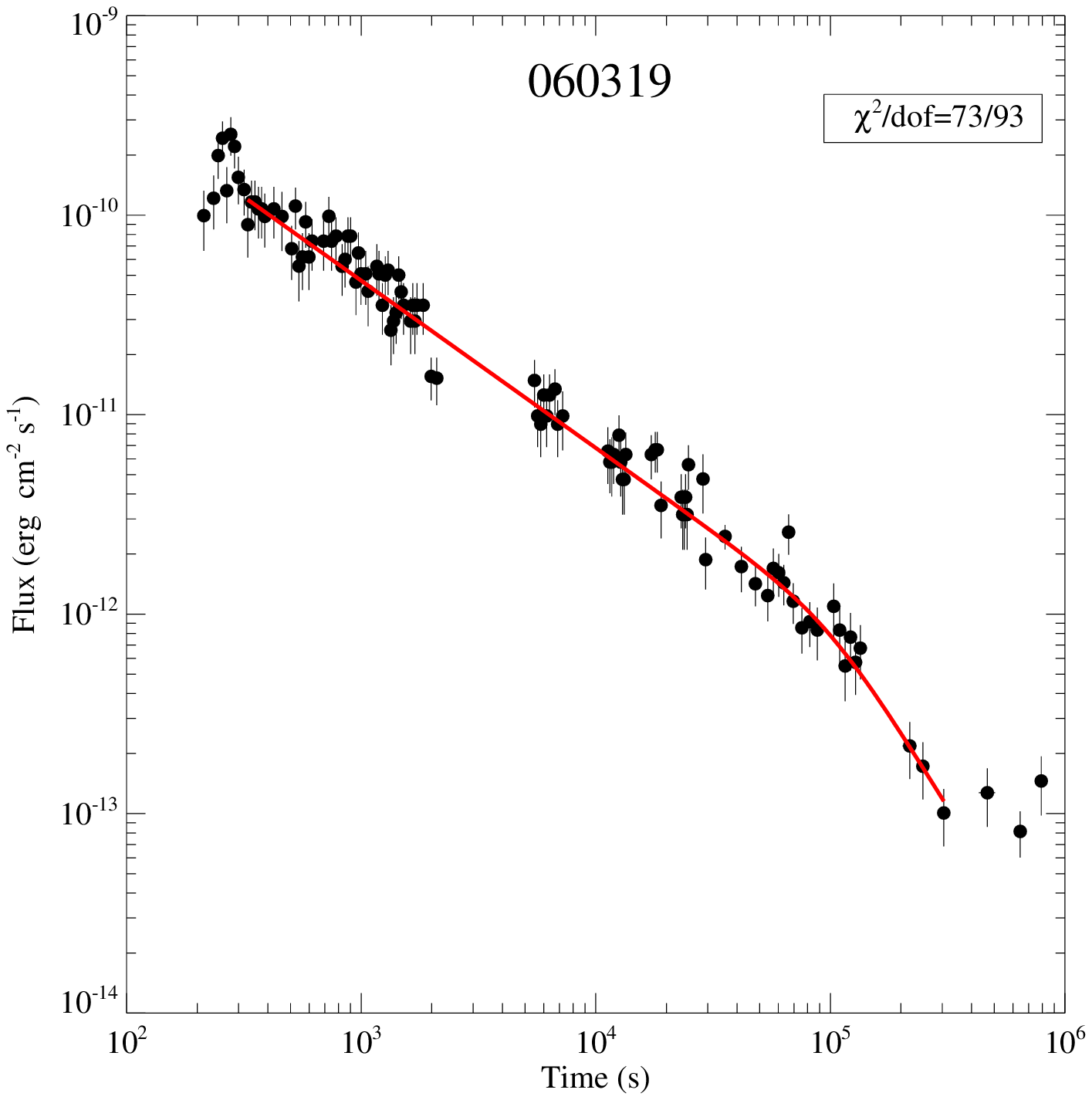}
\includegraphics[angle=0,scale=0.35]{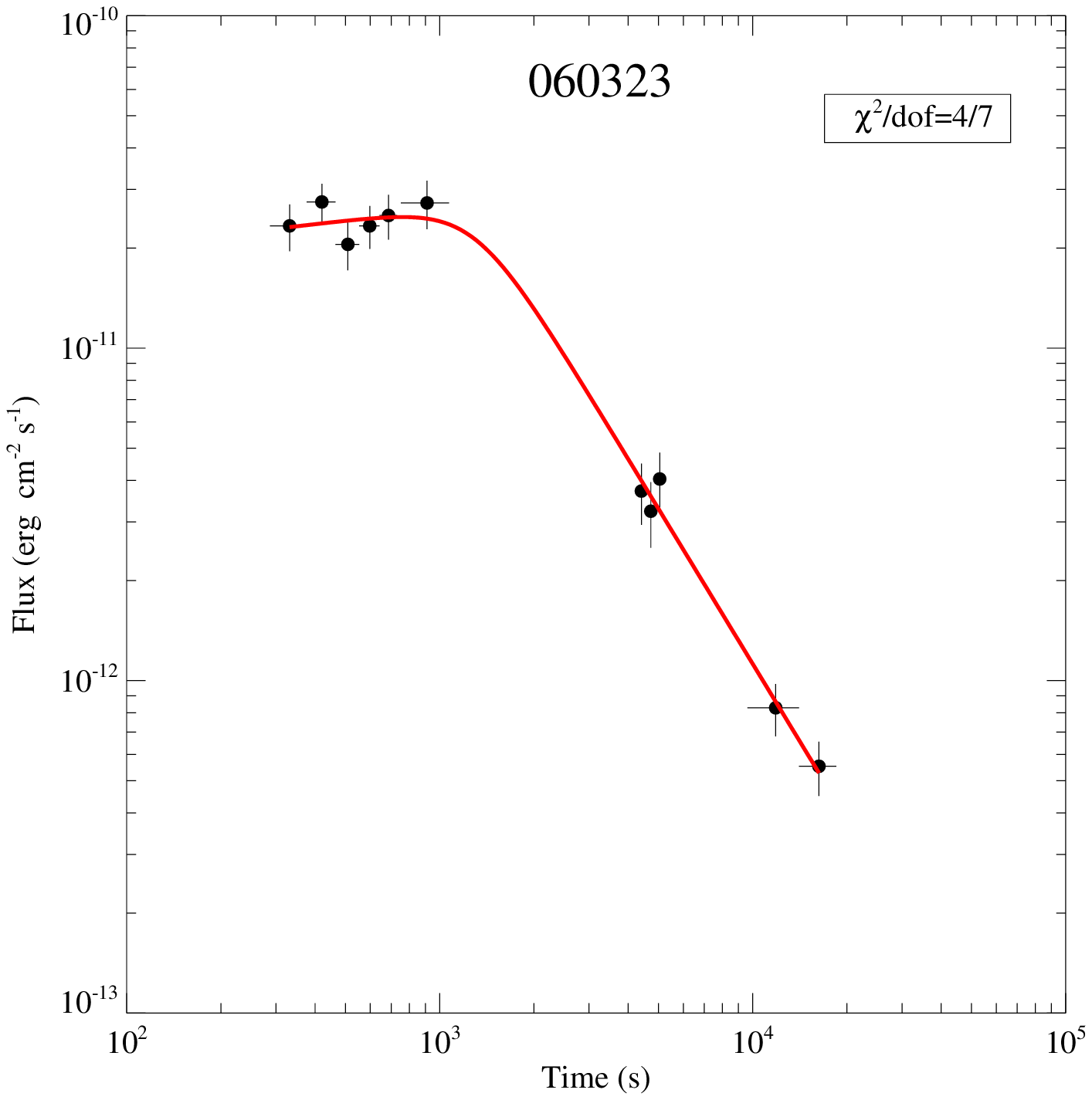}
\includegraphics[angle=0,scale=0.35]{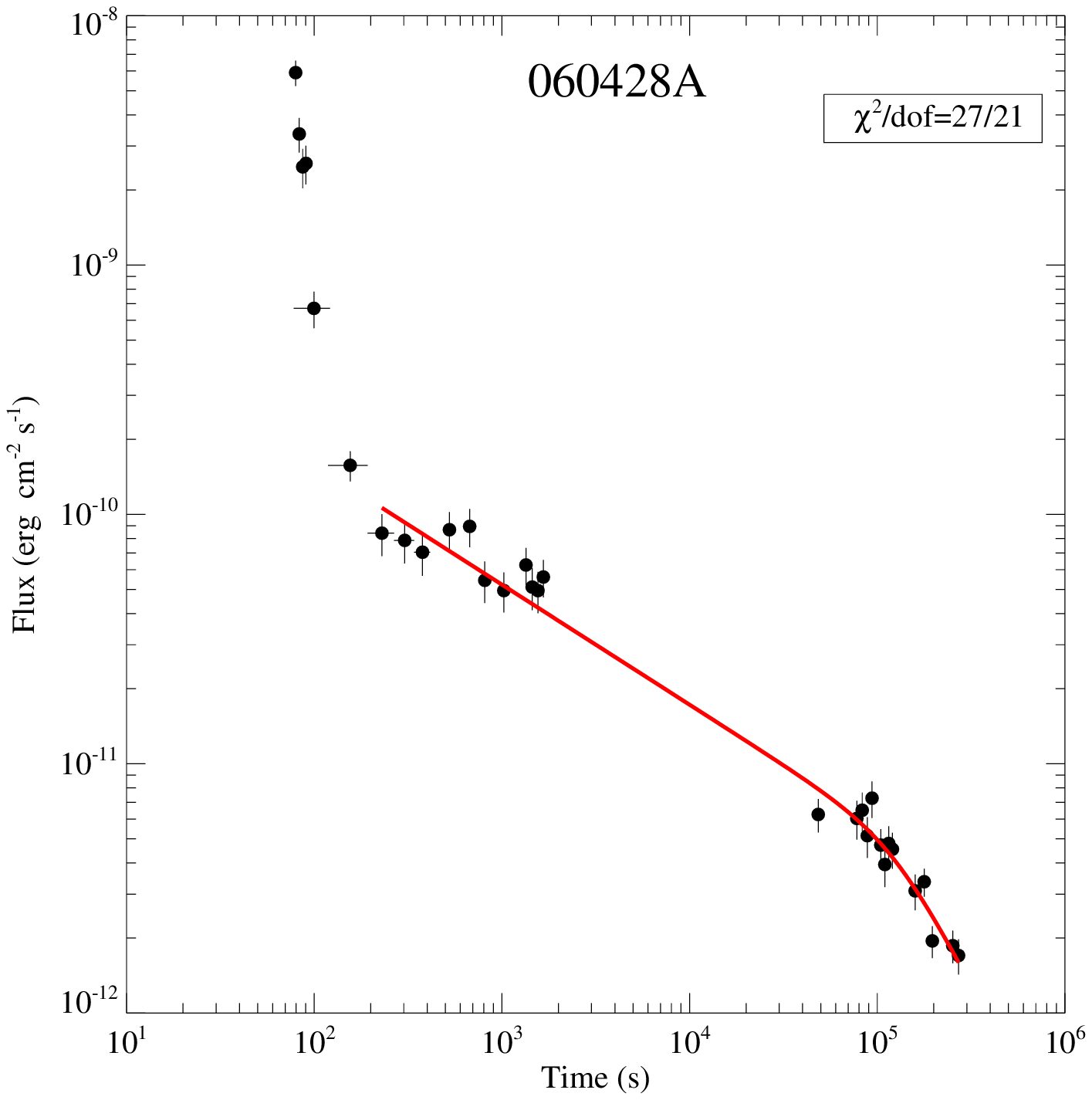}
\includegraphics[angle=0,scale=0.35]{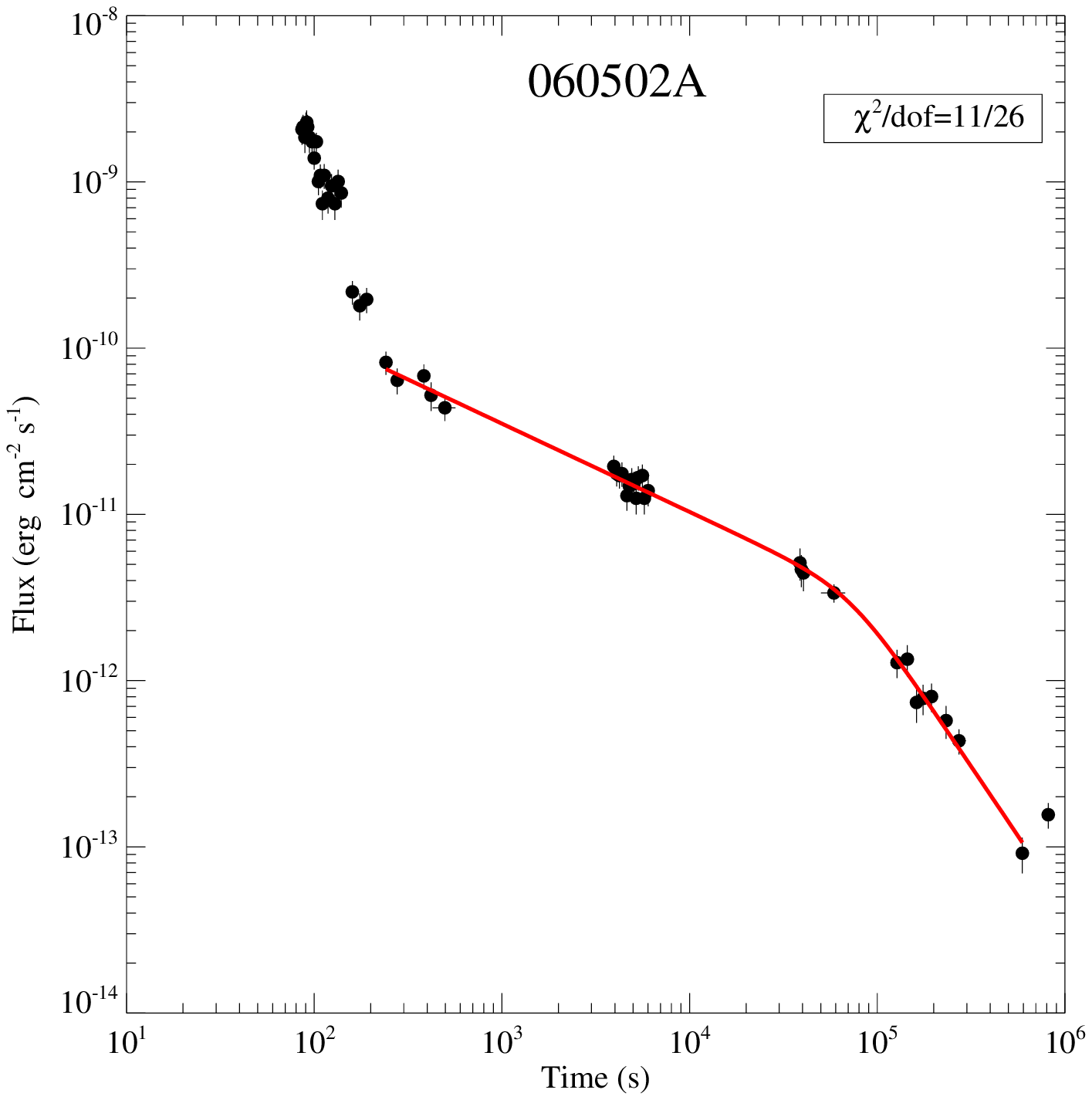}
\includegraphics[angle=0,scale=0.35]{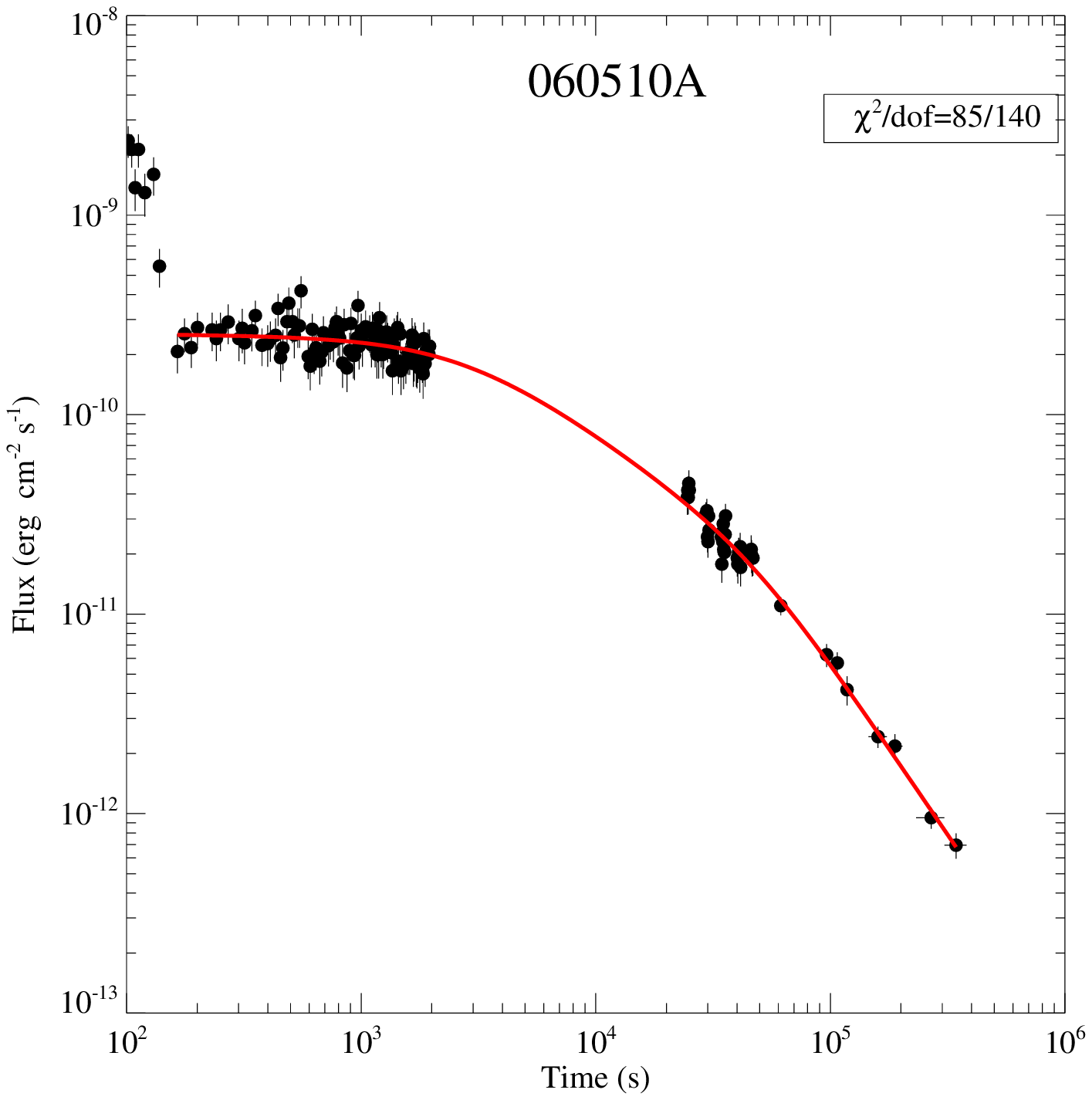}
\includegraphics[angle=0,scale=0.35]{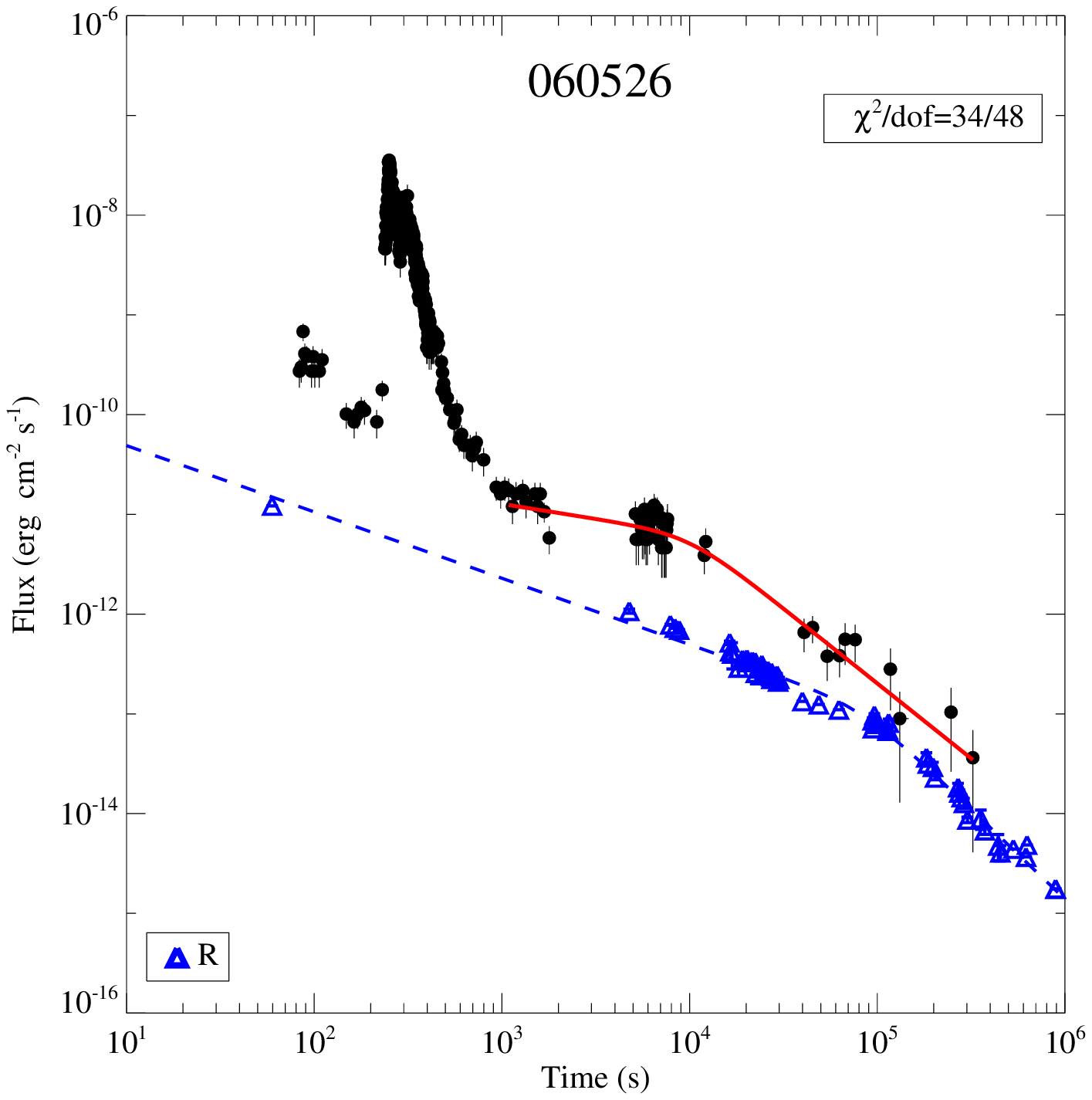}
\includegraphics[angle=0,scale=0.35]{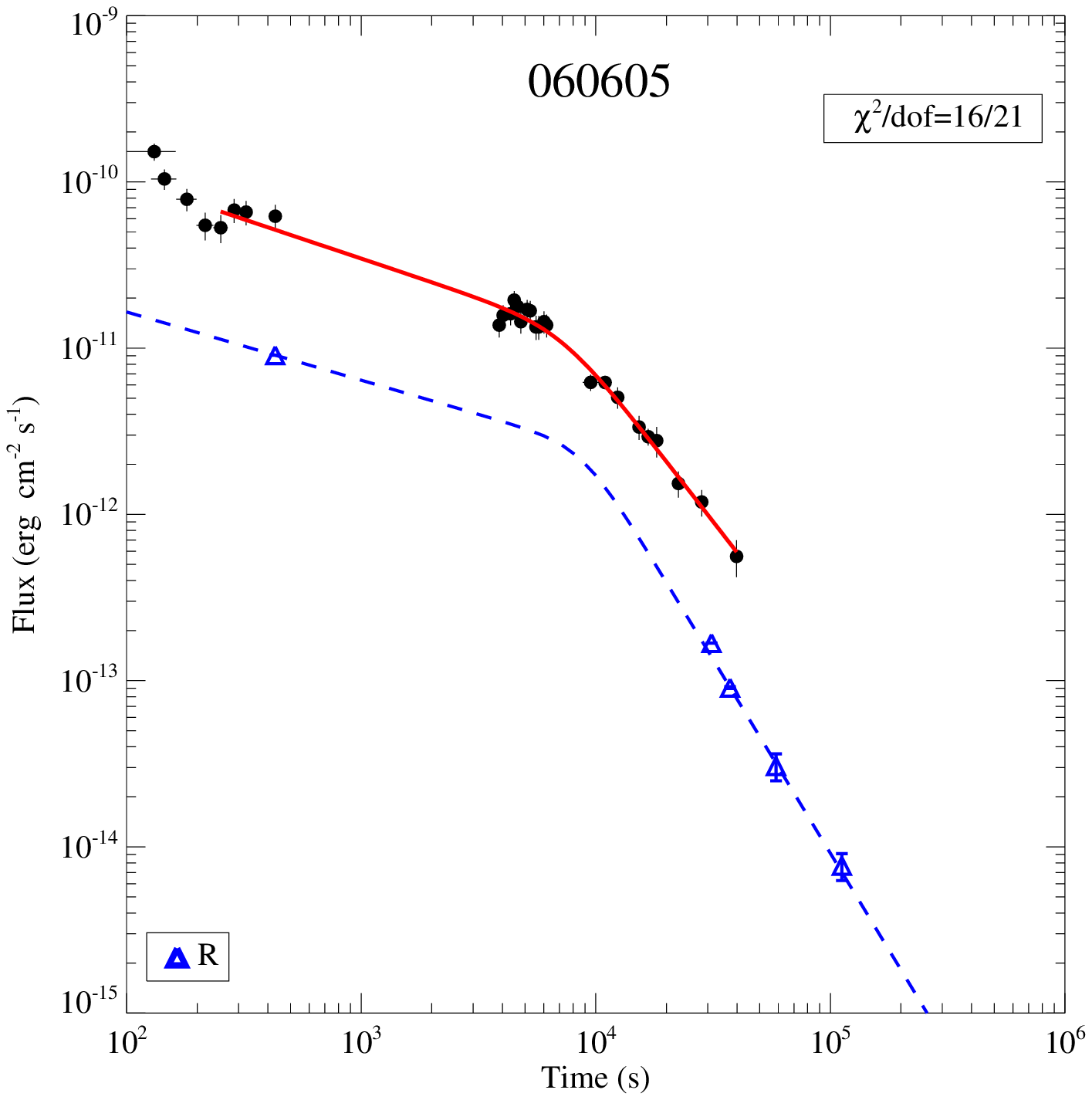}
\includegraphics[angle=0,scale=0.35]{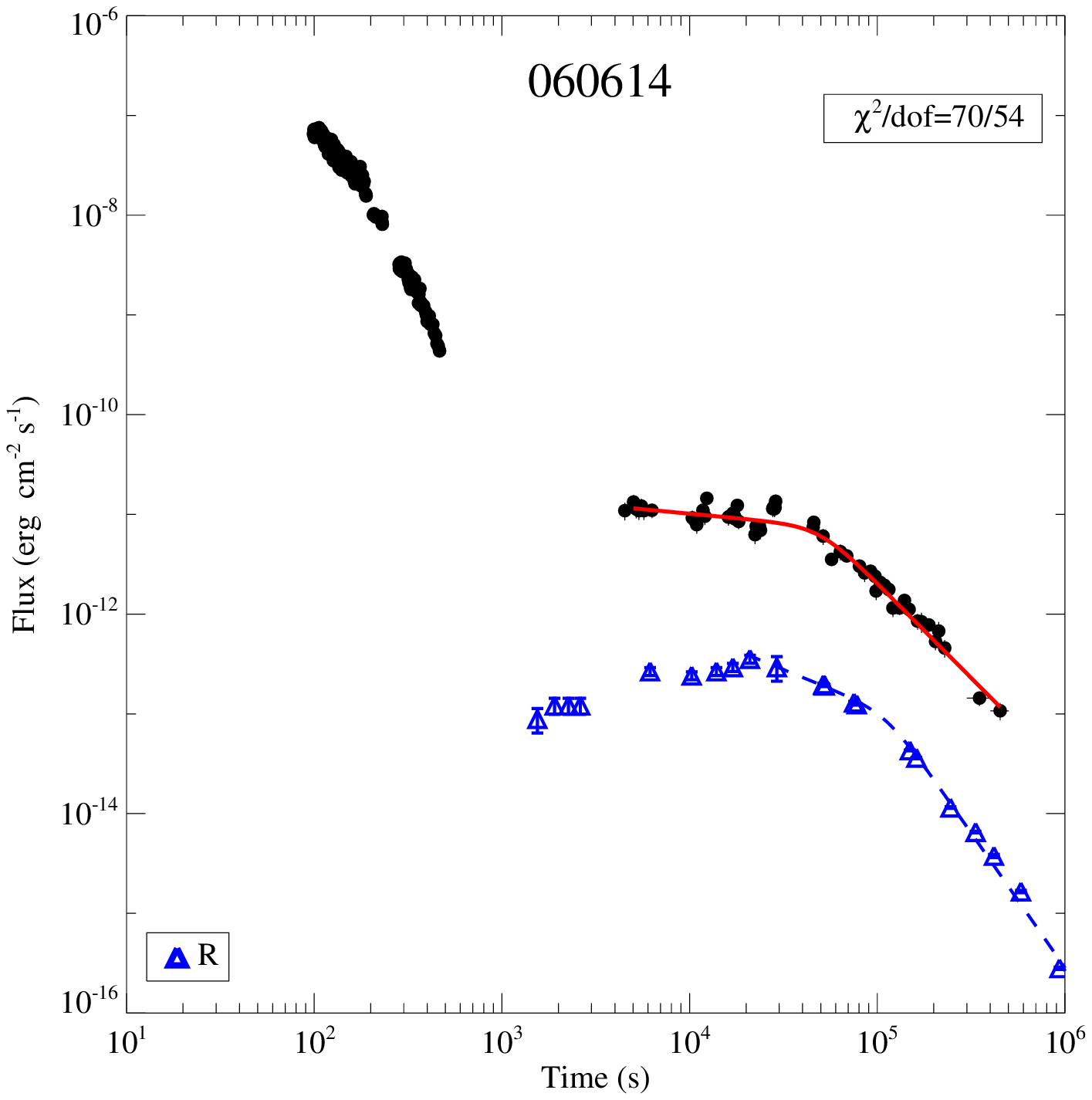}
\includegraphics[angle=0,scale=0.35]{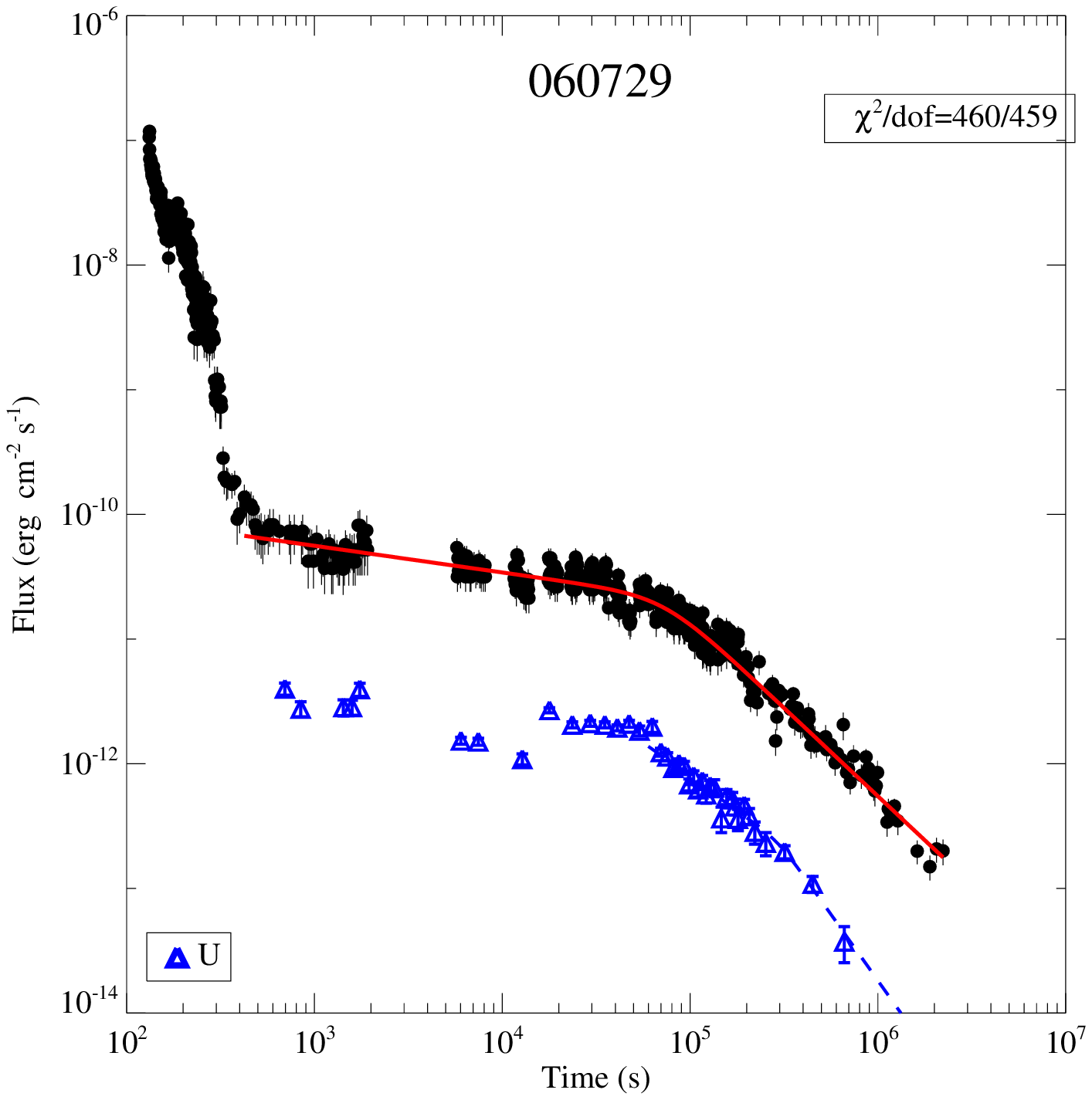}
\hfill
\includegraphics[angle=0,scale=0.35]{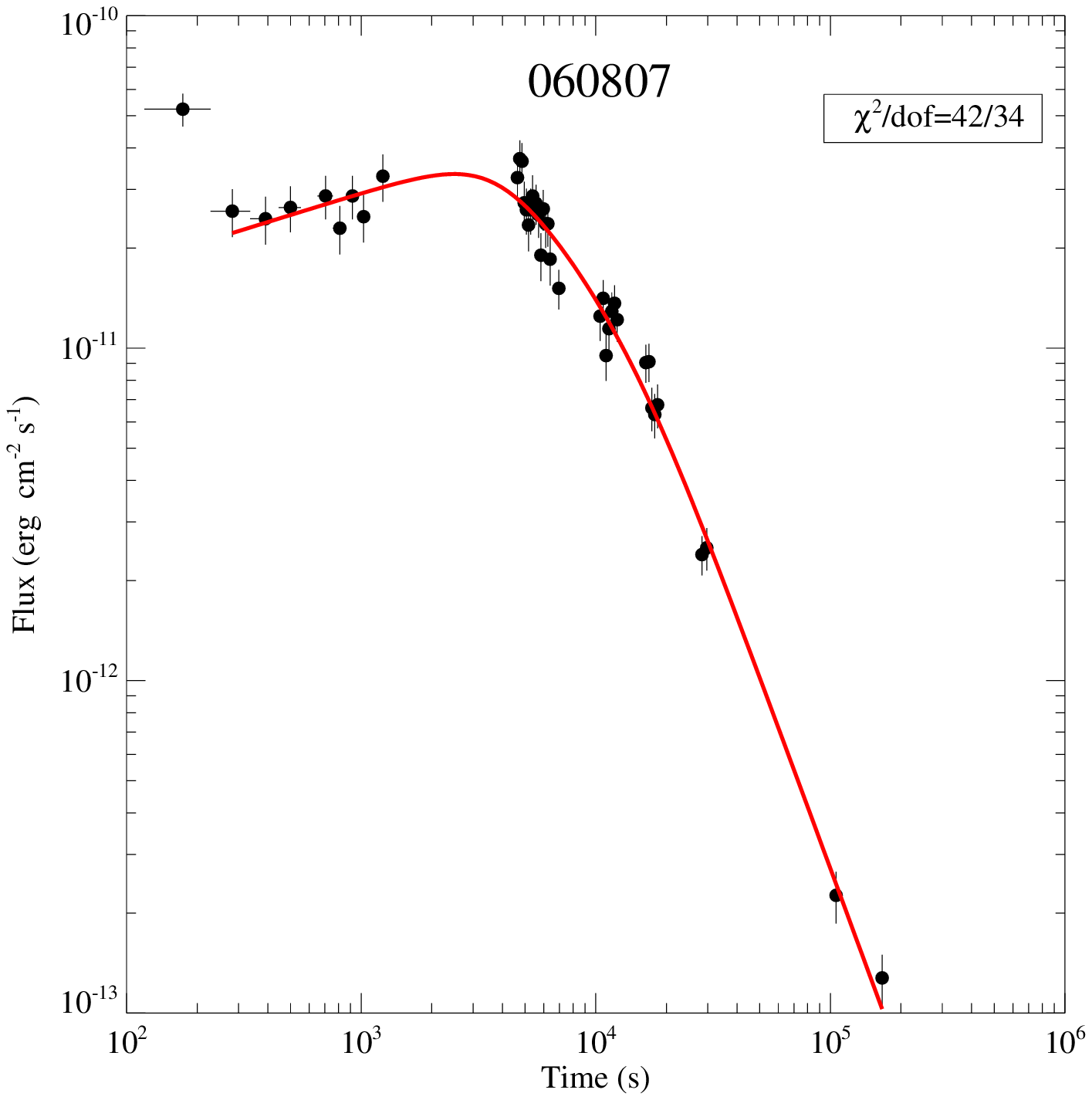}
\center{Fig.2---  continued}
\end{figure*}

\begin{figure*}
\includegraphics[angle=0,scale=0.35]{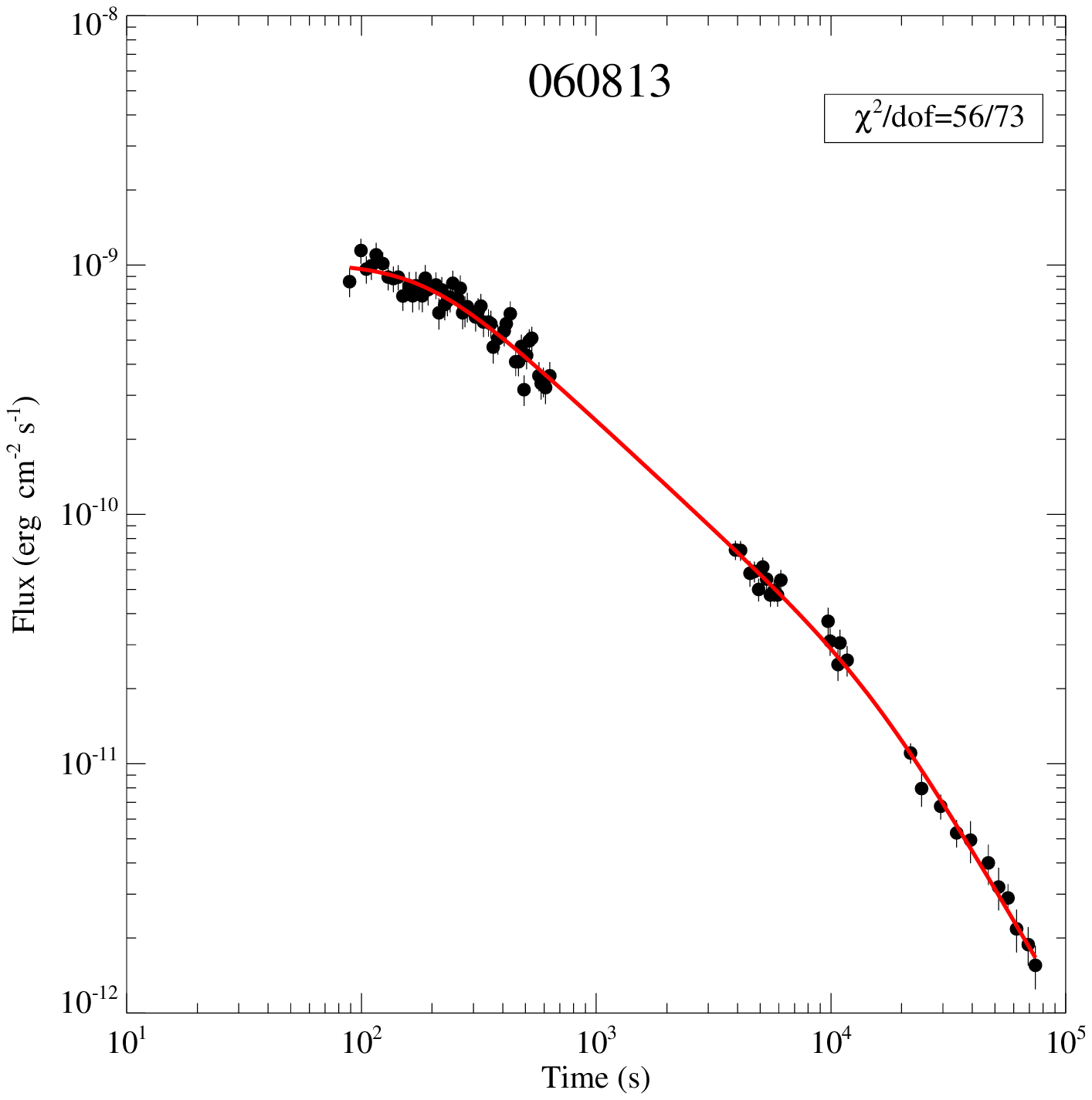}
\includegraphics[angle=0,scale=0.35]{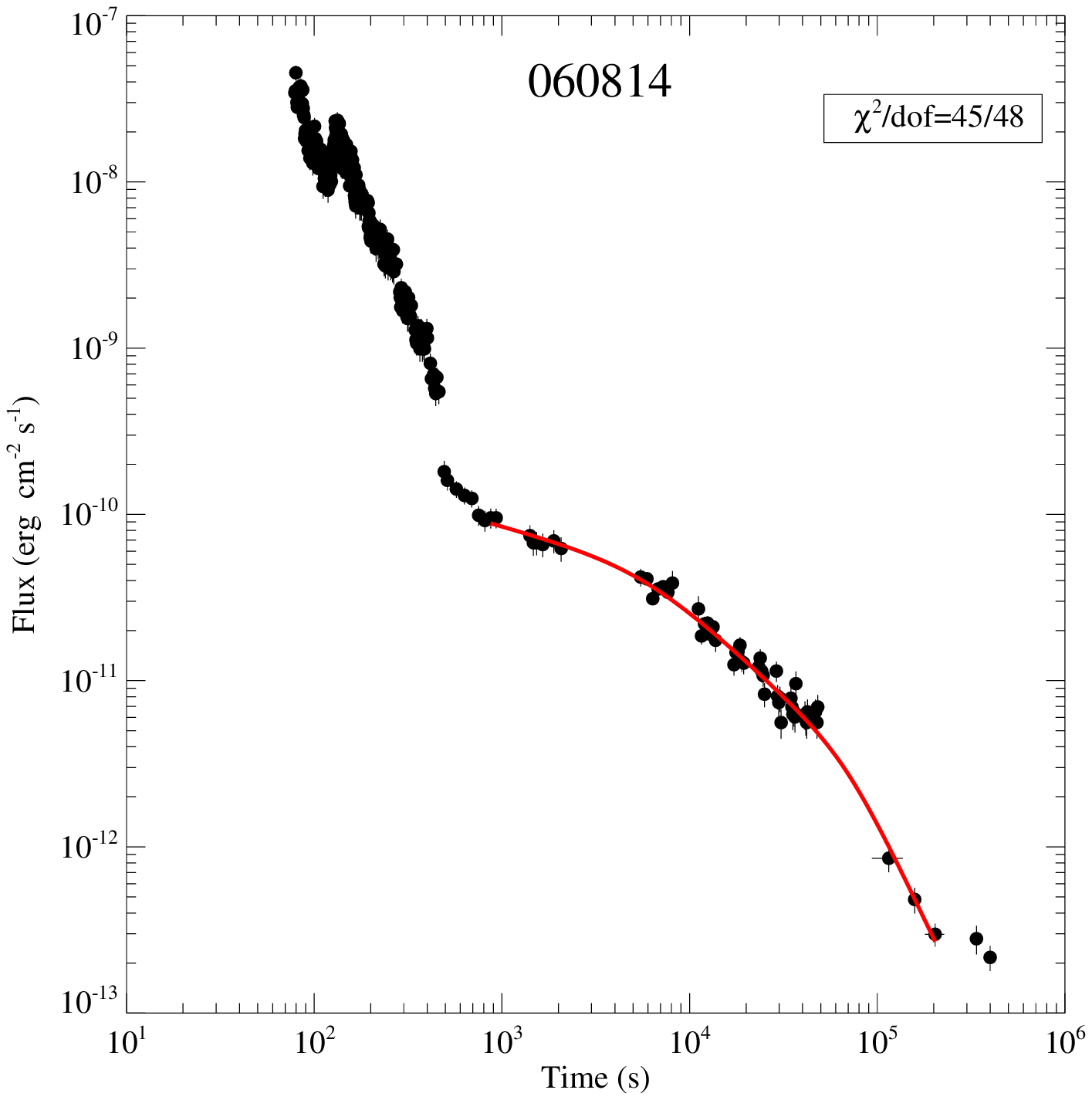}
\includegraphics[angle=0,scale=0.35]{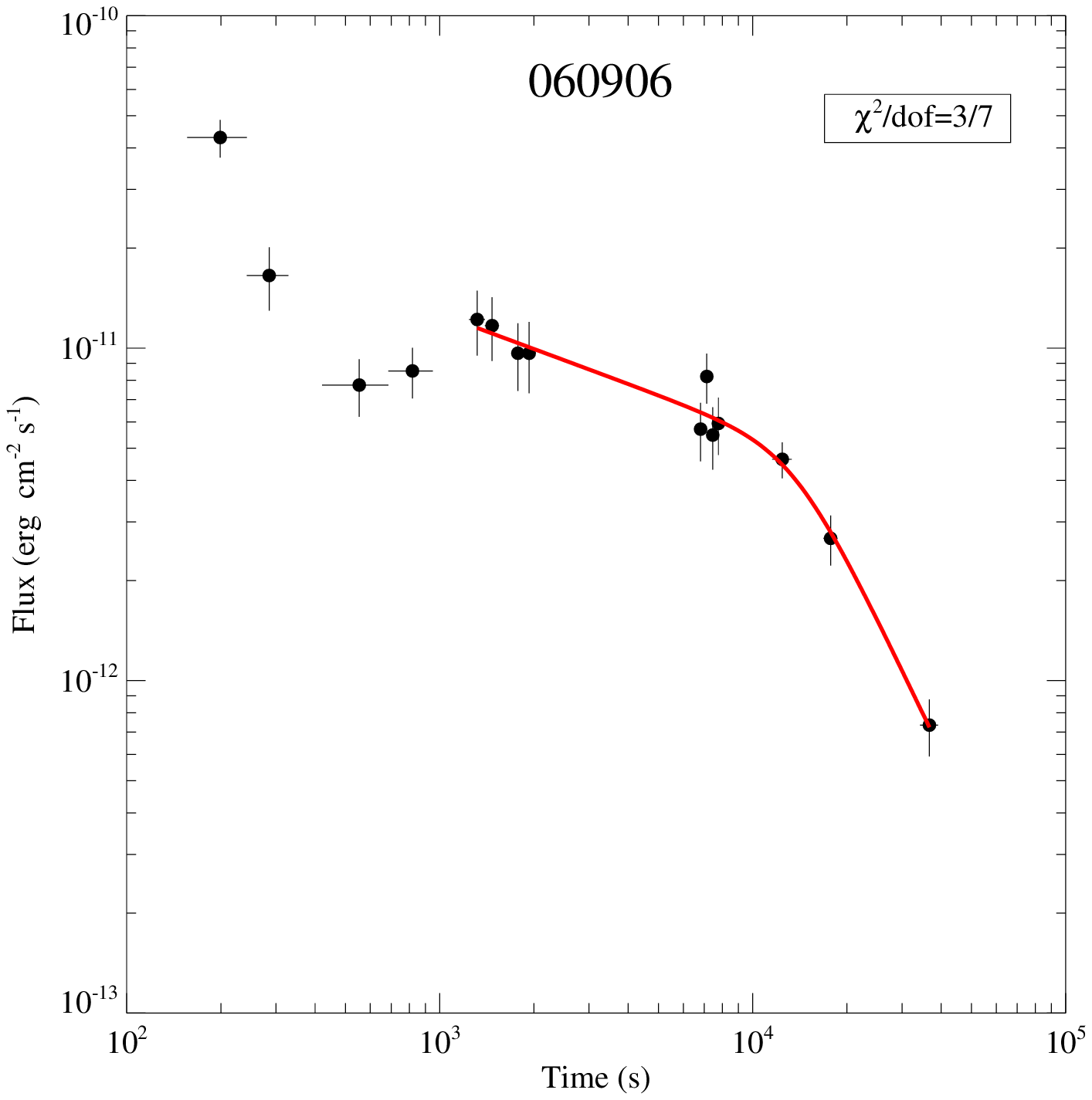}
\includegraphics[angle=0,scale=0.35]{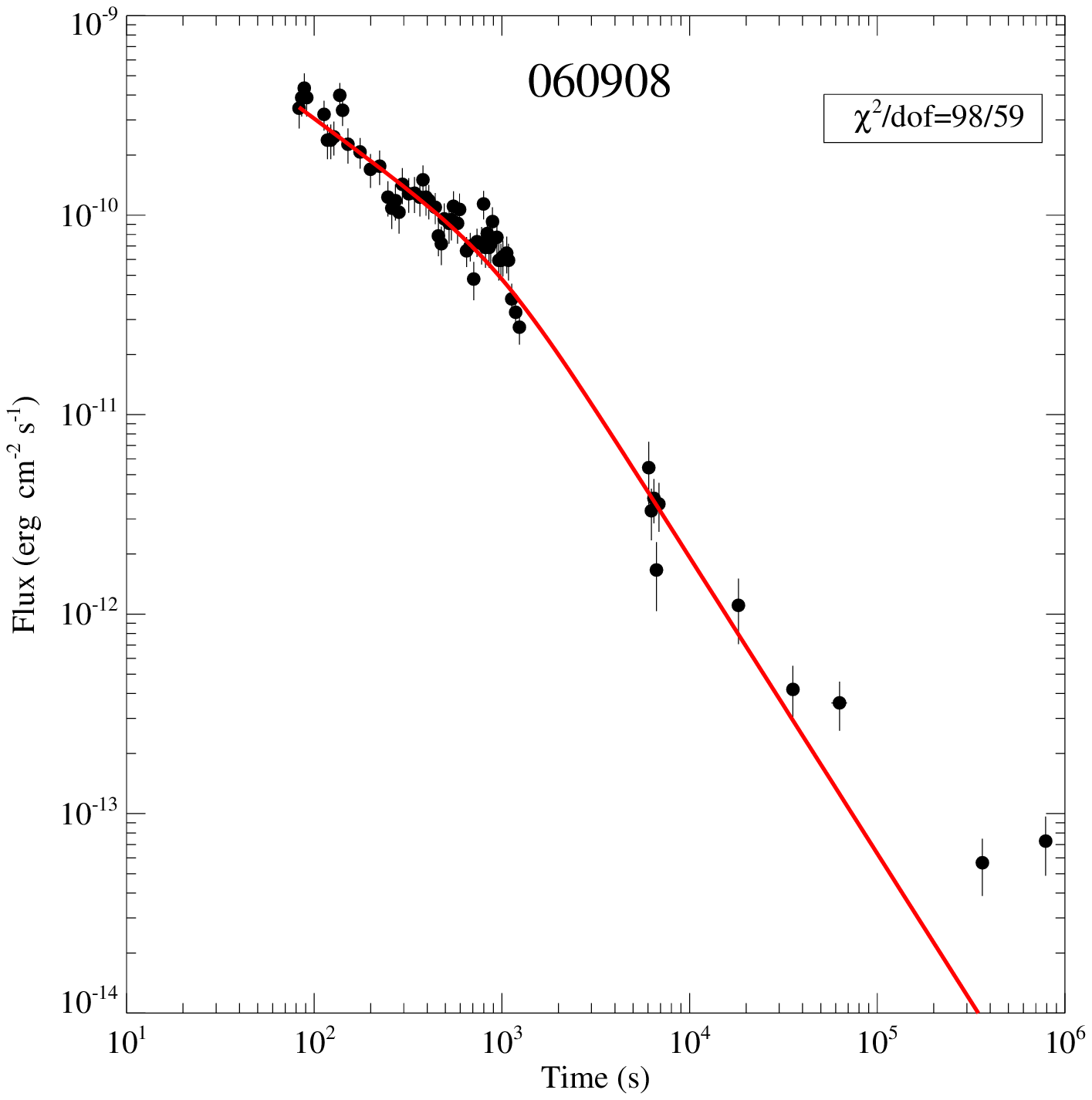}
\includegraphics[angle=0,scale=0.35]{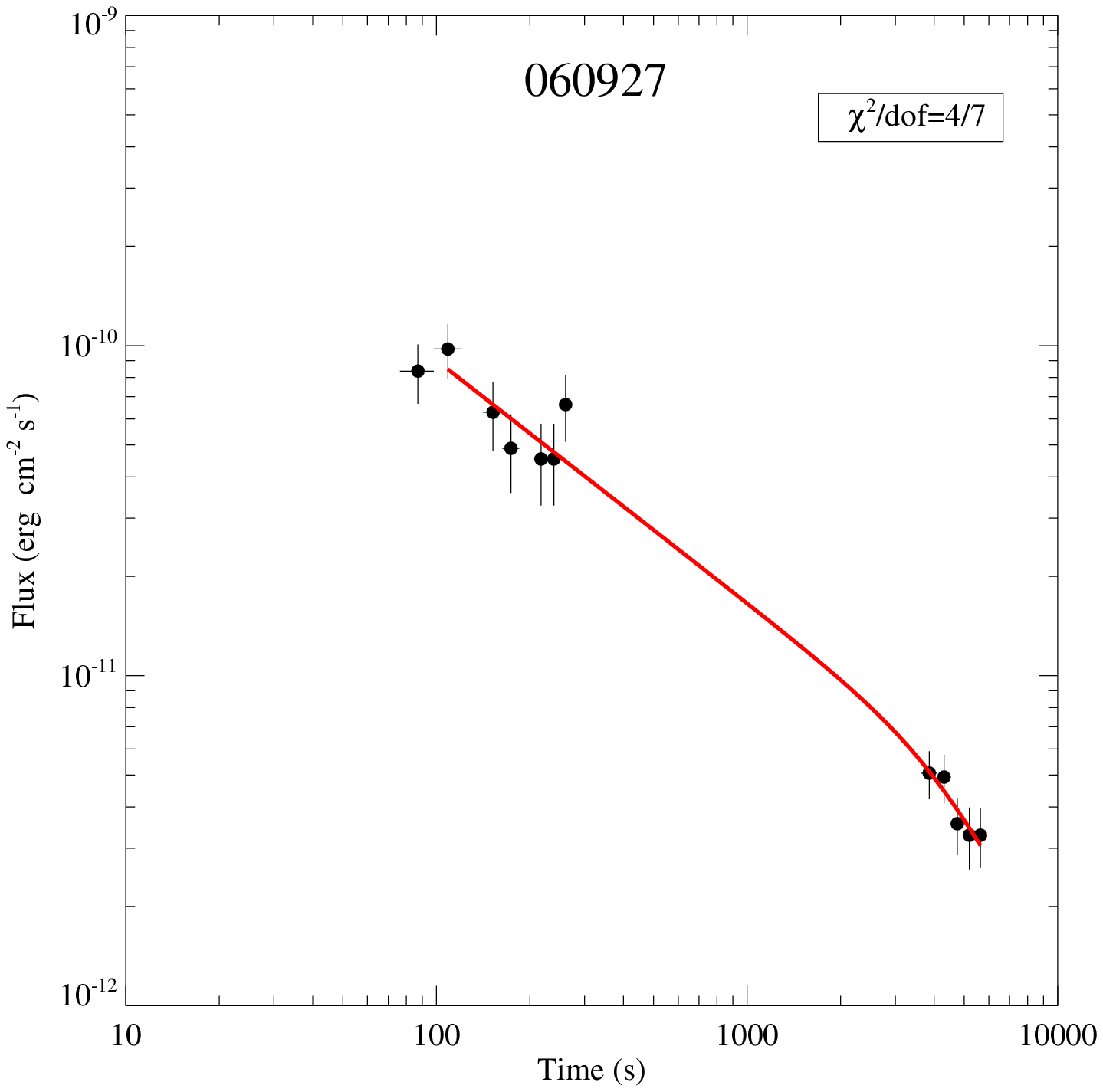}
\includegraphics[angle=0,scale=0.35]{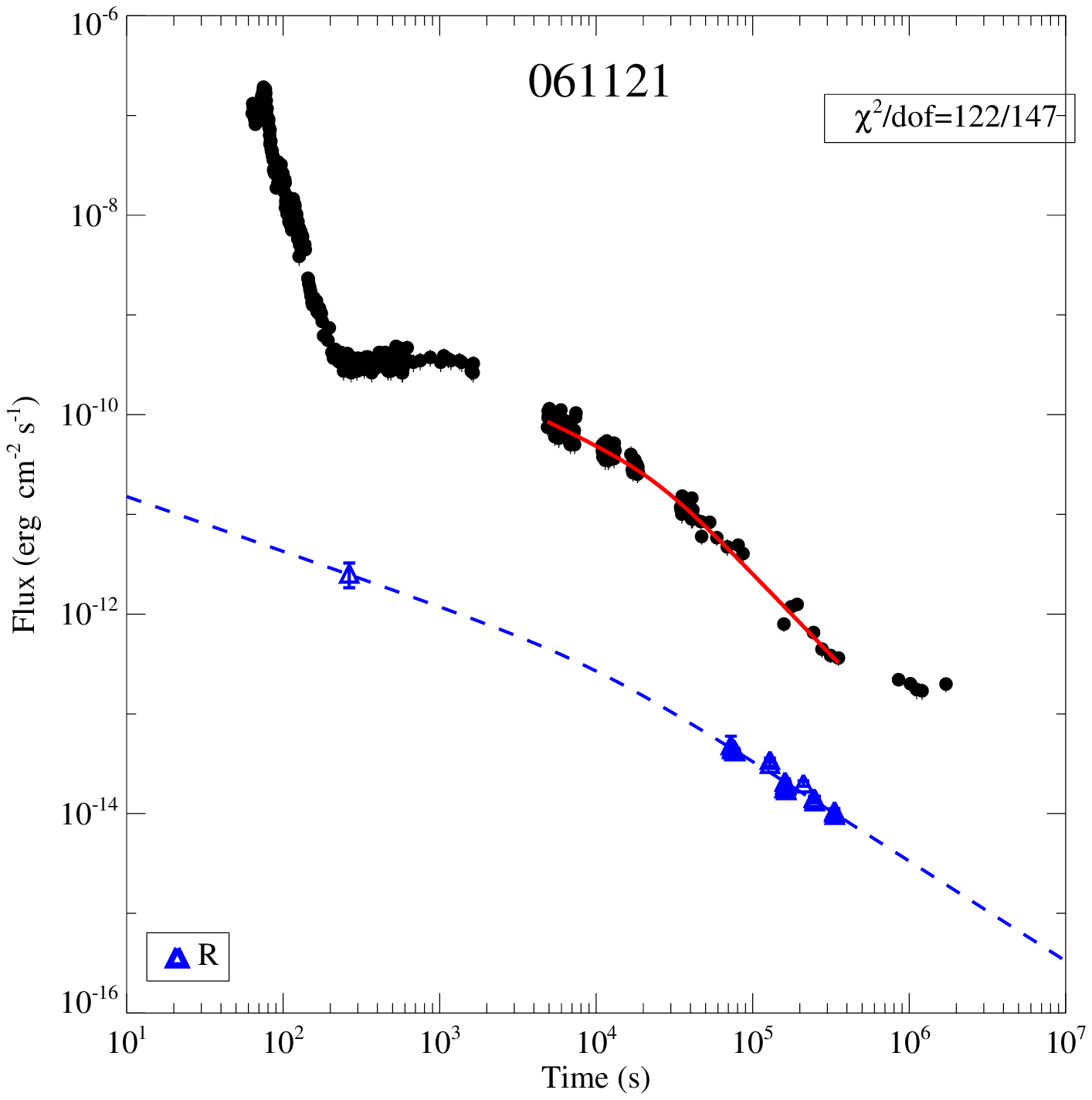}
\includegraphics[angle=0,scale=0.35]{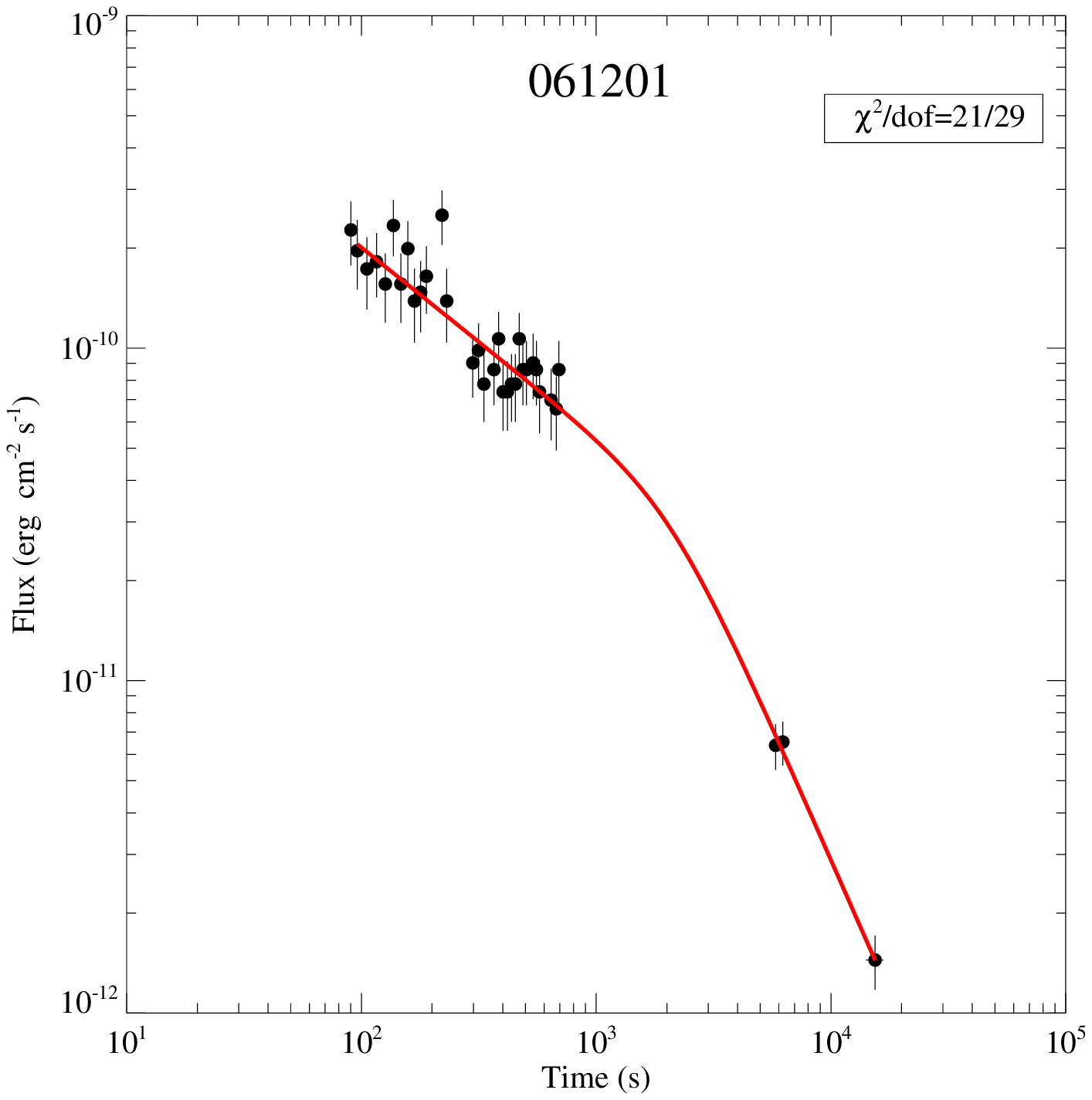}
\includegraphics[angle=0,scale=0.35]{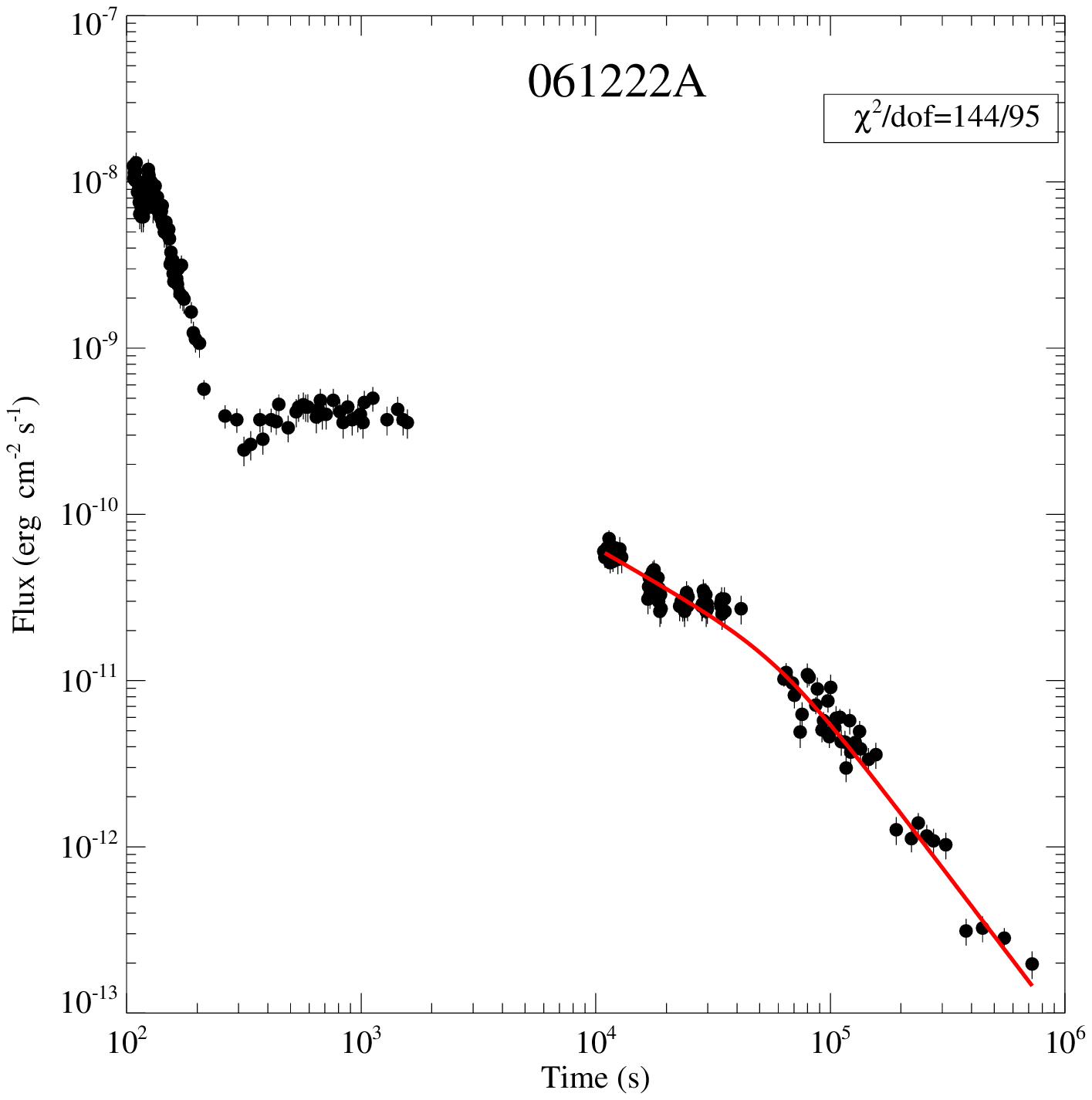}
\includegraphics[angle=0,scale=0.35]{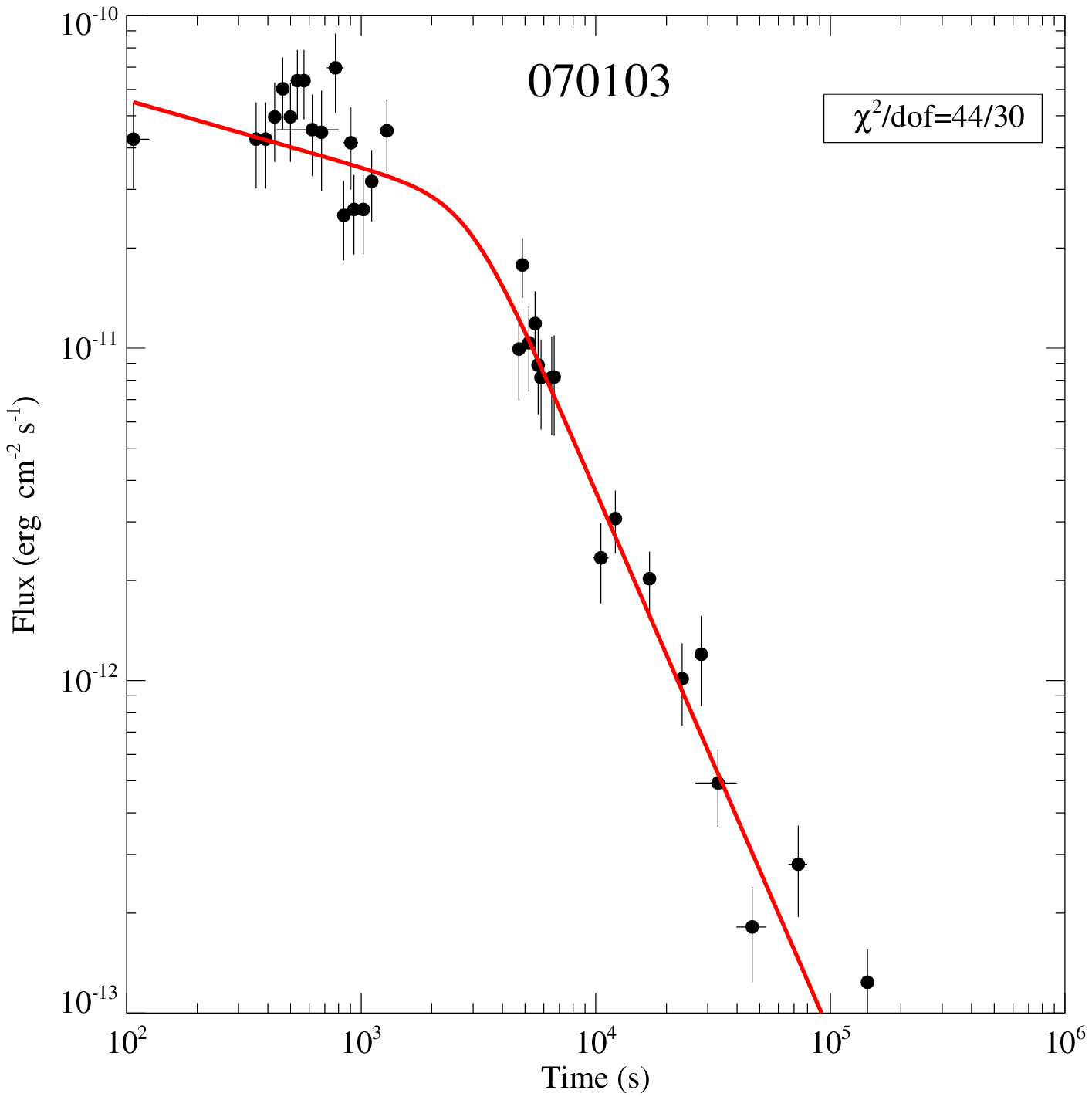}
\includegraphics[angle=0,scale=0.35]{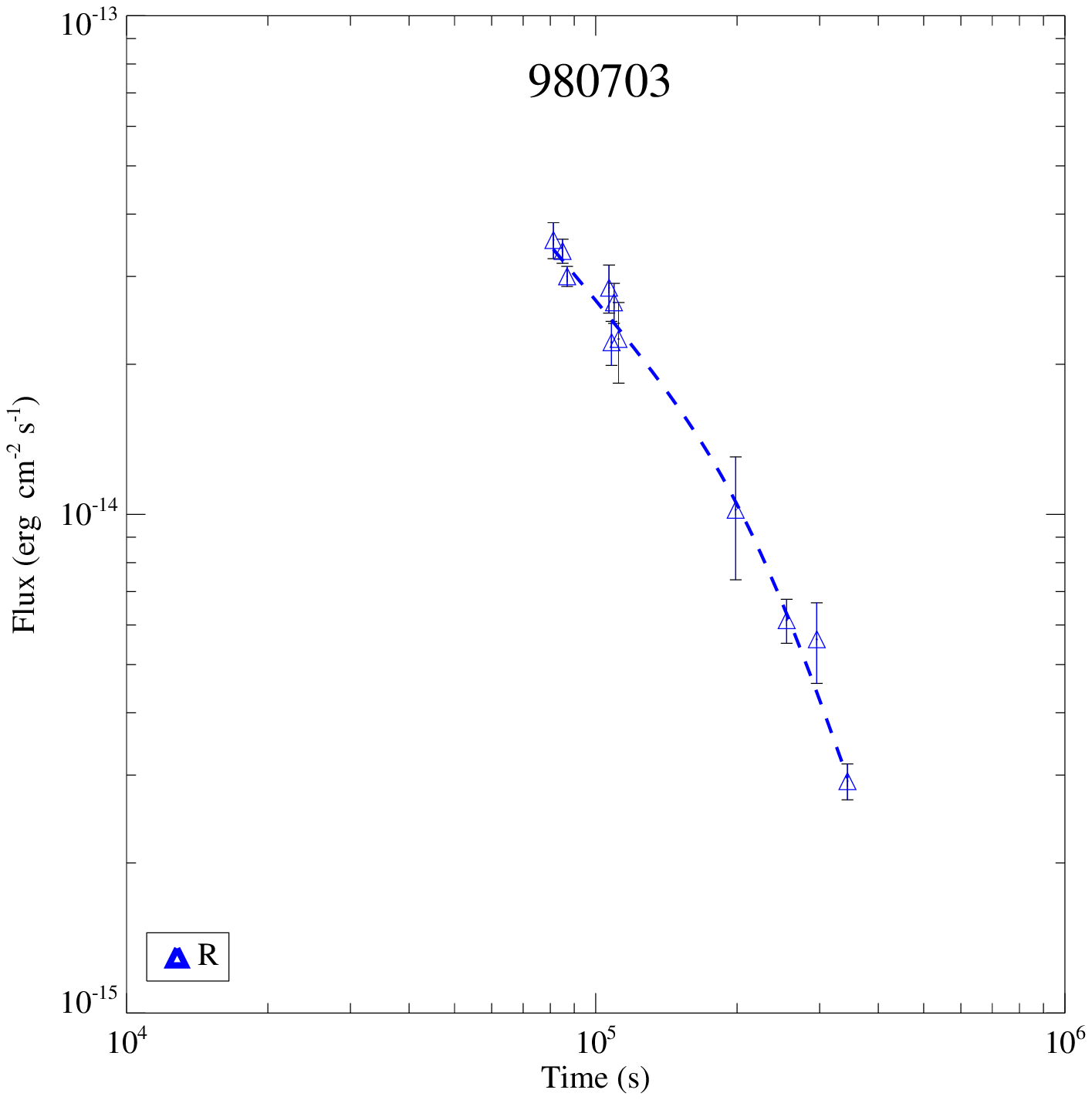}
\includegraphics[angle=0,scale=0.35]{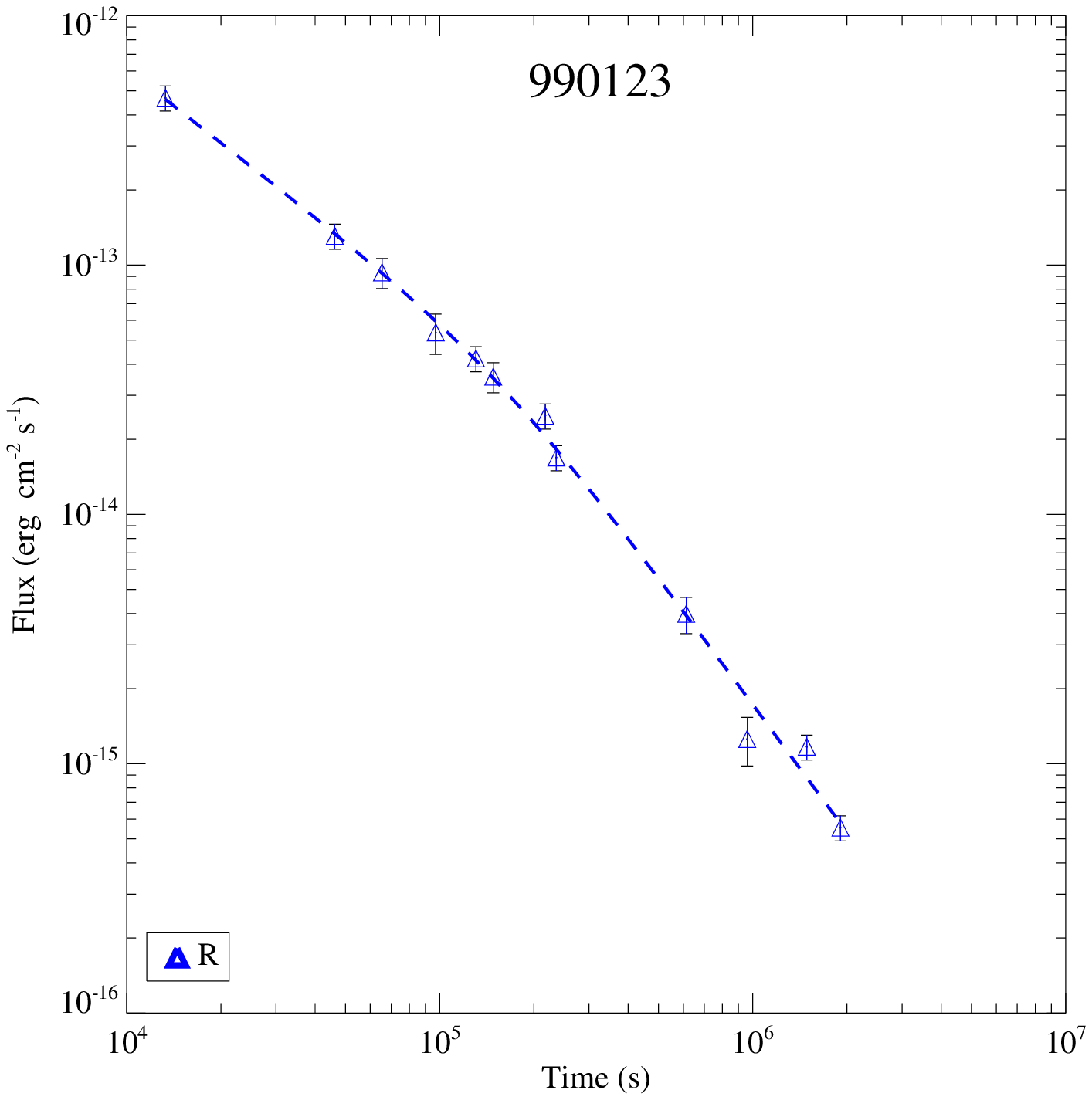}
\hfill
\includegraphics[angle=0,scale=0.35]{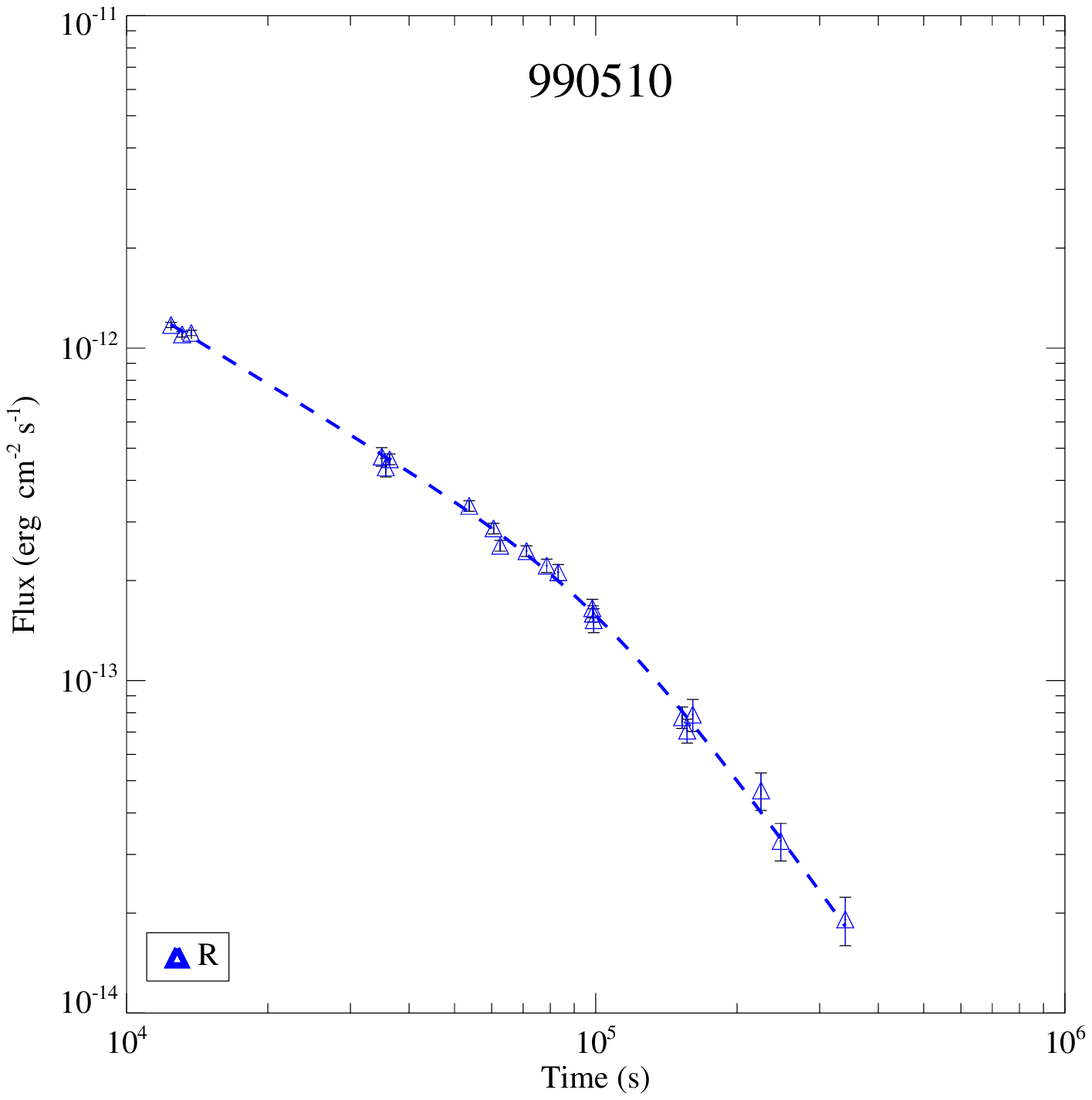}
\center{Fig.2---  continued}
\end{figure*}

\begin{figure*}
\includegraphics[angle=0,scale=0.35]{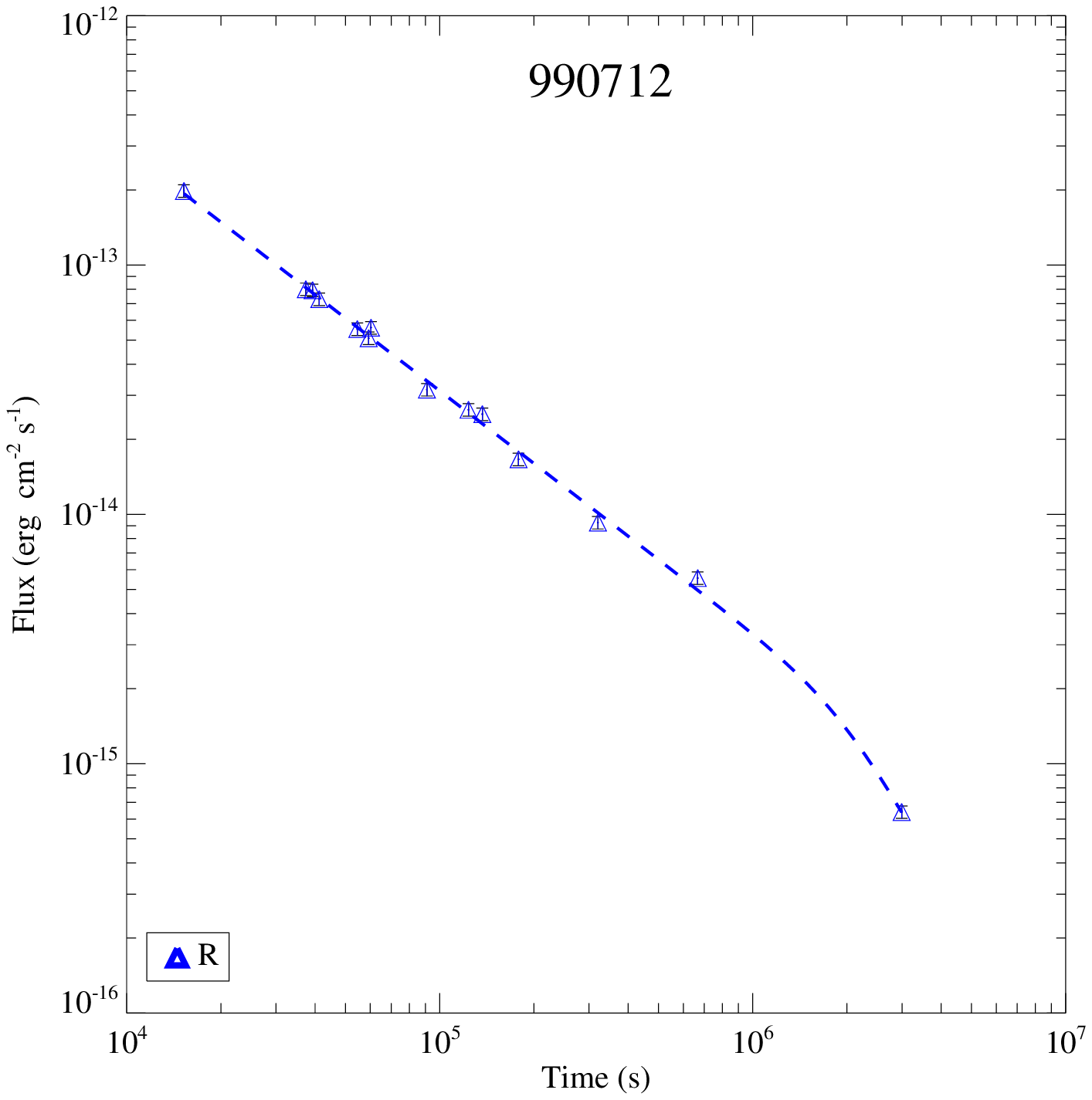}
\includegraphics[angle=0,scale=0.35]{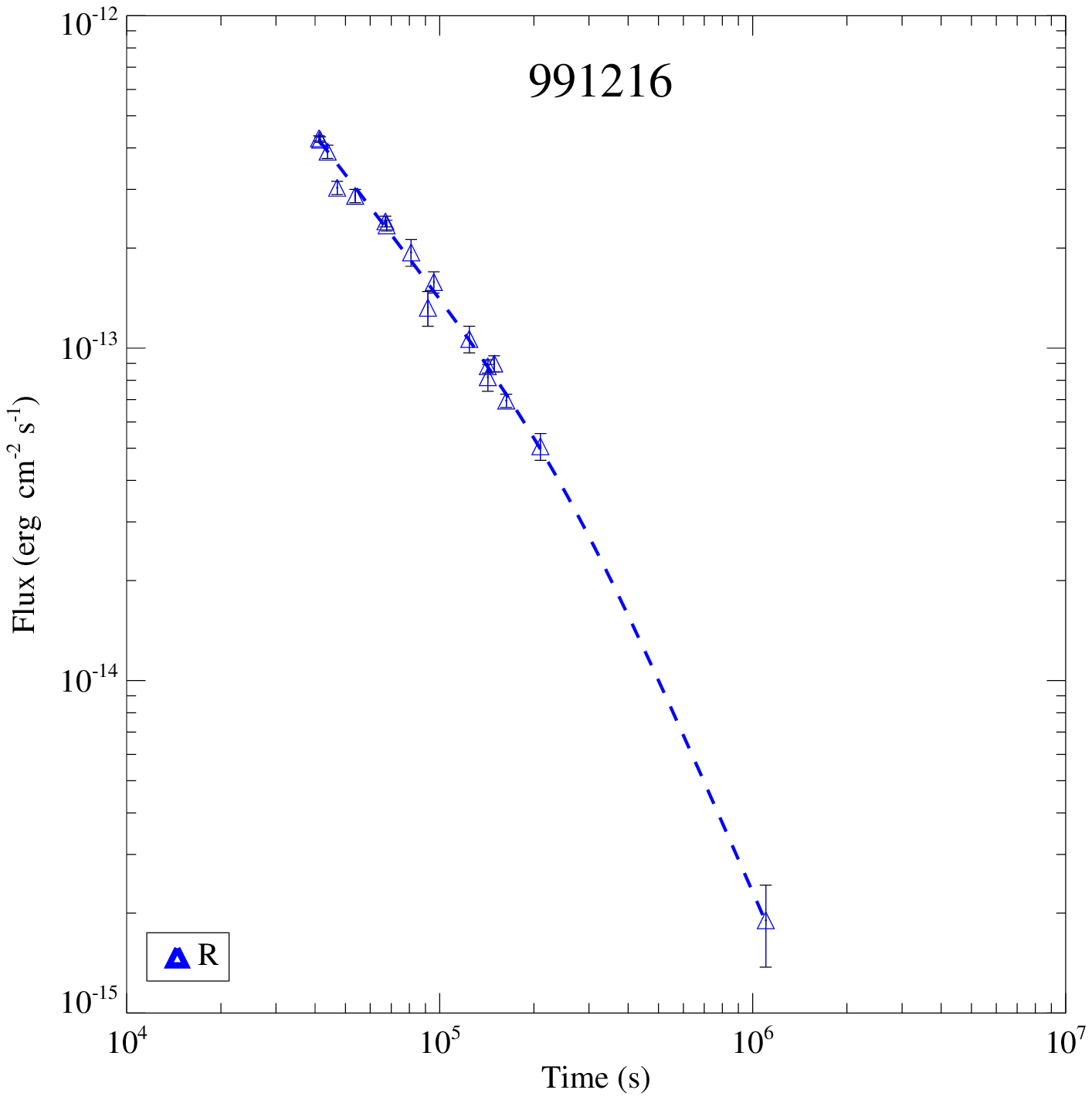}
\includegraphics[angle=0,scale=0.35]{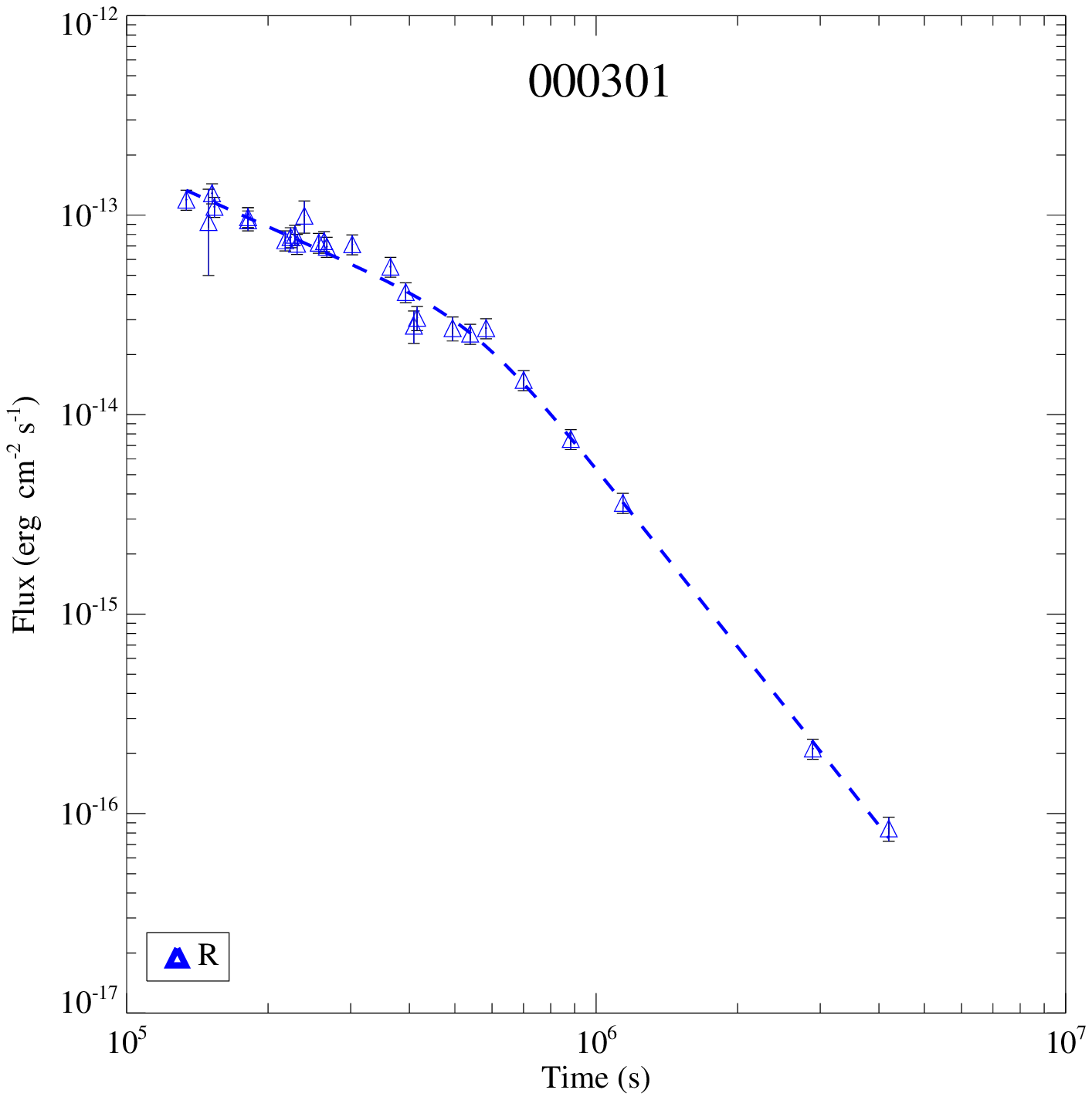}
\includegraphics[angle=0,scale=0.35]{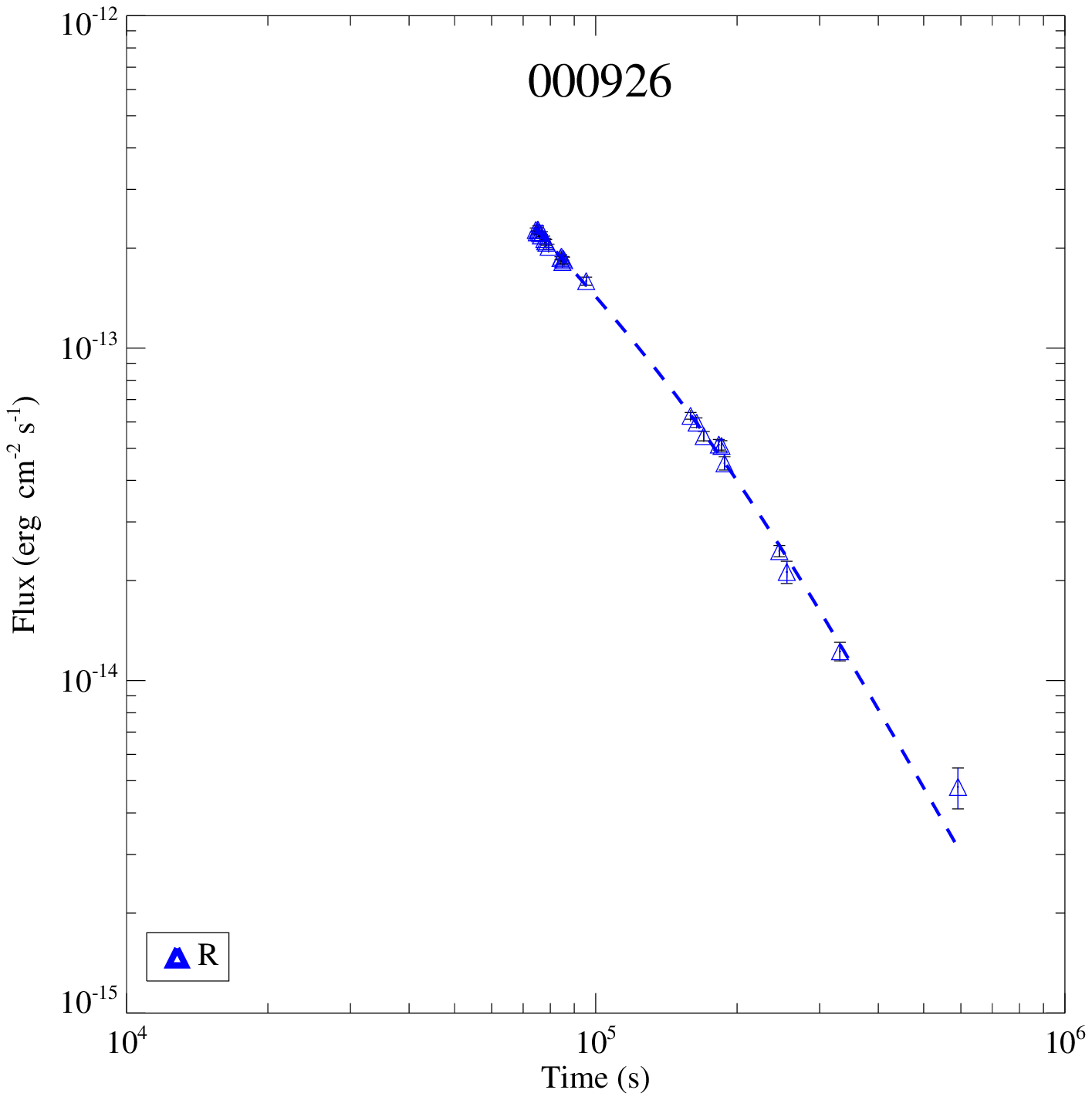}
\includegraphics[angle=0,scale=0.35]{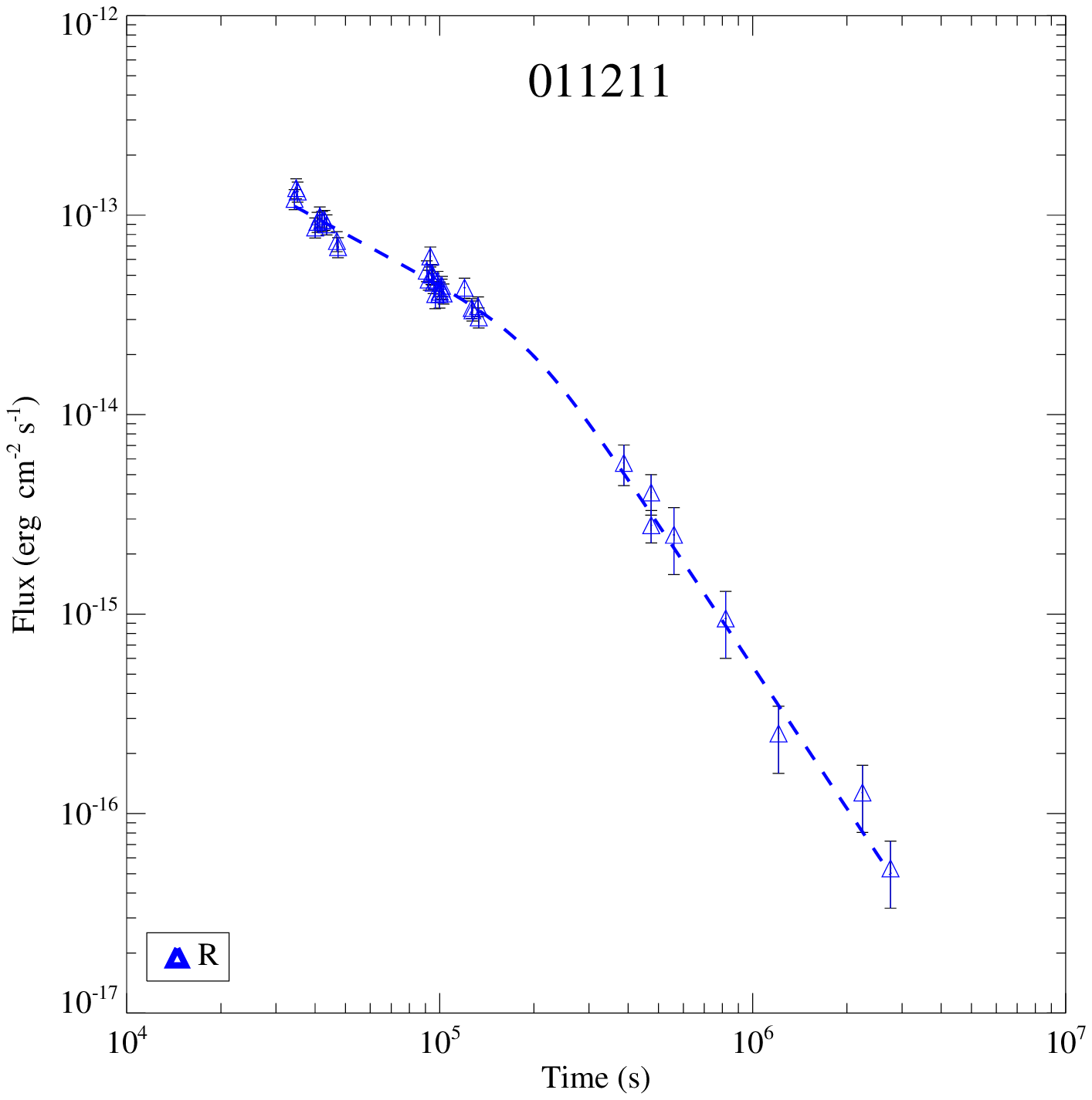}
\includegraphics[angle=0,scale=0.35]{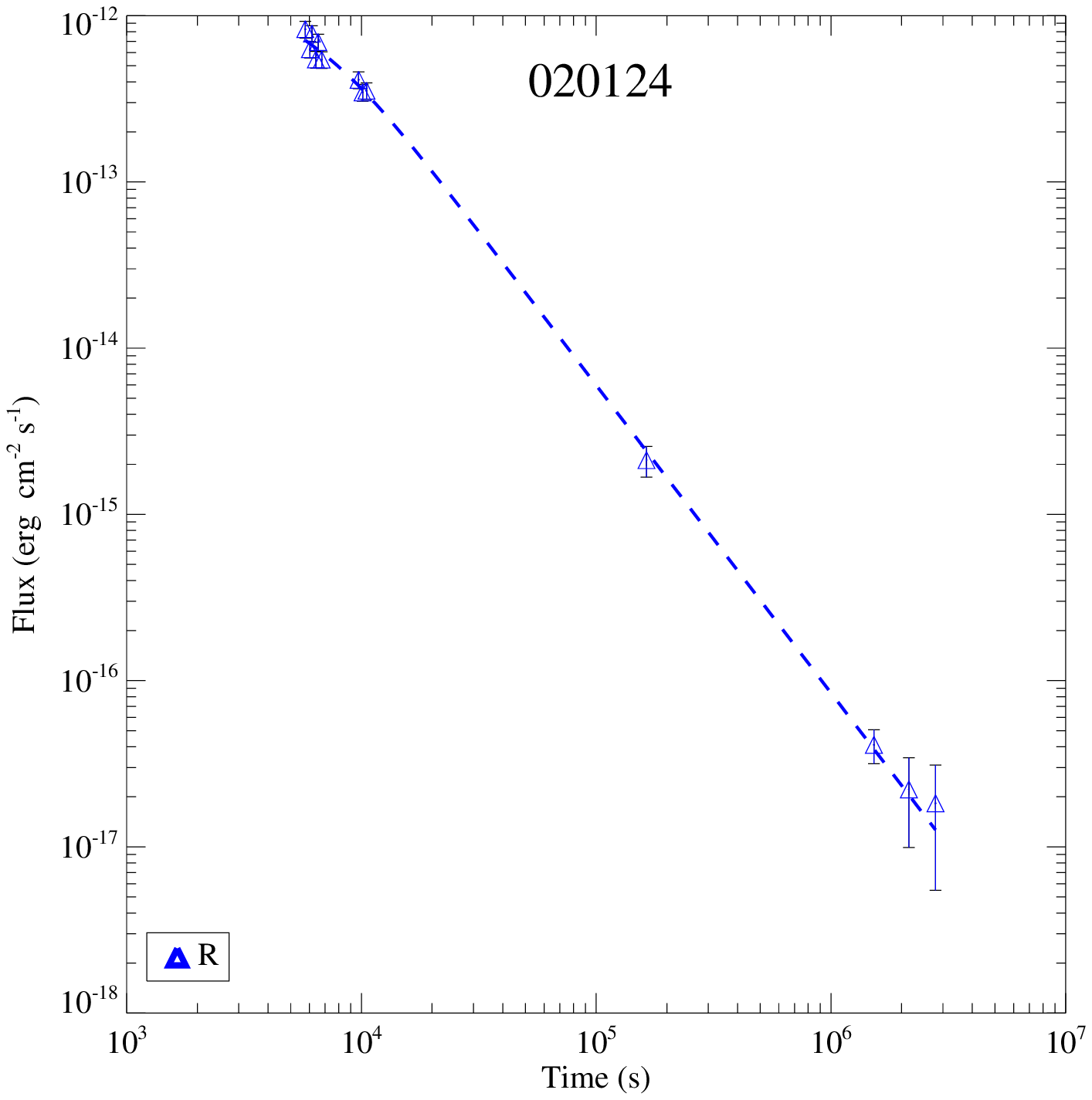}
\includegraphics[angle=0,scale=0.35]{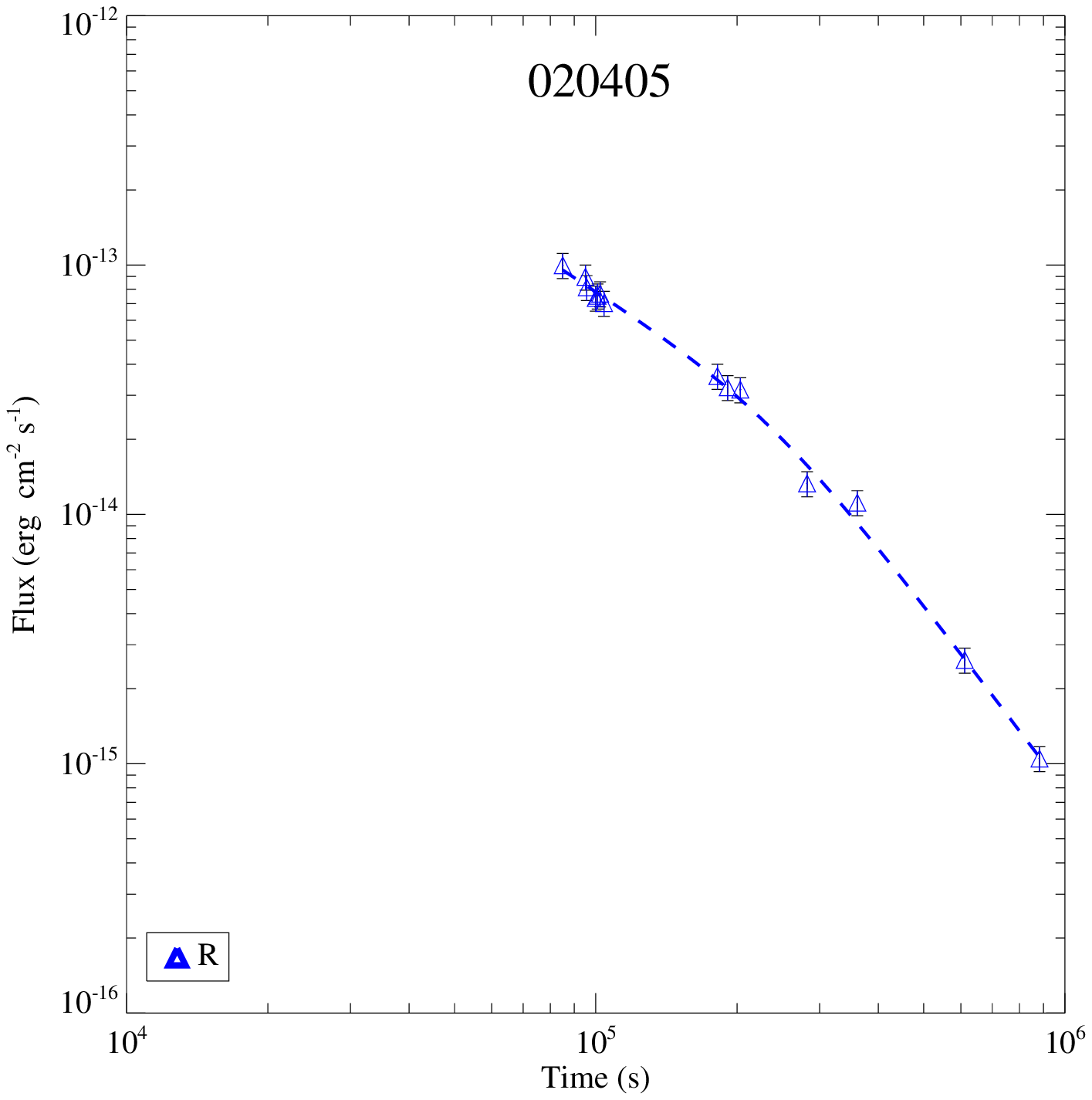}
\includegraphics[angle=0,scale=0.35]{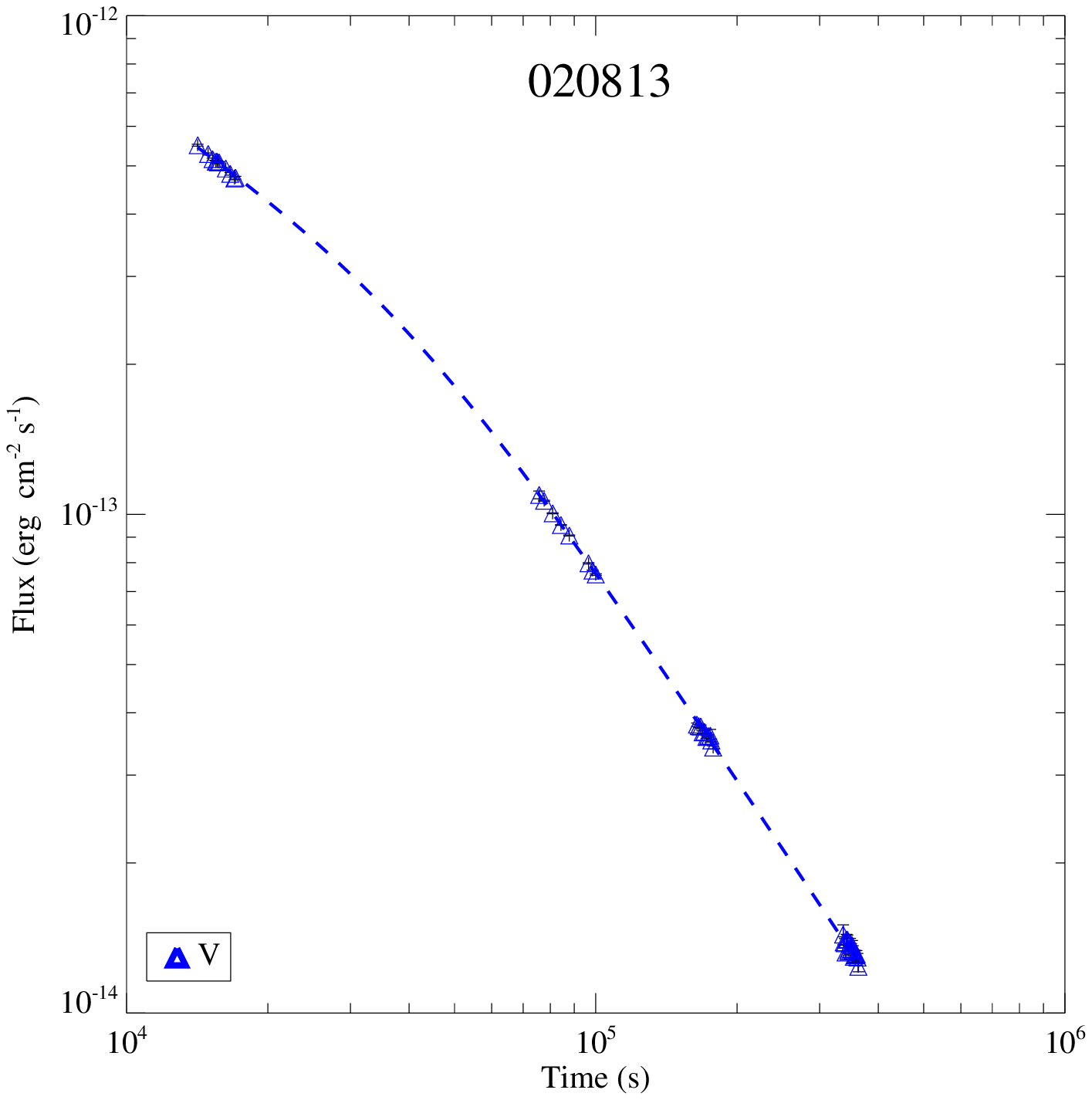}
\includegraphics[angle=0,scale=0.35]{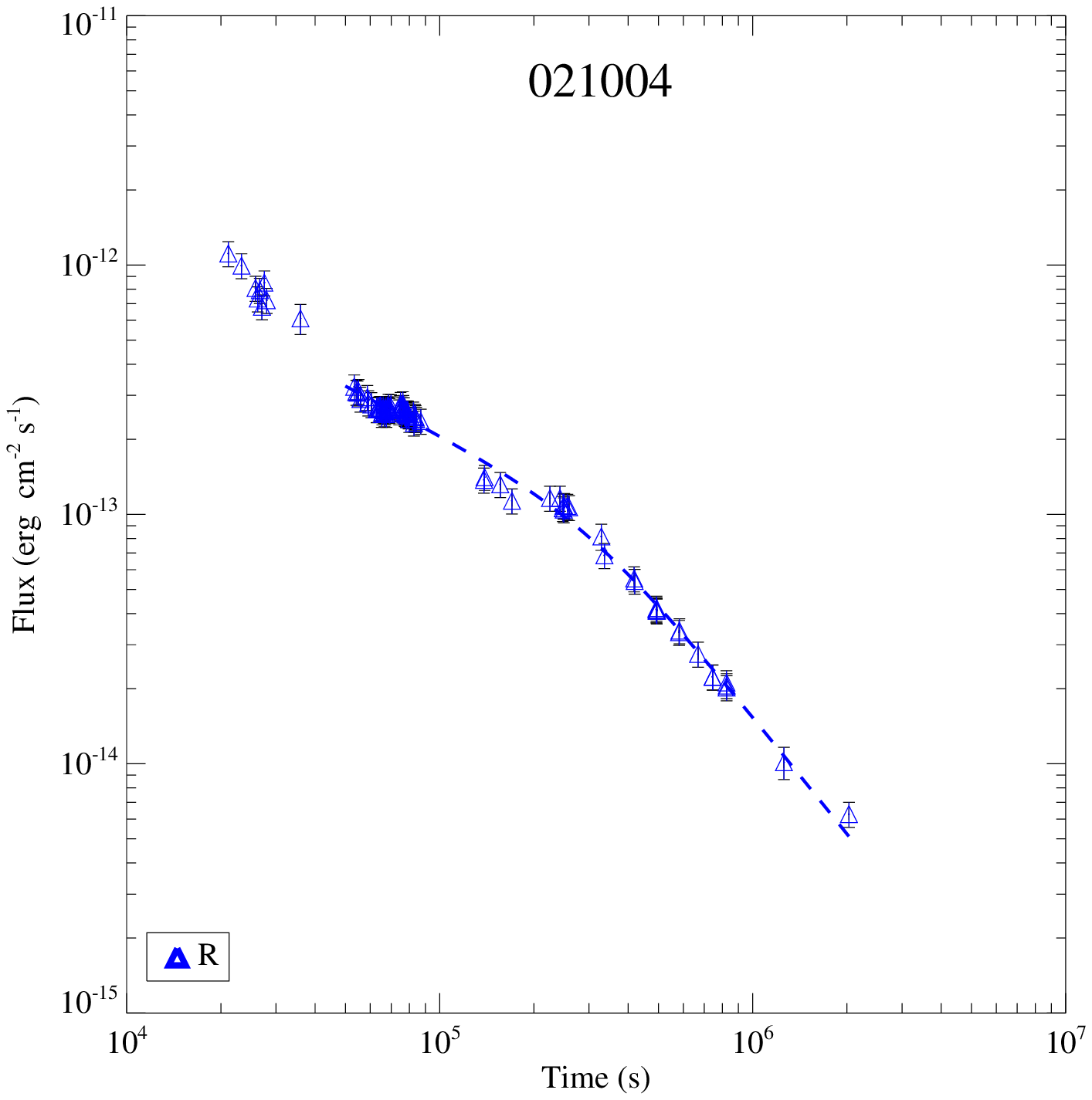}
\includegraphics[angle=0,scale=0.35]{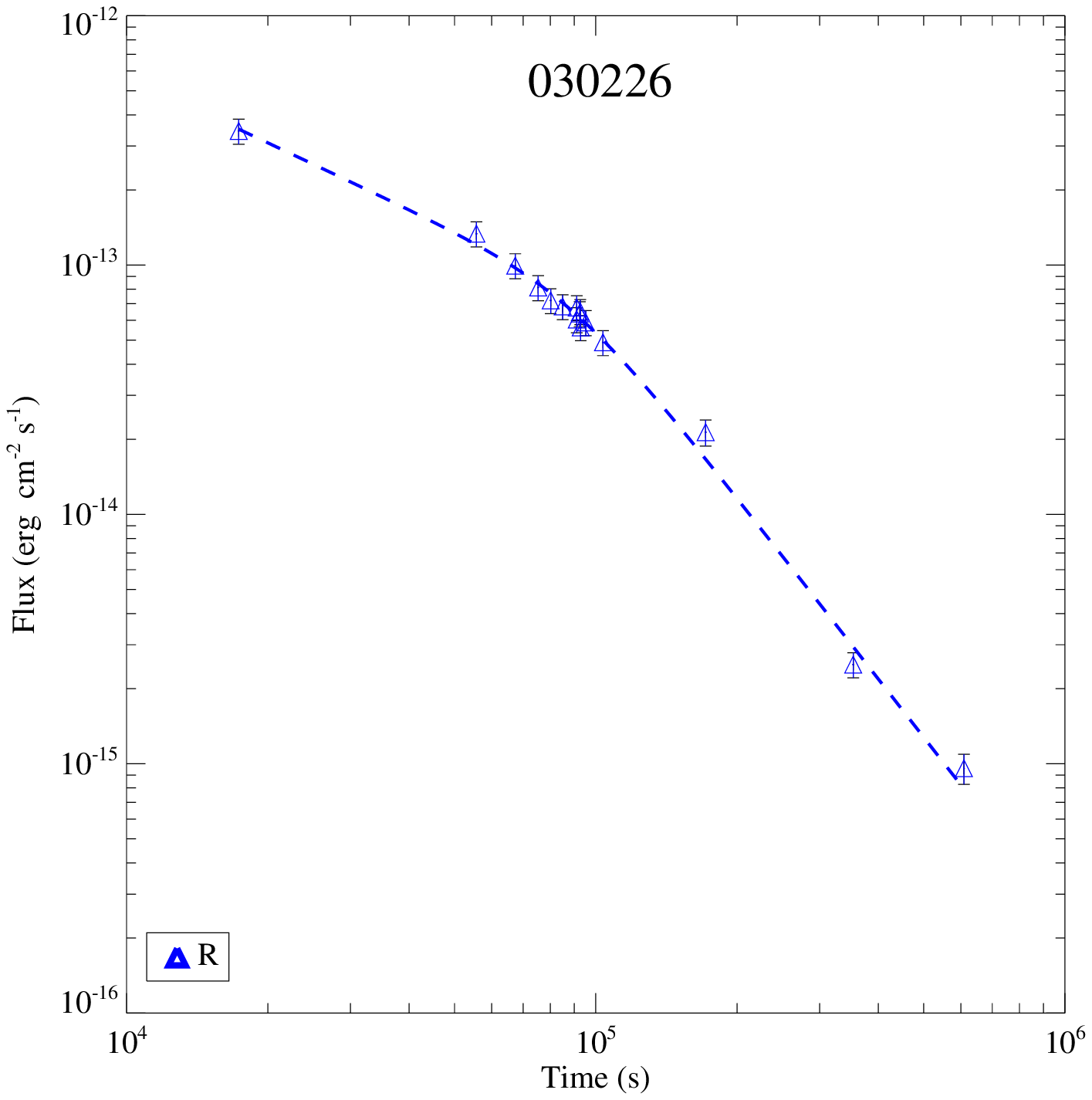}
\includegraphics[angle=0,scale=0.35]{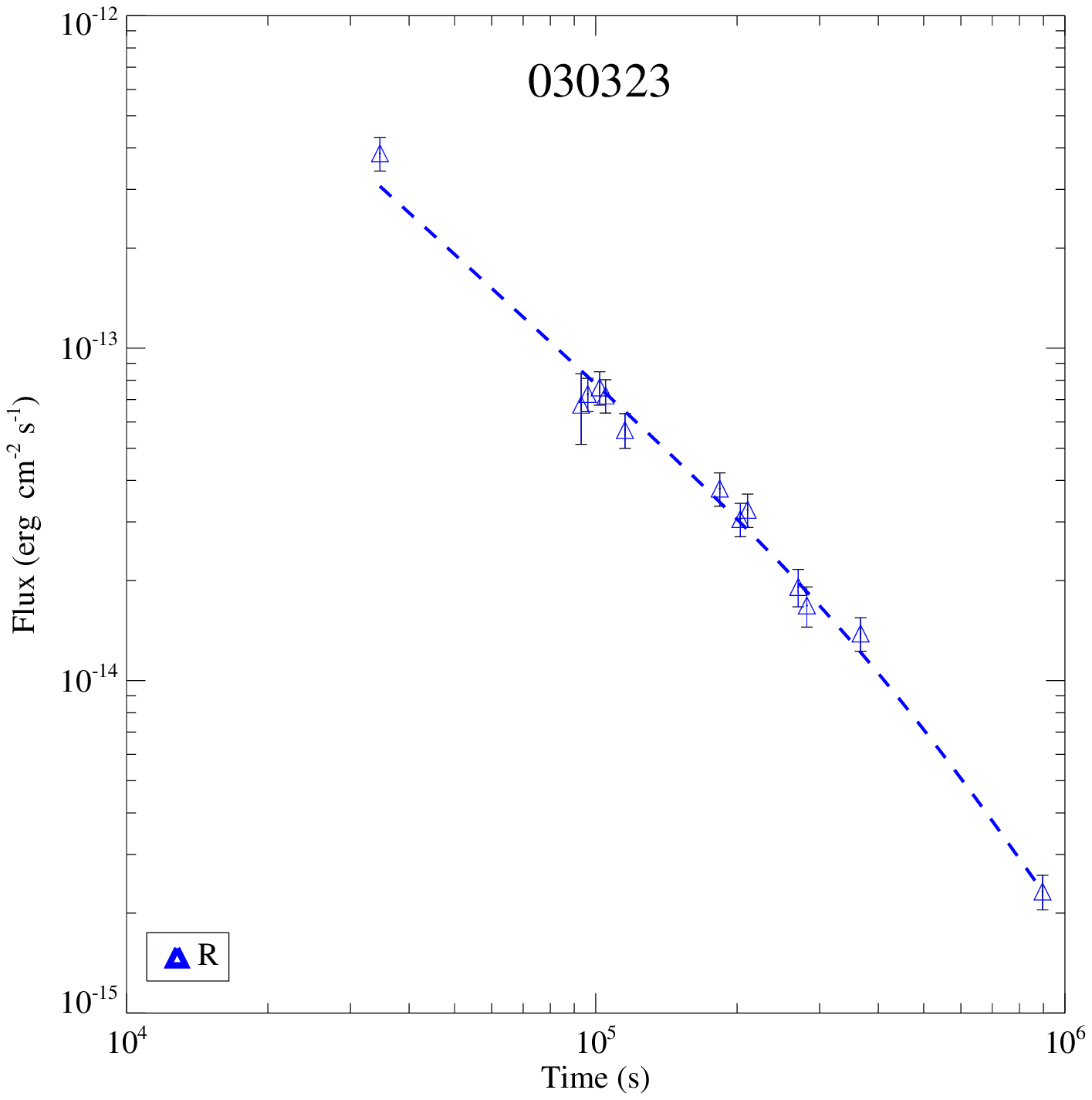}
\hfill
\includegraphics[angle=0,scale=0.35]{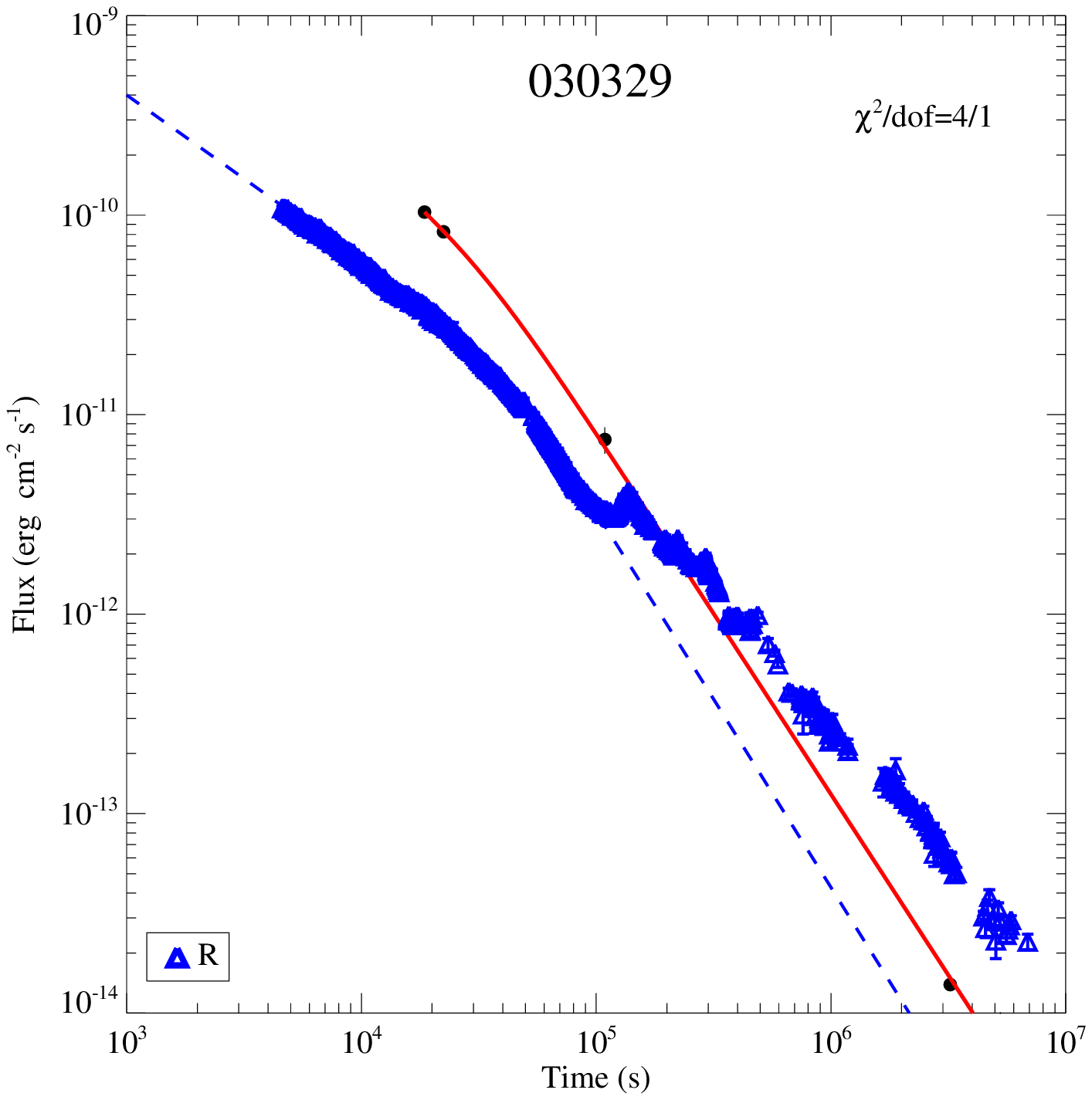}
\center{Fig.2---  continued}
\end{figure*}

\begin{figure*}
\includegraphics[angle=0,scale=0.35]{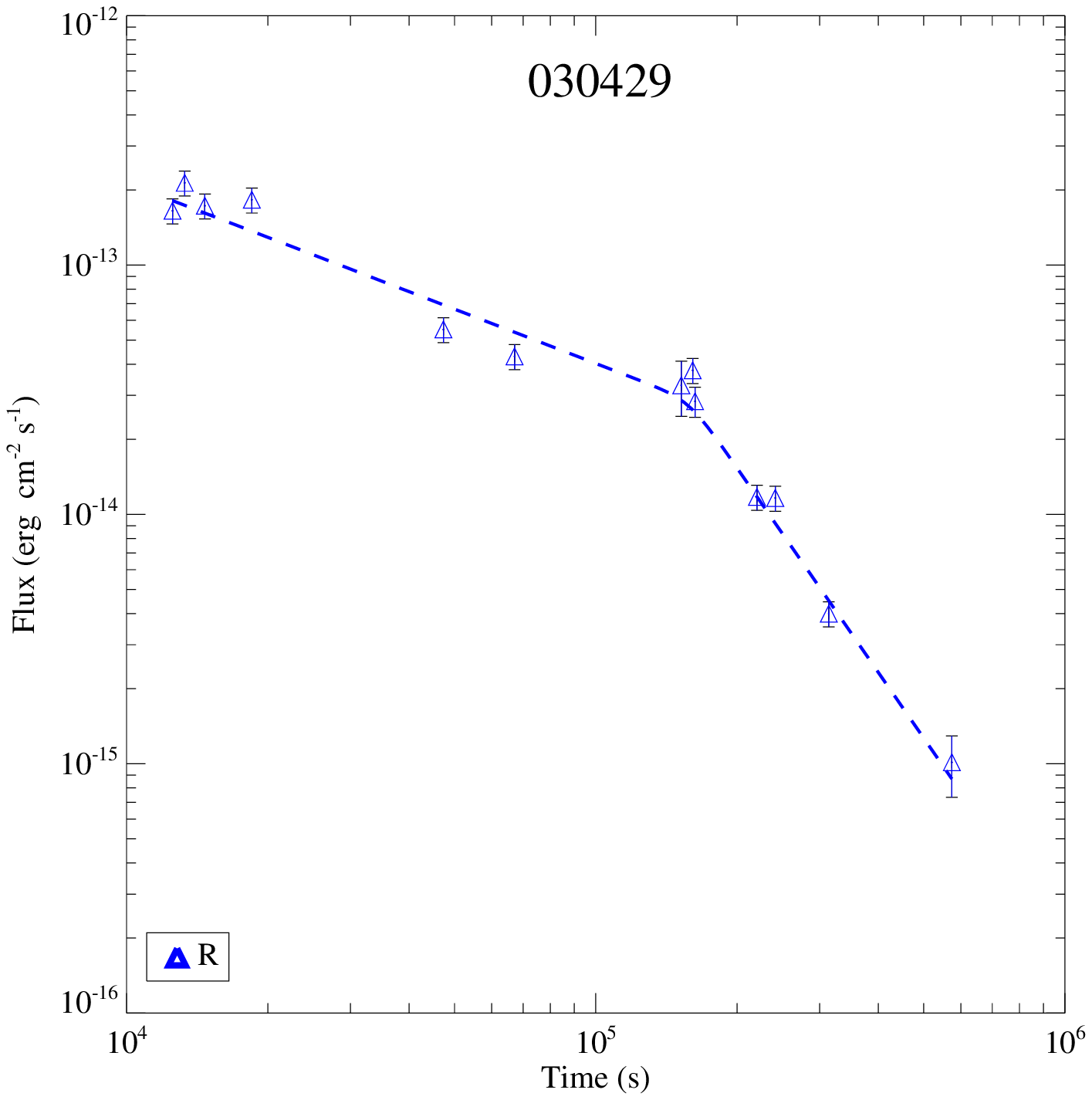}
\includegraphics[angle=0,scale=0.35]{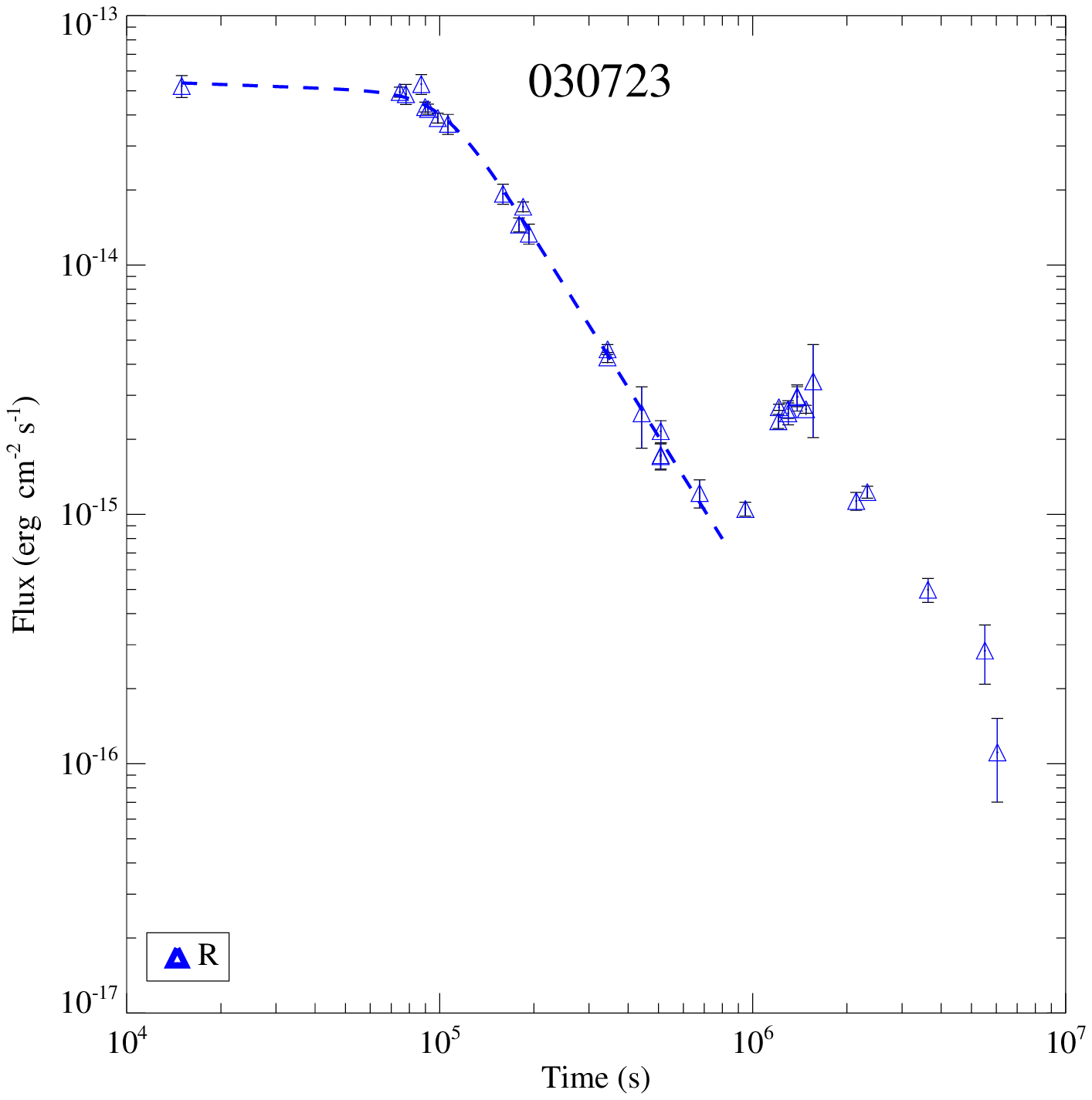}
\hfill \center{Fig. 2---  continued}
\end{figure*}
%
%
%
%

\clearpage
\thispagestyle{empty}
\setlength{\voffset}{-18mm}
\begin{figure*}
\includegraphics[angle=0,scale=0.8]{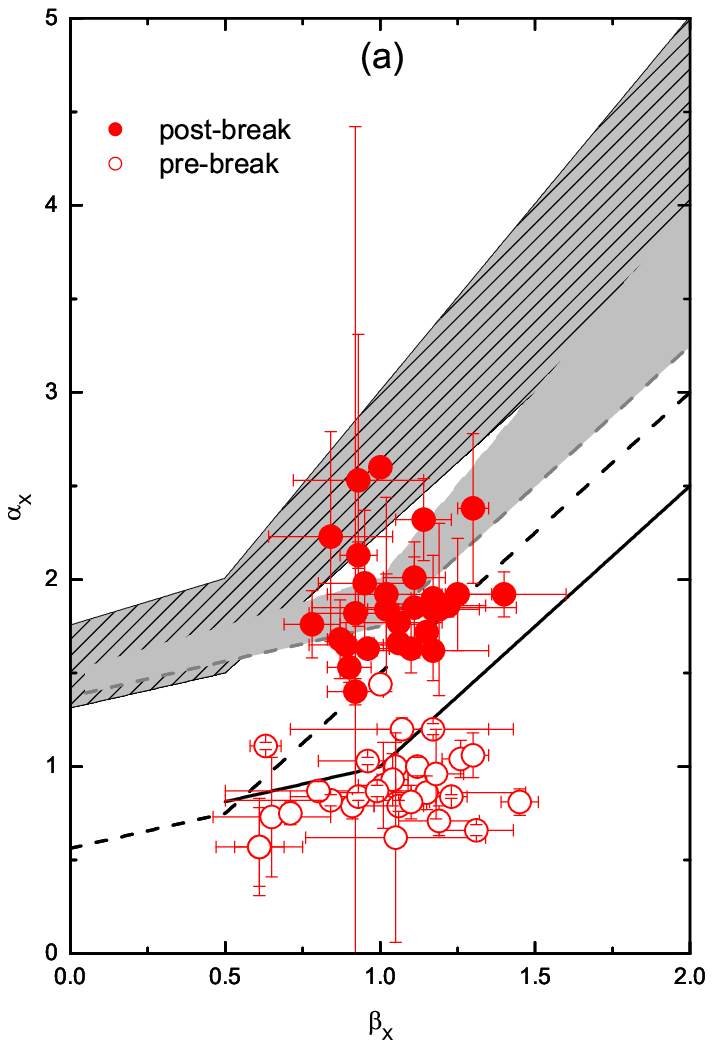}
\includegraphics[angle=0,scale=0.8]{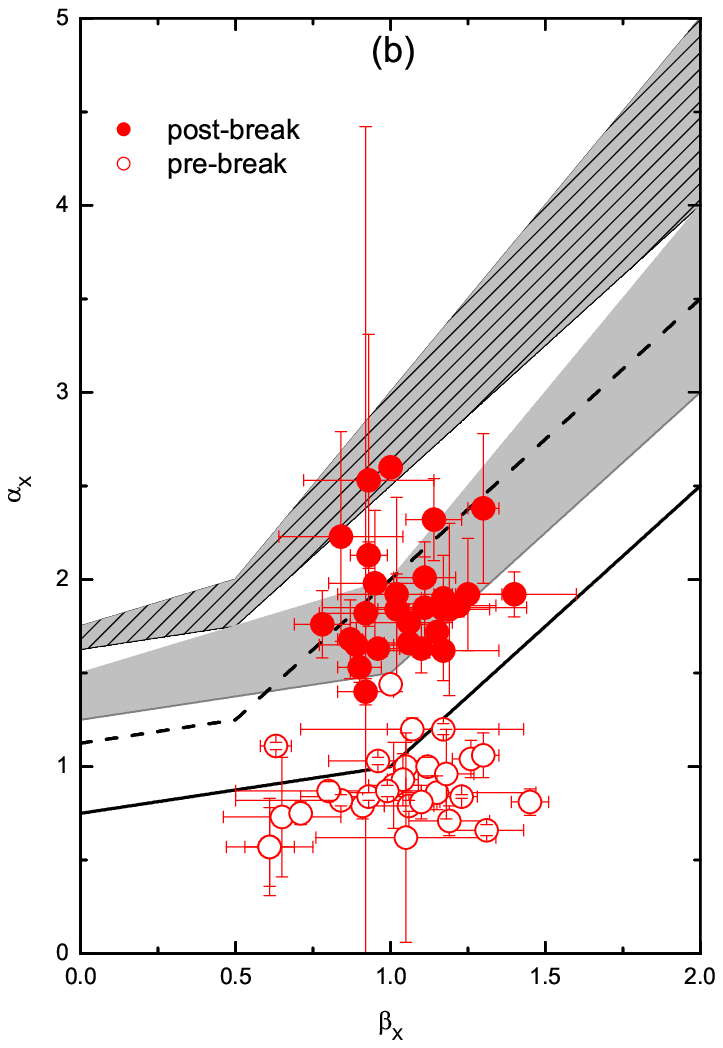}
\includegraphics[angle=0,scale=0.8]{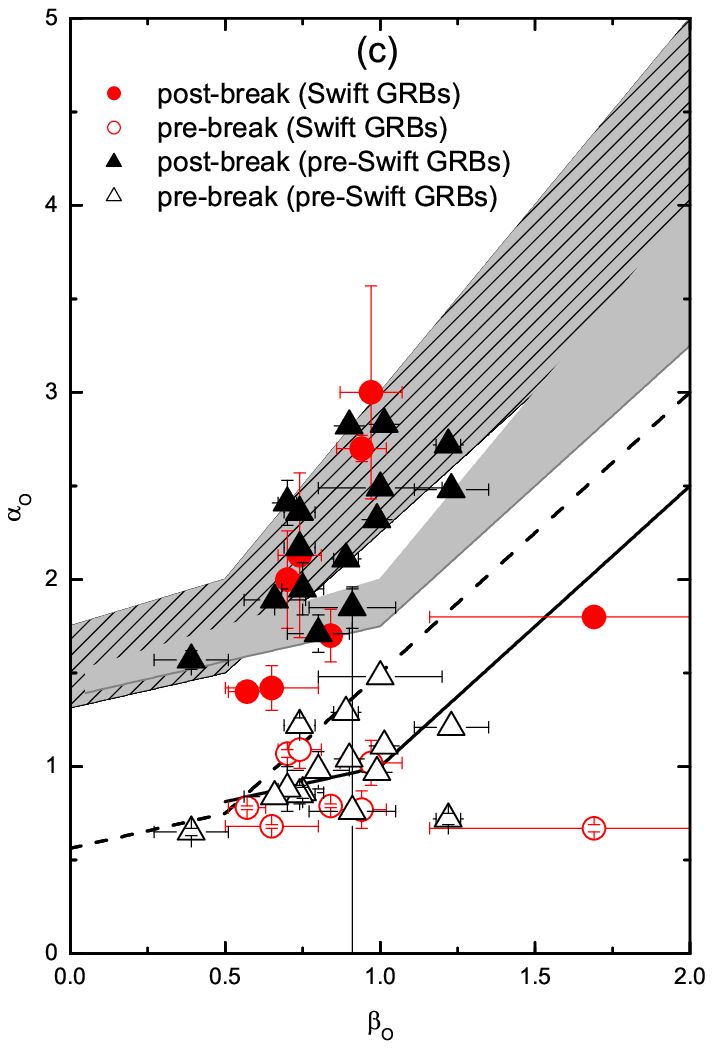}
\hfill
\includegraphics[angle=0,scale=0.8]{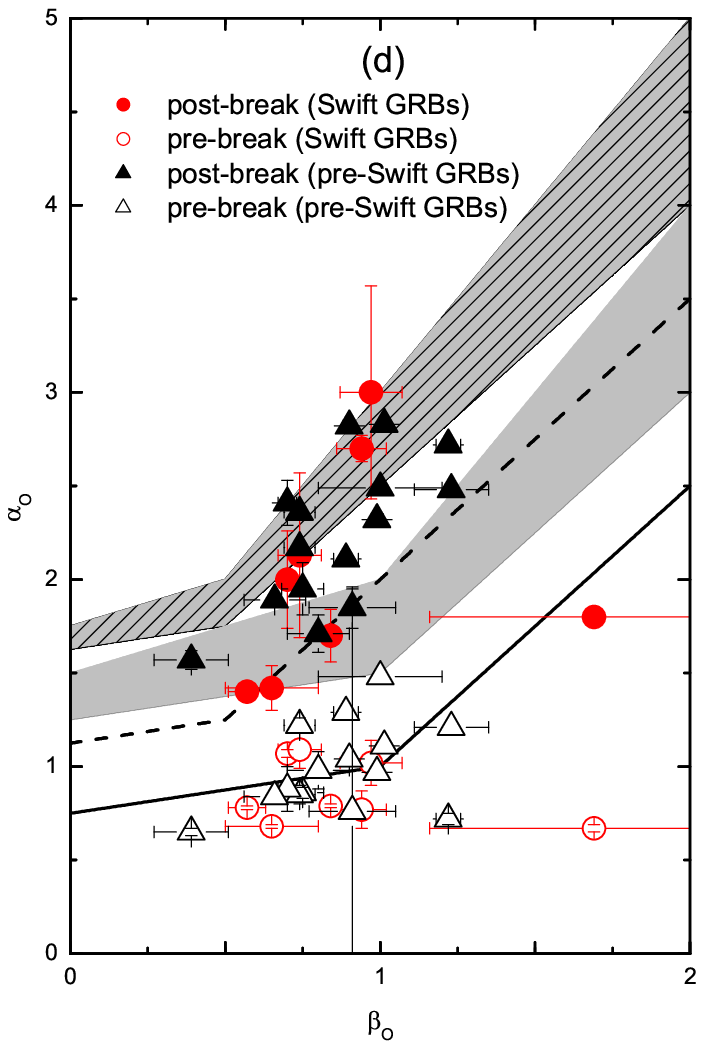}
\caption{XRT and optical data compared for the ``Silver'' jet break candidates
with the closure relations of the forward shock models for emissions in the
spectral regimes (I)$\nu_X>\max(\nu_m,\nu_c)$ and (II) $\nu_m<\nu_X<\nu_c$. The
solid lines and shaded regions indicate the closure relations of the pre- and
post-break segments in the spectral regime I. The lower/upper boundaries of the
regions are defined with the closure relation without/with taking the jet
sideways expansions into account. Similarly, the dashed lines and shaded regions
filled with lines are for the emission in the spectral regime II. The open
symbols represent the data for the pre-break segments, and the filled symbols for
the post-break segment. The circles stand for the {\em Swift} GRBs, and triangles
for the pre-{\em Swift} GRBs. {\em Panel (a)}: ISM, XRT data; {\em Panel (b)}:
wind, XRT data; {\em Panel (c)}: ISM, optical data; {\em Panel (d)}: wind,
optical data.}\label{Fig_model}
\end{figure*}
\clearpage
\setlength{\voffset}{0mm}

\begin{figure}
\epsscale{0.8} \plotone{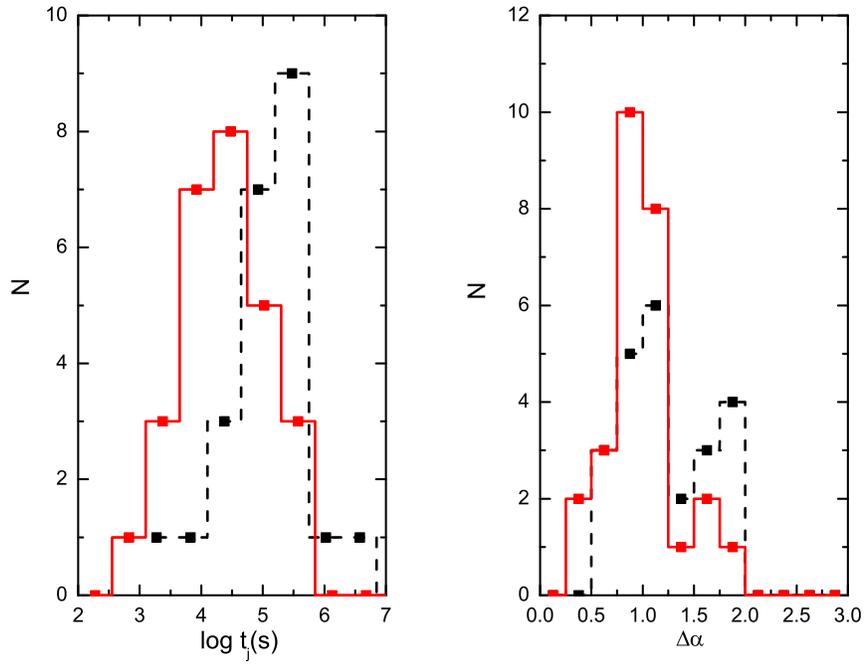} \caption{Comparison of the distributions of $t_j$
and $\Delta \alpha$ for the XRT data (solid lines) and the optical data (dashed
lines).} \label{Fig_Ek_p_01}
\end{figure}

\begin{figure}
\epsscale{0.8} \plotone{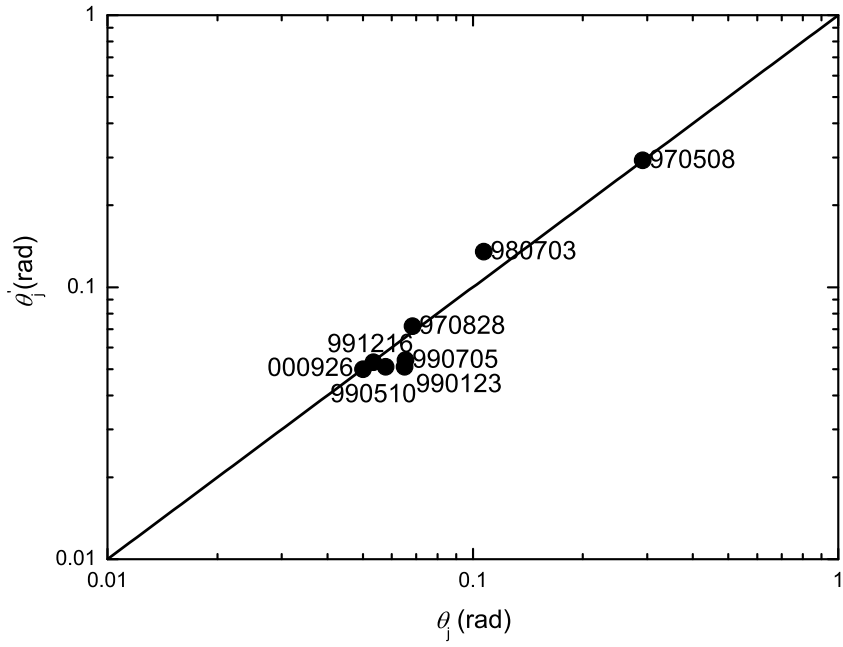} \caption{ Comparison of the $\theta_j$ derived
from the X-ray afterglow with that from the prompt gamma-ray emission (from Frail
et al. 2001). The line is $\theta_j=\theta_j^{'}$.} \label{Fig_5}
\end{figure}

\begin{figure}
\epsscale{0.8} \plotone{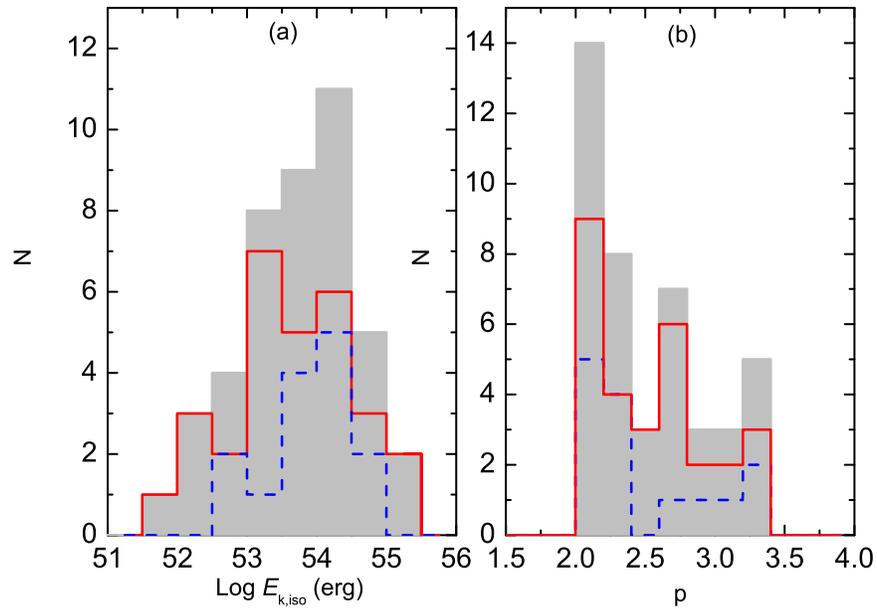} \caption{Comparisons of the distributions of
$E_{\rm K,iso}$ (panel a) and $p$ (panel b) for {\em Swift} GRBs (solid lines)
with that of the pre-{\em Swift} GRBs (dashed lines). The shaded columns are for
both pre-{\em Swift} and {\em Swift} GRBs combined. } \label{Fig_Ek_p_02}
\end{figure}

\clearpage

\begin{figure}
\epsscale{1} \plotone{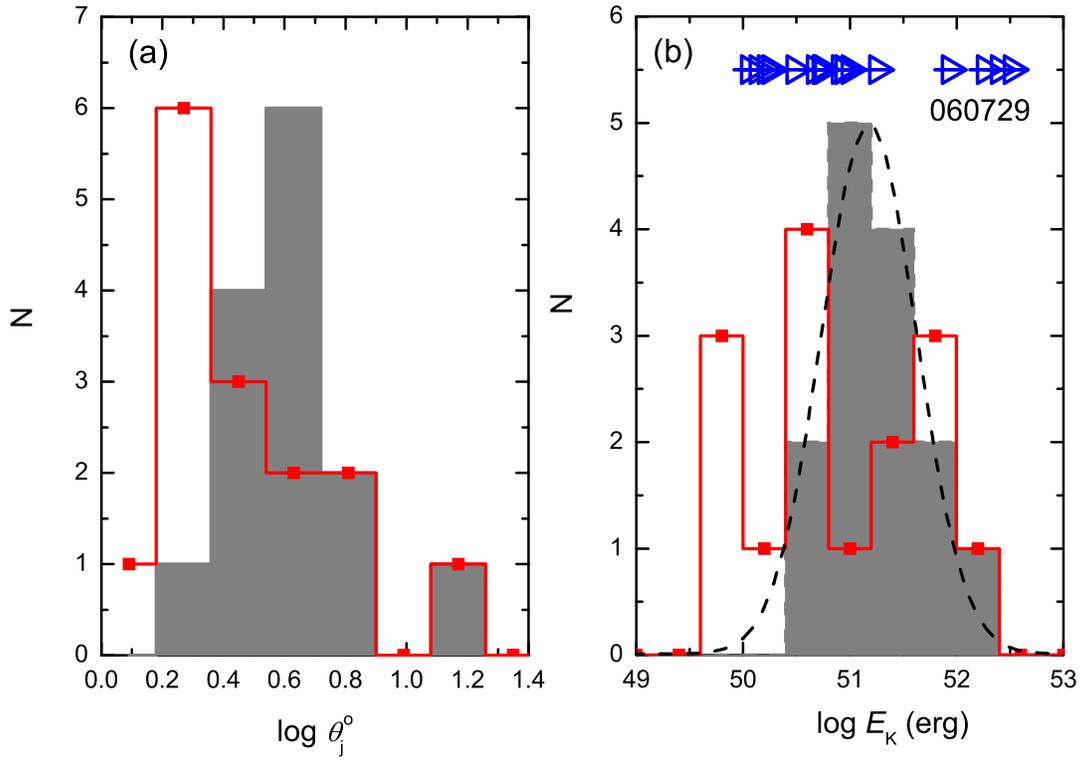} \caption{Comparison of the $E_{\rm K}$ distribution
of {\em Swift} GRBs with that of the pre-{\em Swift} GRBs (shaded columns). The
lower limits of $E_{\rm K}$ derived from the XRT observations are marked as open
triangles. The dashed line is the Gaussian fit to the distribution of $E_K$ of
pre-{\em Swift} GRBs.} \label{Fig_Ek_p_03}
\end{figure}

\clearpage
\begin{figure}
\epsscale{1} \plotone{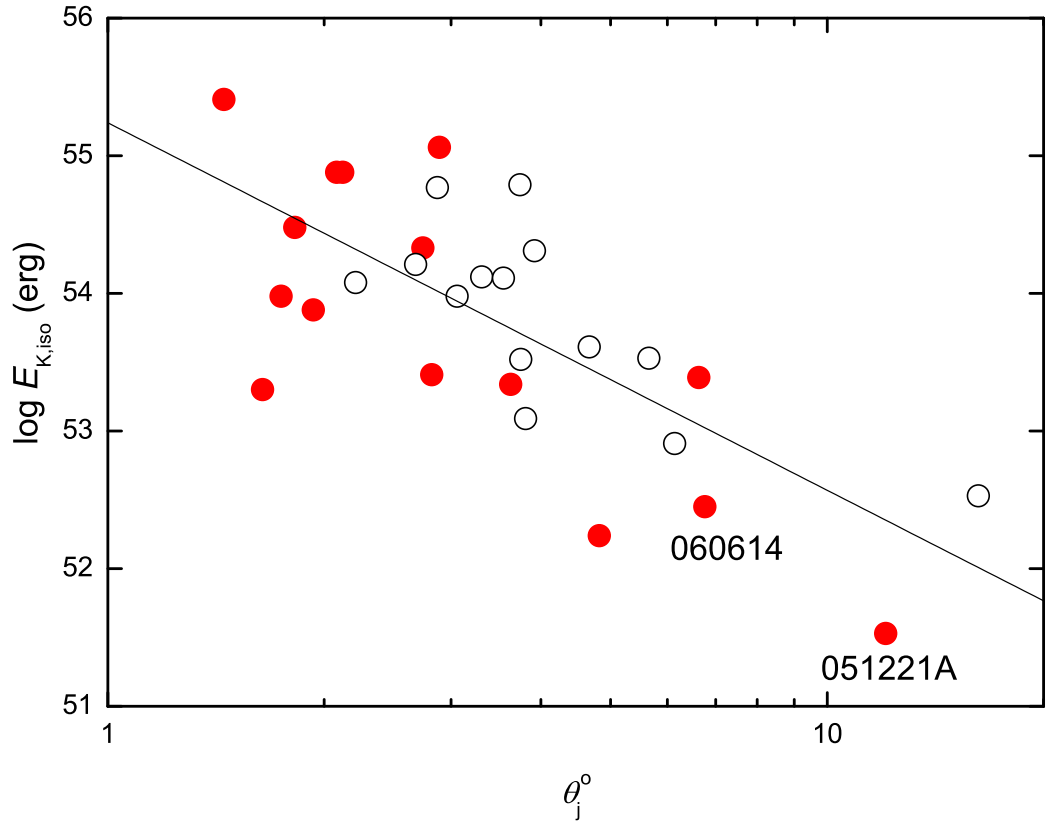} \caption{The $E_{\rm K,iso}$  as a function of
$\theta_j$ for both the pre-{\em Swift} (open circles) and {\em Swift} GRBs
(solid circles). The solid line is the best fit for both the pre-{\em Swift} and
{\em Swift} GRBs.} \label{Fig_Ek_p_04}
\end{figure}


\begin{thebibliography}{}
\bibitem[Aoki et al.(2006)]{2006GCN..4703....1A} Aoki, K., Hattori, T.,Kawabata, K.~S., \& Kawai, N.\ 2006, GCN, 4703, 1
\bibitem[Barthelmy et al.(2005)]{2005Natur.438..994B} Barthelmy, S.~D., et al.\ 2005, \nat, 438, 994
\bibitem[Berger \& Gladders(2006)]{2006GCN..5170....1B} Berger, E., \& Gladders, M.\ 2006, GCN, 5170, 1
\bibitem[Berger \& Mulchaey(2005)]{2005GCN..3122....1B} Berger, E., \& Mulchaey, J.\ 2005, GCN, 3122, 1
\bibitem[Berger \& Soderberg(2005)]{2005GCN..4384....1B} Berger, E., \& Soderberg, A.~M.\ 2005, GCN, 4384, 1
\bibitem[Berger et al.(2001)]{2001ApJ...556..556B} Berger, E., et al.\ 2001, \apj, 556, 556
\bibitem[Berger et al.(2003)]{2003ApJ...590..379B} Berger, E., Kulkarni, S.~R., \& Frail, D.~A.\ 2003, \apj, 590, 379
\bibitem[Berger et al. (2005a)]{2005Natur.438..988B} Berger, E. et al. 2005a, Nature, 438, 988
\bibitem[Berger et al. (2005b)]{2005ApJ...634..501B} Berger, E. et al. 2005b, ApJ, 634, 501
\bibitem[Berger et al.(2005c)]{2005GCN..3368....1B} Berger, E., Cenko, S.~B., Steidel, C., Reddy, N., \& Fox, D.~B.\ 2005c, GCN, 3368, 1
\bibitem[Bloom et al.(2003)]{2003ApJ...594..674B} Bloom, J.~S., Frail, D.~A., \& Kulkarni, S.~R.\ 2003, \apj, 594, 674
\bibitem[Bloom et al.(2006a)]{2006GCN..5217....1B} Bloom, J.~S., Foley,R.~J., Koceveki, D., \& Perley, D.\ 2006a, GCN, 5217, 1
\bibitem[Bloom et al.(2006b)]{2006GCN..5826....1B} Bloom, J.~S., Perley,D.~A., \& Chen, H.~W.\ 2006b, GCN, 5826, 1
\bibitem[Blustin et al.(2006)]{2006ApJ...637..901B} Blustin, A.~J., et al.\ 2006, \apj, 637, 901
\bibitem[Bromm \& Loeb(2002)]{2002ApJ...575..111B} Bromm, V., \& Loeb, A.\ 2002, \apj, 575, 111
\bibitem[Burrows \& Racusin 2007]{2007astro.ph..2633B} Burrows, D.~N., \& Racusin, J.\ 2007, preprint(astro-ph/200702633)
\bibitem[Burrows et al.(2005a)]{2005SSRv..120..165B} Burrows, D.~N., et al.\ 2005a, Space Science Reviews, 120, 165
\bibitem[Burrows et al. 2005b]{2005Sci...309.1833B} Burrows, D.~N., et al.\ 2005b, Science, 309, 1833
\bibitem[Burrows et al.(2006)]{2006ApJ...653..468B} Burrows, D.~N., et al.\ 2006, \apj, 653, 468
\bibitem[Campana et al.(2007)]{2007A&A...472..395C} Campana, S., Guidorzi, C., Tagliaferri, G., Chincarini, G., Moretti, A., Rizzuto, D., \& Romano, P.\ 2007, \aap, 472, 395
\bibitem[Castro-Tirado et al.(2006)]{2006GCN..5218....1C} Castro-Tirado,A.~J., Amado, P., Negueruela, I., Gorosabel, J., Jelinek, M., \& de Ugarte Postigo, A.\ 2006, GCN, 5218, 1
\bibitem[Cenko et al.(2005)]{2005GCN..3542....1C} Cenko, S.~B., et al. 2005, GCN, 3542, 1
\bibitem[Cenko et al.(2006a)]{2006GCN..4592....1C} Cenko, S.~B., Berger, E., \& Cohen, J.\ 2006a, GCN, 4592, 1
\bibitem[Cenko et al.(2006b)]{2006GCN..5155....1C} Cenko, S.~B., et al. 2006b, GCN, 5155, 1
\bibitem[Chevalier \& Li(2000)]{2000ApJ...536..195C} Chevalier, R.~A., \& Li, Z.-Y.\ 2000, \apj, 536, 195
\bibitem[Chincarini et al.(2007)]{2007astro.ph..2371C} Chincarini, G., et al.\ 2007, ApJ, submitted (arXiv:astro-ph/0702371)
\bibitem[Covino et al.(2006)]{2006astro.ph.12643C} Covino, S., et al. \ 2006, arXiv:astro-ph/0612643
\bibitem[Cucchiara et al.(2006a)]{2006GCN..4729....1C} Cucchiara, A., Fox, D.~B., \& Berger, E.\ 2006, GCN, 4729, 1
\bibitem[Cucchiara et al.(2006b)]{2006GCN..5052....1C} Cucchiara, A., Price, P.~A., Fox, D.~B., Cenko, S.~B., \& Schmidt, B.~P.\ 2006, GCN, 5052, 1
\bibitem[Curran et al.(2007)]{2007arXiv0706.1188C} Curran, P.~A., et al.\ 2007, MNRAS, in press(arXiv:0706.1188)
\bibitem[Dai \& Cheng 2001]{mmm} Dai, Z. G., \& Cheng, K. S. 2001, \apj, 558, L109
\bibitem[Dai \& Lu 1999]{838}Dai, Z. G. \& Lu, T. 1999, ApJ, 519, L155
\bibitem[Dai \& Lu(1998)]{1998A&A...333L..87D} Dai, Z.~G., \& Lu, T.\ 1998, \aap, 333, L87
\bibitem[Dai et al.(2004)]{2004ApJ...612L.101D} Dai, Z.~G., Liang, E.~W., \& Xu, D.\ 2004, \apjl, 612, L101
\bibitem[Dai et al.(2007)]{2007ApJ...658..509D} Dai, X., Halpern, J.~P., Morgan, N.~D., Armstrong, E., Mirabal, N., Haislip, J.~B., Reichart, D.~E., \& Stanek, K.~Z.\ 2007, \apj, 658, 509
\bibitem[D'Elia et al.(2005)]{2005GCN..4044....1D} D'Elia, V., et al.\ 2005, GCN, 4044, 1
\bibitem[D'Elia et al.(2006)]{2006GCN..5637....1D} D'Elia, V., et al.\ 2006, GCN, 5637, 1
\bibitem[Della Valle et al.(2006)]{2006Natur.444.1050D} Della Valle, M., et al.\ 2006, \nat, 444, 1050
\bibitem[Djorgovski et al.(2001)]{2001ApJ...562..654D} Djorgovski, S.~G., Frail, D.~A., Kulkarni, S.~R., Bloom, J.~S., Odewahn, S.~C., \& Diercks, A.\ 2001, \apj, 562, 654
\bibitem[Falcone et al.(2007)]{2007arXiv0706.1564F} Falcone, A.~D., et al.\ 2007, ApJ, submitted (arXiv:0706.1564)
\bibitem[Fox et al.(2005)]{2005Natur.437..845F} Fox, D.~B., et al.\ 2005, \nat, 437, 845
\bibitem[Frail et al.(2000)]{2000ApJ...537..191F} Frail, D.~A., Waxman, E., \& Kulkarni, S.~R.\ 2000, \apj, 537, 191
\bibitem[Frail et al.(2001)]{2001ApJ...562L..55F} Frail, D.~A., et al.\ 2001, \apjl, 562, L55
\bibitem[Fugazza et al.(2006)]{2006GCN..5276....1F} Fugazza, D., et al.\ 2006, GCN, 5276, 1
\bibitem[Fynbo et al.(2005a)]{2005GCN..3136....1F} Fynbo, J.~P.~U., Hjorth,J., Jensen, B.~L., Jakobsson, P., Moller, P., \& Naranen, J.\ 2005a, GCN, 3136, 1
\bibitem[Fynbo et al.(2005b)]{2005GCN..3176....1F} Fynbo, J.~P.~U., et al.\ 2005b, GCN, 3176, 1
\bibitem[Fynbo et al.(2005c)]{2005GCN..3749....1F} Fynbo, J.~P.~U., et al.\ 2005c, GCN, 3749, 1
\bibitem[Fynbo et al.(2006)]{2006Natur.444.1047F} Fynbo, J.~P.~U., et al.\ 2006, \nat, 444, 1047
\bibitem[Gal-Yam et al.(2006)]{2006Natur.444.1053G} Gal-Yam, A., et al.\ 2006, \nat, 444, 1053
\bibitem[Gehrels et al. 2004]{2004ApJ...611.1005G} Gehrels, N., et al.\ 2004, \apj, 611, 1005
\bibitem[Gehrels et al.(2006)]{2006Natur.444.1044G} Gehrels, N., et al.\ 2006, \nat, 444, 1044
\bibitem[Ghirlanda et al.(2004a)]{2004ApJ...613L..13G} Ghirlanda, G., Ghisellini, G., Lazzati, D., \& Firmani, C.\ 2004a, \apjl, 613, L13
\bibitem[Ghirlanda et al.(2004b)]{2004ApJ...616..331G} Ghirlanda, G., Ghisellini, G., \& Lazzati, D.\ 2004b, \apj, 616, 331
\bibitem[Gou et al. (2004)]{2004ApJ...604..508G} Gou, L. J., M\'esz\'aros, P., Abel, T., Zhang, B. 2002, ApJ, 604, 508
\bibitem[Grupe et al.(2006)]{2006ApJ...653..462G} Grupe, D., Burrows, D.~N., Patel, S.~K., Kouveliotou, C., Zhang, B., M{\'e}sz{\'a}ros, P., Wijers, R.~A.~M., \& Gehrels, N.\ 2006, \apj, 653, 462
\bibitem[Harrison et al.(1999)]{1999ApJ...523L.121H} Harrison, F.~A., et al.\ 1999, \apjl, 523, L121
\bibitem[Huang, Gou, Dai, \& Lu(2000)]{2000ApJ...543...90H} Huang, Y.~F.,Gou, L.~J., Dai, Z.~G., \& Lu, T.\ 2000, \apj, 543, 90
\bibitem[Ioka et al.(2005)]{2005ApJ...631..429I} Ioka, K., Kobayashi, S., \& Zhang, B.\ 2005, \apj, 631, 429
\bibitem[Ioka et al.(2006)]{2006AA...458....7}Ioka, K., Toma, K., Yamazaki, R., \& Nakamura, T.\ 2006, \aap, 458, 7
\bibitem[Jakobsson et al.(2006a)]{2006AA...447..897J} Jakobsson, P., et al.\ 2006a, \aap, 460, L13
\bibitem[Jakobsson et al.(2006b)]{2006GCN..5320....1J} Jakobsson, P., Vreeswijk, P., Fynbo, J.~P.~U., Hjorth, J., Starling, R., Kann, D.~A., \& Hartmann, D.\ 2006a, GCN, 5320, 1
\bibitem[Jakobsson et al.(2006c)]{2006GCN..5617....1J} Jakobsson, P., Levan, A., Chapman, R., Rol, E., Tanvir, N., Vreeswijk, P., \& Watson, D.\ 2006b, GCN, 5617, 1
\bibitem[Jakobsson et al.(2006e)]{2006A&A...447..897J} Jakobsson, P., et al.\ 2006e, \aap, 447, 897
\bibitem[Jaunsen et al.(2006)]{2006GCN..6010....1T} Jaunsen, A.~0., et al. \ 2006, GCN, 6010, 1
\bibitem[Jin \& Fan(2007)]{2007MNRAS.378.1043J} Jin, Z.~P., \& Fan, Y.~Z.\ 2007, \mnras, 378, 1043
\bibitem[Kelson \& Berger(2005)]{2005GCN..3101....1K} Kelson, D., \& Berger, E.\ 2005, GCN, 3101, 1
\bibitem[King et al.(2007)]{2007MNRAS.374L..34K} King, A., Olsson, E., \& Davies, M.~B.\ 2007, \mnras, 374, L34
\bibitem[Kobayashi \& Zhang(2007)]{2007ApJ...655..973K} Kobayashi, S., \& Zhang, B.\ 2007, \apj, 655, 973
\bibitem[Kocevski \& Butler(2007)]{2007arXiv0707.4478K} Kocevski, D., \& Butler, N.\ 2007, arXiv:0707.4478
\bibitem[Kouveliotou et al.(1993)]{1993ApJ...413L.101K} Kouveliotou, C., Meegan, C.~A., Fishman, G.~J., Bhat, N.~P., Briggs, M.~S., Koshut, T.~M., Paciesas, W.~S., \& Pendleton, G.~N.\ 1993, \apjl, 413, L101
\bibitem[Kumar \& Panaitescu(2000)]{2000ApJ...541L...9K} Kumar, P., \& Panaitescu, A.\ 2000, \apjl, 541, L9
\bibitem[Lamb(2000)]{2000PhR...333..505L} Lamb, D.~Q.\ 2000, \physrep, 333, 505
\bibitem[Ledoux et al.(2005)]{2005GCN..3860....1L} Ledoux, C., et al.\ 2005, GCN, 3860, 1
\bibitem[Liang \& Zhang(2005)]{2005ApJ...633..611L} Liang, E., \& Zhang, B.\ 2005, \apj, 633, 611
\bibitem[Liang et al. (2004)]{875} Liang, E. W., Dai, Z. G., Wu, X. F. 2004, ApJ, 606, L29
\bibitem[Liang et al. 2006]{2006ApJ...646..351L} Liang, E.~W., et al.\ 2006, \apj, 646, 351
\bibitem[Liang et al.(2007)]{2007arXiv0705.1373L} Liang, E.-W., Zhang, B.-B., \& Zhang, B.\ 2007, ApJ, in press(arXiv:0705.1373) (Paper II)
\bibitem[Lin et al.(2004)]{2004ApJ...605..819L} Lin, J.~R., Zhang, S.~N., \& Li, T.~P.\ 2004, \apj, 605, 819
\bibitem[Malesani et al.(2007)]{2007A&A...473...77M} Malesani, D., et al.\ 2007, \aap, 473, 77
\bibitem[M{\'e}sz{\'a}ros 2006]{2006RPPh...69.2259M} M{\'e}sz{\'a}ros, P.\ 2006, Reports of Progress in Physics, 69, 2259
\bibitem[M{\'e}sz{\'a}ros(2002)]{2002ARA&A..40..137M} M{\'e}sz{\'a}ros, P.\ 2002, \araa, 40, 137
\bibitem[Mangano et al. 2007]{2007arXiv0704.2235M} Mangano, V., et al.\ 2007, ArXiv e-prints, 704, arXiv:0704.2235
\bibitem[Meszaros \& Rees(1993)]{1993ApJ...405..278M} M\'{e}sz\'{a}ros, P., \& Rees, M.~J.\ 1993, \apj, 405, 278
\bibitem[Moderski et al.(2000)]{2000ApJ...529..151M} Moderski, R., Sikora, M., \& Bulik, T.\ 2000, \apj, 529, 151
\bibitem[Molinari et al.(2007)]{2007A&A...469L..13M} Molinari, E., et al.\ 2007, \aap, 469, L13
\bibitem[Mondardini et al. (2006)]{2006ApJ...648.1125M} Monfardini, A. et al. 2006, ApJ, 648, 1125
\bibitem[Nousek et al. 2006]{2006ApJ...642..389N} Nousek, J.~A., et al.\ 2006, \apj, 642, 389
\bibitem[O'Brien et al. 2006]{2006ApJ...647.1213O} O'Brien, P.~T., et al.\ 2006a, \apj, 647, 1213
\bibitem[O'Brien et al.(2006)]{2006NJPh....8..121O} O'Brien, P.~T., Willingale, R., Osborne, J.~P., \& Goad, M.~R.\ 2006b, New Journal of Physics, 8, 121
\bibitem[Panaitescu \& Kumar(2002)]{2002ApJ...571..779P} Panaitescu, A., \& Kumar, P.\ 2002, \apj, 571, 779
\bibitem[Panaitescu \& M{\'e}sz{\'a}ros(1999)]{1999ApJ...526..707P} Panaitescu, A., \& M{\'e}sz{\'a}ros, P.\ 1999, \apj, 526, 707
\bibitem[Panaitescu et al.(2006)]{2006MNRAS.369.2059P} Panaitescu, A., M{\'e}sz{\'a}ros, P., Burrows, D., Nousek, J., Gehrels, N., O'Brien, P., \& Willingale, R.\ 2006, \mnras, 369, 2059
\bibitem[Panaitescu(2005)]{2005MNRAS.363.1409P}Panaitescu, A.\ 2005, \mnras, 363, 1409
\bibitem[Panaitescu(2007a)]{2007MNRAS.380..374P} Panaitescu, A.\ 2007, \mnras, 380, 374
\bibitem[Piran(2005)]{2005RvMP...76.1143P} Piran, T.\ 2005, Reviews of Modern Physics, 76, 1143
\bibitem[Rhoads(1999)]{1999ApJ...525..737R} Rhoads, J.~E.\ 1999, \apj, 525,737
\bibitem[Rol et al.(2006)]{2006GCN..5555....1R} Rol, E., Jakobsson, P., Tanvir, N., \& Levan, A.\ 2006, GCN, 5555, 1
\bibitem[Sako, Harrison, \& Rutledge(2005)]{2005ApJ...623..973S} Sako, M., Harrison, F.~A., \& Rutledge, R.~E.\ 2005, \apj, 623, 973
\bibitem[Sari et al. 1998]{1998ApJ...497L..17S} Sari, R., Piran, T., \& Narayan, R.\ 1998, \apjl, 497, L17
\bibitem[Sari et al. 1999]{1999ApJ...519L..17S} Sari, R., Piran, T., \& Halpern, J.~P.\ 1999, \apjl, 519, L17
\bibitem[Sato et al.(2007)]{2007ApJ...657..359S} Sato, G., et al.\ 2007, \apj, 657, 359
\bibitem[Shen et al.(2006)]{2006MNRAS.371.1441S} Shen, R., Kumar, P., \& Robinson, E.~L.\ 2006, \mnras, 371, 1441
\bibitem[Soderberg et al.(2005)]{2005GCN..4186....1S} Soderberg, A.~M., Berger, E., \& Ofek, E.\ 2005, GCN, 4186, 1
\bibitem[Still et al.(2006)]{2006GCN..5226....1S} Still, A., et al.\ 2006, GCN, 5226, 1
\bibitem[Thoene et al.(2006)]{2006GCN..5373....1T} Thoene, C.~C., et al.\ 2006, GCN, 5373, 1
\bibitem[Thoene et al.(2007)]{2007GCN..6663....1T}Thoene, C. C., Perley, D. A. , \& Bloom, J. S. 2007, GCN 6663
\bibitem[Tominaga et al.(2007)]{2007ApJ...657L..77T} Tominaga, N., et al. \ 2007, \apjl, 657, L77
\bibitem[Troja et al.(2007)]{2007ApJ...665..599T} Troja, E., et al.\ 2007, \apj, 665, 599
\bibitem[Wang \& Dai(2006)]{2006MNRAS.368..371W} Wang, F.~Y., \& Dai, Z.~G.\ 2006, \mnras, 368, 371
\bibitem[Wei \& Lu(2000)]{2000ApJ...541..203W} Wei, D.~M., \& Lu, T.\ 2000, \apj, 541, 203
\bibitem[Wei \& Lu(2002a)]{2002} Wei, D.~M., \& Lu, T.\ 2002a, \mnras, 332,994
\bibitem[Wei \& Lu(2002b)]{2002AA}Wei, D. M. \& Lu, T. 2002, A\&A, 381, 731
\bibitem[Wijers\& Galama (1999)]{913} Wijers, R. A. M. J.\& Galama, T. J. 1999, ApJ, 523, 177
\bibitem[Willingale et al.(2004)]{2004MNRAS.349...31W} Willingale, R., Osborne, J.~P., O'Brien, P.~T., Ward, M.~J., Levan, A., \& Page, K.~L.\ 2004, \mnras, 349, 31
\bibitem[Willingale et al.(2007)]{2007ApJ...662.1093W} Willingale, R., et al.\ 2007, \apj, 662, 1093
\bibitem[Woosley \& Bloom(2006)]{2006ARA&A..44..507W_01} Woosley, S.~E., \& Bloom, J.~S.\ 2006, \araa, 44, 507
\bibitem[Wu et al. 2004]{917} Wu, X. F., Dai, Z. G., Liang, E. W. 2004, ApJ, 615, 359
\bibitem[Yamazaki et al.(2006)]{2006MNRAS.369..311Y} Yamazaki, R., Toma, K., Ioka, K., \& Nakamura, T.\ 2006, \mnras, 369, 311
\bibitem[Yost et al.(2003)]{2003ApJ...597..459Y} Yost, S.~A., Harrison, F.~A., Sari, R., \& Frail, D.~A.\ 2003, \apj, 597, 459
\bibitem[Zhang (2006)]{2006Natur.444.1010Z} Zhang, B.\ 2006, \nat, 444, 1010
\bibitem[Zhang \& M{\'e}sz{\'a}ros 2004]{2004IJMPA..19.2385Z} Zhang, B., \& M{\'e}sz{\'a}ros, P.\ 2004, International Journal of Modern Physics A, 19, 2385
\bibitem[Zhang 2007]{2007ChJAA...7....1Z} Zhang, B.\ 2007, Chinese Journal of Astronomy and Astrophysics, 7, 1
\bibitem[Zhang et al. (2004)]{2004ApJ...601L.119Z} Zhang, B., Dai, X., Lloyd-Ronning, N.~M., \& M{\'e}sz{\'a}ros, P.\ 2004, \apjl, 601, L119
\bibitem[Zhang et al. 2006]{2006ApJ...642..354Z} Zhang, B., Fan, Y.~Z., Dyks, J., Kobayashi, S., M{\'e}sz{\'a}ros, P., Burrows, D.~N., Nousek, J.~A., \& Gehrels, N.\ 2006, \apj, 642, 354
\bibitem[Zhang et al. 2007a]{2007ApJ...655..989Z} Zhang, B., et al.\ 2007a, \apj, 655, 989
\bibitem[Zhang et al. 2007b]{2007ApJ...655L..25Z} Zhang, B., Zhang, B.-B., Liang, E.-W., Gehrels, N., Burrows, D.~N., \& M{\'e}sz{\'a}ros, P.\ 2007b, \apjl, 655, L25
\bibitem[Zhang et al.(2007c)]{2007ApJ...666.1002Z} Zhang, B.-B., Liang, E.-W., \& Zhang, B.\ 2007c, \apj, 666, 1002

\end{thebibliography}
\end{document}